\documentclass[11pt]{article}
 
\usepackage[margin=1.in]{geometry}     

\usepackage{amsmath,amsthm,amsfonts,amssymb,amscd}
\usepackage[mathscr]{euscript}
\usepackage[dvipsnames]{xcolor}
	\usepackage{soul}

\usepackage[subfigure]{tocloft}
	\usepackage[toc,titletoc]{appendix}
\usepackage[square,numbers]{natbib}
	\usepackage{etoolbox,xstring}
	\makeatletter
	\patchcmd{\NAT@test}{\else\NAT@nm}{\else\NAT@nmfmt{\NAT@nm}}{}{}
	\let\NAT@up\scshape
	\makeatother
	\makeatletter
	\patchcmd{\NAT@test}{\else\NAT@nm}{\else\NAT@nmfmt{\NAT@nm}}{}{}
	\renewcommand{\NAT@nmfmt}{\expandafter\aliNAT@nmfmt\expandafter}
	\newcommand\aliNAT@nmfmt[1]{{%
  	\noexpandarg
  	\def~{}%
  	\edef\temp#1\edef\temp{\detokenize\expandafter{\temp}}%
 	 \begingroup\edef\x{\endgroup
   	 \noexpand\StrSubstitute{\temp}{\detokenize{etal}}}\x
    	{\textnormal{et\nobreakspace al}}[\tempa]%
  	\textsc{\tempa}}}
	\makeatother
\usepackage[T1]{fontenc}
\usepackage{titlesec}
	\titleformat{\section}{\large\bfseries\scshape}{\thesection.}{0.5em}{\centering}
	\titleformat{\subsection}{\large\bfseries}{\thesubsection.}{0.5em}{}
	\titleformat{\subsubsection}[runin]{\normalsize\bfseries}{\thesubsubsection.}{0.5em}{}[.]
\usepackage{fancyhdr}
\usepackage{bm}
\usepackage{bbm}
\usepackage{array}
	\newcolumntype{P}[1]{>{\centering\arraybackslash}p{#1}}		
	\newcolumntype{M}[1]{>{\centering\arraybackslash}m{#1}}
	\usepackage{multirow}
	\usepackage{booktabs}
\usepackage{float}
\usepackage{tikz}
	\usetikzlibrary{decorations.pathreplacing,angles,quotes,patterns,arrows}
	\usetikzlibrary{arrows.meta}

	\tikzset{interpolation/.pic = {
\draw[step=0.7cm,line width=0.35mm,black, fill=black!10!white] (0,0) grid (3.5,3.5) rectangle (0,0);
\draw[blue,dashed,line width=0.35mm] (0,0) -- (3.5,0) -- (3.5,3.5) -- (0,3.5) -- (0,0);
\draw[blue,dashed,line width=0.35mm] (0.75, 0) .. controls(1.375,1.25) .. (0.75, 3.5);
\draw[blue,dashed,line width=0.35mm] (1.5, 0) .. controls(1.75,1.25) .. (1.375, 3.5);
\draw[blue,dashed,line width=0.35mm] (2.25, 0) .. controls(2.05,1.25) .. (2.25, 3.5);
\draw[blue,dashed,line width=0.35mm] (2.875, 0) .. controls(2.375,1.25) .. (2.75, 3.5);
\draw[blue,dashed,line width=0.35mm] (0, 0.5) .. controls(2,1) .. (3.5, 0.5);
\draw[blue,dashed,line width=0.35mm] (0, 1.25) .. controls(2,1.375) .. (3.5, 1.125);
\draw[blue,dashed,line width=0.35mm] (0, 1.875) .. controls(2,1.6) .. (3.5, 1.875);
\draw[blue,dashed,line width=0.35mm] (0, 2.5) .. controls(2,2) .. (3.5,2.25);
\draw[-{Latex[length=2mm]},red,line width=0.25mm] (-0.2,0)--(-0.2,1.0);
\draw[-{Latex[length=2mm]},red,line width=0.25mm] (3.7,1.0)--(3.7,0.0);
\draw [-{Latex[length=2mm]},red,line width=0.25mm] (0,3.65) to [out=90,in=90, looseness=1.25] (0.75,3.6);
\draw [-{Latex[length=2mm]},red,line width=0.25mm] (0.75,-0.15) to [out=270,in=270, looseness=1.25] (1.5,-0.1);
\draw [-{Latex[length=2mm]},red,line width=0.25mm] (1.375,3.65) to [out=90,in=90, looseness=1.15] (2.25,3.6);
\draw [-{Latex[length=2mm]},red,line width=0.25mm] (2.25,-0.15) to [out=270,in=270, looseness=1.5] (2.875,-0.1);
\draw [-{Latex[length=2mm]},red,line width=0.25mm] (2.75,3.65) to [out=90,in=90, looseness=1.25] (3.5,3.6);
\filldraw [blue] (0,0) circle (2pt);\filldraw [blue] (0,0.5) circle (2pt);\filldraw [blue] (0,1.25) circle (2pt);\filldraw [blue] (0,1.875) circle (2pt);\filldraw [blue] (0,2.5) circle (2pt);\filldraw [blue] (0,3.5) circle (2pt);
\filldraw [blue] (0.75,0) circle (2pt);\filldraw [blue] (1.1,0.775) circle (2pt);\filldraw [blue] (1.22,1.32) circle (2pt);\filldraw [blue] (1.19,1.71) circle (2pt);\filldraw [blue] (1.08,2.23) circle (2pt);\filldraw [blue] (0.75,3.5) circle (2pt);
\filldraw [blue] (1.5,0) circle (2pt);\filldraw [blue] (1.66,0.85) circle (2pt);\filldraw [blue] (1.68,1.32) circle (2pt);\filldraw [blue] (1.66,1.68) circle (2pt);\filldraw [blue] (1.6,2.13) circle (2pt);\filldraw [blue] (1.375,3.5) circle (2pt);
\filldraw [blue] (2.25,0) circle (2pt);\filldraw [blue] (2.13,0.865) circle (2pt);\filldraw [blue] (2.1,1.31) circle (2pt);\filldraw [blue] (2.11,1.67) circle (2pt);\filldraw [blue] (2.14,2.1) circle (2pt);\filldraw [blue] (2.25,3.5) circle (2pt);
\filldraw [blue] (2.875,0) circle (2pt);\filldraw [blue] (2.58,0.8) circle (2pt);\filldraw [blue] (2.49,1.29) circle (2pt);\filldraw [blue] (2.49,1.69) circle (2pt);\filldraw [blue] (2.53,2.11) circle (2pt);\filldraw [blue] (2.75,3.5) circle (2pt);
\filldraw [blue] (3.5,0) circle (2pt);\filldraw [blue] (3.5,0.5) circle (2pt);\filldraw [blue] (3.5,1.125) circle (2pt);\filldraw [blue] (3.5,1.875) circle (2pt);\filldraw [blue] (3.5,2.25) circle (2pt);\filldraw [blue] (3.5,3.5) circle (2pt);
}}

\usepackage{graphicx,subfigure}
	\usepackage{wrapfig}
	\usepackage[font=footnotesize,labelfont=bf]{caption}
\usepackage{mathtools}
\usepackage{stmaryrd}
\usepackage{MnSymbol}
\usepackage{algorithm}
\usepackage{enumitem}
\usepackage{hyperref}
\usepackage[overload]{empheq}
	\hypersetup{colorlinks=True, linkcolor=blue, urlcolor=blue, citecolor=red}
\usepackage{cleveref}
\usepackage{upgreek}

\theoremstyle{plain}

\theoremstyle{definition}

\newtheorem{remark}{\scshape Remark}

\def\divv{\operatorname{div}}
\def\Omegaref{\Omega_{\mathrm{ref}}}
\def\Tref{\mathcal{T}_{\mathrm{ref}}}
\def\Vref{\mathcal{V}_{\mathrm{ref}}}
\def\xmin{x_{\mathrm{min}}}
\def\xmax{x_{\mathrm{max}}}

\def\ue{\mathsf{u}}
\def\pe{\mathsf{p}}
\def\Ee{\mathsf{E}}

\def\ze{\mathsf{z}}
\def\fe{\mathsf{f}}
\def\Qe{\mathsf{Q}}

\def\Ge{\mathsf{G}}
\def\Fe{\mathsf{F}}

\title{\Large\textbf{\textsc{A fast dynamic smooth adaptive meshing scheme with applications to compressible flow}}}
\author{
 {\normalsize \textbf{\textsc{Raaghav Ramani}}}
 \vspace{-.05 in}
\\{\small Department of Mathematics}
\vspace{-.05 in}
\\{\small University of California}
\vspace{-.05 in}
\\{\small Davis, CA 95616 USA}
\vspace{-.05 in}
\\{\small\textit{\url{rramani@math.ucdavis.edu}}} \and
 {\normalsize \textbf{\textsc{Steve Shkoller}}}
 \vspace{-.05 in}
\\{\small Department of Mathematics}
\vspace{-.05 in}
\\{\small University of California}
\vspace{-.05 in}
\\{\small Davis, CA 95616 USA}
\vspace{-.05 in}
\\{\small\textit{\url{shkoller@math.ucdavis.edu}}}
}

\date{\today}

\begin{document}

% make title and toc
\maketitle

\begin{abstract}
We develop a fast-running smooth adaptive meshing (SAM) algorithm for dynamic 
curvilinear mesh generation, which is based on a fast
solution strategy of the time-dependent 
Monge-Amp\`{e}re (MA) equation,  $\det \nabla \psi(x,t) = \Ge \circ\psi (x,t)$.  
The novelty of our approach is a new so-called \emph{perturbation formulation} of MA, which constructs the 
solution map $\psi$ via 
composition of a sequence of near-identity deformations of a reference mesh. 
Then, we formulate a new version of the deformation method \cite{DaMo1990} that results in a 
simple, fast, and high-order accurate numerical scheme and a dynamic SAM algorithm that is 
of optimal complexity when 
applied to time-dependent mesh generation for solutions to hyperbolic systems
such as the Euler equations of gas dynamics.  
We perform a series of challenging 2$D$ and 3$D$ mesh 
generation experiments for grids with large deformations, and demonstrate that SAM is 
able to produce smooth meshes 
comparable to state-of-the-art 
solvers \cite{Delzanno2008,Chacon2011}, while running approximately 200 times faster. 
The SAM algorithm is then coupled to a simple Arbitrary Lagrangian Eulerian (ALE) scheme for 
2$D$ gas dynamics. 
Specifically, we implement the $C$-method \cite{RaReSh2019a,RaReSh2019b} and develop a 
new ALE interface 
tracking algorithm for contact discontinuities. 
We perform numerical experiments for both the  Noh  implosion problem as well as a
 classical Rayleigh-Taylor instability problem. Results confirm that 
low-resolution  simulations using our SAM-ALE algorithm compare favorably 
with high-resolution uniform mesh runs.
\end{abstract}

{\small
\tableofcontents}

%Section for Introduction 
\section{Introduction} \label{sec:intro}

The efficiency of smooth moving-mesh  methods for numerical simulations of 
gas dynamics\footnote{Moving-mesh simulations  are often referred to as {\it adaptive simulations} and we shall use this terminology herein.} 
and related systems has been investigated in recent years 
\cite{AzIvTa2003,ZeBoTa2005,HeTa2012,HeTa2012b,PaSh2016,LuHuQi2019,DuTa2021,LiDuTa2022,DuTa2022};
however, to the best of our knowledge, compelling evidence of  the gain in 
efficiency relative to  fixed uniform-mesh simulations in multiple space dimensions has rarely been provided.
In a recent result \cite{LiDuTa2022},  the authors demonstrate
that low-resolution adaptive simulations are roughly 2-6 times faster than high-resolution uniform simulations of comparable 
quality; most results in this area focus on novel solution methodologies but not on the ultimate speed-up that may be gained by the algorithms that they produce. 
The papers cited above focus on one half of the moving-mesh methodology, namely,  the numerical discretization 
of the physical PDEs. They develop state-of-the-art high-resolution shock-capturing techniques, but use 
 well-established and somewhat standard meshing algorithms.   Our point-of-view is that it is essential to simultaneously develop both numerical methods for hyperbolic systems (for discontinuous solutions) as well as   novel 
meshing strategies.\footnote{This philosophy is in agreement with 
 \cite{PaSh2016}, in which the authors state that the main obstacle in their moving-mesh simulations 
is the lack of  a simple, robust, and efficient algorithm for dynamic and smooth 
adaptive mesh generation, particularly in 3$D$ geometries, and for multi-phase flows with unstable interfaces.}

Herein, we propose a novel and fast\footnote{We will demonstrate that our SAM algorithm is the first to be able to solve classical Rayleigh-Taylor problems on
coarse, but adaptive, grids faster than simulations on uniform grids.}   Smooth Adaptive Meshing 
(SAM) algorithm for  multi-$D$ simulations requiring mesh adaptivity.  
We present 
adaptive-simulation speed-up results for two classical but 
extremely challenging gas dynamic problems: the Noh shock implosion,  and the (highly unstable) Rayleigh-Taylor (RT) test.
For the Noh problem, our adaptive simulations are free of the numerical anomalies that are present in almost all reported results, while running approximately  6 times faster than a comparable uniform-mesh simulation.  
The  ten-fold speed-up
provided by SAM  for  the RT problem  is, to the best of our knowledge, the first of its kind.\footnote{Most attempts 
at using moving-mesh adaptivity to numerically simulate the RT instability 
result in runs that prematurely blow-up due 
to mesh tangling, meaning that those algorithms are not
sufficiently stable to provide a competitive speed-up factor.  Recent papers  \cite{Loubere2010,BaMaRiRiSh2016} instead focus on 
novel and sophisticated meshing techniques with the goal of simply simulating  the RT instability until the final simulation time without the code crashing; however, these meshing algorithms are currently too expensive to 
provide speed-up over uniform-mesh simulations. 
}

\subsection{Mesh refinement for multi-$D$ gas dynamics}

It is by now well-known that static uniform meshes are both inaccurate and inefficient at representing 
the dynamically evolving and interacting small-scale structures that appear in solutions to nonlinear 
conservation laws in multiple space dimensions.  
Adaptive mesh refinement (AMR) via $h$-adaptivity is the most well-developed refinement 
technique and is used in many commercial codes
\cite{ATHENA2008,ENZO2014,FLASH2000,RAGE2008}. 
However, the dyadic refinement at the heart of AMR schemes results in an artificially discontinuous
transition from coarse-scale to fine-scale representation of 
numerical solutions on AMR meshes.     
Several theoretical and numerical 
studies \cite{BergerColella1989,VichnevetskyTurner1991,LongThuburn2011} have 
demonstrated the spurious 
reflection, refraction, and scattering of waves that propagate across discontinuously refined grids.  
Many problems in gas dynamics, such as strong blast waves, self-similar implosions, and 
unstable contact discontinuities are 
extremely sensitive to small perturbations; spurious wave reflections produce corrupted numerical 
solutions, with the anomalies persisting, or even worsening, as the AMR mesh is 
globally refined \cite{FLASH2000,Timmes2005}.     

On the other hand, Lagrangian-type schemes are well-known to produce highly 
distorted or {tangled} meshes i.e. some cells in the grid are 
non-convex or have folded over, at which point the simulation breaks down. 
Arbitrary Lagrangian Eulerian (ALE) methods aim to mitigate the problem of mesh tangling.
\emph{Indirect} ALE methods are somewhat \emph{ad hoc}, and current rezoning strategies are 
heuristic in nature \cite{Knupp2002,Loubere2010}.  In this work, we consider the \emph{direct} ALE approach, in which 
an adaptive mesh is generated directly without any initial Lagrangian phase or subsequent mesh rezoning.

\subsubsection{Adaptive mesh redistribution}
Our SAM scheme falls under the category of $r$-refinement schemes, 
or {adaptive mesh redistribution} methods. In contrast to Lagrangian-rezone methods, a grid is generated via 
a user-prescribed {monitor function} which determines the grid size and orientation. High-resolution 
representation of numerical solutions is obtained by defining the monitor function appropriately, e.g., using solution derivatives. 
Moreover, the adaptive grids can be 
generated to {align} with the geometry of evolving fronts \cite{Huang2005}, and to naturally 
capture self-similar dynamics or scale-invariant structures \cite{BuHuRu1996,BuLePi2001}.  

Historically, the first $r$-refinement methods were based on the variational approach, examples of which include the 
equipotential \cite{Winslow1966}, variable diffusion \cite{Winslow1981}, cost function \cite{BrSa1982}, and harmonic 
mapping \cite{Dvinsky1991} methods.  
The variational approach also currently appears to be the method of choice for use in direct ALE schemes, several of which
employ the popular MMPDE framework \cite{HuRu1998,LiTaZh2001}. 
These variational methods, however, require the accurate numerical 
solution of a coupled set of $d$ complicated nonlinear 
auxiliary PDEs in $\mathbb{R}^d$, for which simple, fast, and accurate
algorithms are in general not available.  For these reasons, among others, $r$-refinement methods have 
yet to become incorporated into large scale established hydrodynamics codes.  
See, for example,  \cite{HuangRussell2010,BuHuRu2009,Delzanno2008,Chacon2011} and the 
references therein for thorough reviews of $r$-adaptive methods and their associated 
difficulties.

%The auxiliary PDEs governing the adaptive mesh usually have intimate connections with fundamental problems in differential geometry.  
%Historically, the first $r$-adaptive methods were based on the variational approach,  which 
%provide a solid mathematical foundation.  
%These methods generate a mesh by minimizing a convex combination of 
%weighted functionals measuring the smoothness, orthogonality, and volume variation of the 
%grid. Important examples of the variational approach include the 
%equipotential \cite{Winslow1966}, variable diffusion \cite{Winslow1981}, cost function \cite{BrSa1982}, and harmonic mapping \cite{Dvinsky1991} methods, among others.  
%However, such schemes require the numerical solution of a system of $d$ coupled nonlinear equations for mesh generation in $\mathbb{R}^d$, which can be very difficult to obtain with 
%high accuracy. 
%They also sometimes require several user-provided parameters that need to be fine-tuned for each problem, and an inappropriate 
%choice of parameter can lead to non-convex functionals and ill-posed problems 
%\cite{KrThHa1986}.   The use of multiple grid property measures 
%in the minimization functional means that none of the measures are individually minimized, which can potentially lead to skewed or tangled grids.  Elliptic grid generation methods 
%\cite{ThThMa1974,Knupp1996} resolve some of these issues, but still require the 
%solution of $d$ coupled nonlinear elliptic equations.  

\subsubsection{Prescribing the Jacobian determinant}

The fundamental guiding principle for smooth adaptive mesh generation is control of
the local cell volume of the adaptive grid. In the time-dependent multi-$D$ setting, we assume that we have a
 given smooth positive \emph{target Jacobian function} $\Ge(y,t)$ describing the size of the cells in the moving target adaptive mesh.  
 We then seek to construct a
diffeomorphism $\psi(x,t)$ mapping a fixed reference mesh to the target mesh by requiring that 
$\det \nabla \psi (x,t)= \Ge(\psi(x,t),t)$.   
A semi-discretization in time $t = t_k$, where $k$ is the time-index, yields a sequence of 
nonlinear elliptic equations of Monge-Amp\`{e}re (MA) type 
\begin{equation}\label{MA-intro-1}
\det \nabla \psi_k (x) = \Ge_k (\psi_k(x))  \,,
\end{equation}
where each $\Ge_k$ is again a given positive target Jacobian function. 

Solutions to the MA equation are unique in 1$D$.
For dimension $d \geq 2$, however, the single scalar 
MA equation is insufficient to uniquely determine $\psi$. 
The question then becomes how to choose a particular solution $\psi$ that is in 
some sense optimal.  One such choice that has received a great deal of attention in recent years is the Monge-Kantorovich (MK) 
formulation based on optimal transport, in which a map $\psi$ is (uniquely \cite{Brenier1991,Caffarelli1990}) constructed to minimize 
the $L^2$ displacement $||\psi(x) - x ||_{L^2}$.  
This is attractive from a numerical perspective, since smaller grid velocities can reduce interpolation and other numerical errors \cite{LiPe1997}.

On the other hand, the MK formulation results in a fully nonlinear second order elliptic equation, whose numerical 
solution is difficult to obtain.  One approach is 
to consider a parabolized formulation by introducing an artificial time variable $\tau$ then iterating until a steady state is reached 
\cite{SuWiRu2011,BrBuPiCu2014,McCoBu2018,WeBrBuCu2016}. In this case, the 
Jacobian constraint is only satisfied in the asymptotic limit $\tau \to \infty$, and many iterations may be 
required to obtain a sufficiently accurate solution, particularly 
for target meshes with large deformations.  
An alternative, fully nonlinear approach using preconditioned Newton-Krylov solvers is designed in \cite{Delzanno2008,Chacon2011}, leading 
to a robust, scalable algorithm that is, to the best of our knowledge, the state-of-the-art in the field 
(see also the recent papers \cite{BuRuWa2015,BuMcCo2018}).  
However, the Newton-Krylov iterative approach is still relatively slow for our ultimate goal of efficient
adaptive gas dynamics simulations; specifically, its implementation in our ALE scheme (to be described below) leads to 
adaptive mesh simulations with computational runtimes greater than would otherwise be
obtained with a uniformly high-resolution mesh, thereby defeating the purpose of 
using an adaptive meshing scheme in the first place.

\subsection{Fast Smooth Adaptive Meshing}
In contrast to the MK approach, we 
construct a map $\psi_k$ satisfying \eqref{MA-intro-1}  
with the aim of optimizing for the efficiency of the resulting numerical algorithm, which we refer to as SAM.   
The key to our fast SAM algorithm is a new \emph{perturbation formulation} of \eqref{MA-intro-1} 
along with a new formulation and 
implementation of the {deformation method} \cite{DaMo1990}.

Specifically, the perturbation formulation constructs each map  $\psi_{k+1}$ as the image of the map 
$\psi_k$ acting on a \emph{near identity deformation} $\delta \psi_{k+1} \approx \mathrm{id} $ of a fixed 
 reference mesh $\Omegaref$. 
The formulation on $\Omegaref$ is crucial, since it enables the use of, at each time-step $t_k$, the 
same numerical solvers for the mesh PDEs\footnote{This is in contrast with other methods \cite{Liao1995,Grajewski2010} which require finite-element 
solvers with costly recalculation (at each time-step of a dynamic simulation) of the mass and stiffness 
matrices, as well as complicated interpolation procedures.}.  
This, in turn, produces a code with a simple modular structure so that the basic mesh 
redistribution procedure is developed entirely in the static setting on $\Omegaref$, then 
``bootstrapped'' to form a dynamic scheme.  
The same principle also yields an algorithm
for efficiently generating smooth meshes with very large zoom-in factors, which allows us to obtain 
high-resolution representation of small-scale structures with few total number of mesh points. 

The mesh redistribution algorithm we propose is a new version of the deformation 
 method \cite{LiaoAnderson1992,LiaoLiuJi1998,Liao2001}, which 
constructs a solution to the nonlinear Monge-Amp\`{e}re equation 
via a single elliptic solve for a linear Poisson problem, along with the solution of a 
system of transport equations for a flowmap $\eta(x,\tau)$ between 
pseudo-time $\tau = 0$ and $\tau=1$. 
There are at least two advantages of this 
new deformation method: the first is that the algorithm can be made fully automated with no user-prescribed 
parameters; the second is that 
costly and often complicated interpolation procedures are not required.  
We design a simple, fast, stable, and high-order accurate method 
using an efficient spectral solver with boundary smoothing for the Poisson equation, and 
standard RK4 time integration with high-order linear upwind differencing for the transport equations. 
A key implication of our numerical design choices is a consistency between the stability conditions for the 
transport problem in SAM  and the physical time-step in an ALE gas dynamics simulation.  
As we shall demonstrate, this consistency results in a dynamic SAM algorithm with 
optimal complexity for hyperbolic systems.

Our SAM algorithm is approximately 200 times faster than the MK 
nonlinear solvers \cite{Delzanno2008,Chacon2011}, and the 
computed numerical solutions exhibit both higher accuracy  as well as better convergence 
rates under global mesh refinement.  
We perform a number of challenging mesh generation experiments designed to 
replicate flows with high vorticity and large deformations, and demonstrate that the meshes 
produced with our dynamic SAM scheme are 
smooth and accurate. 
For example, we are able to generate smooth moving meshes that resolve around a complex 
3$D$ swirling helical-type curve at $256^3$ resolution with only a serial implementation on a laptop 
computer and without any specific and sophisticated algorithmic 
optimizations (see \Cref{subsec:numerics-3D-helical}).  

\subsection{Application to ALE gas dynamics}

To demonstrate the efficacy of our SAM scheme in practical applications, we formulate a
simple coupled SAM-ALE method for 2$D$ gas dynamics.  
Several moving-mesh methods for the 2$D$ Euler system have been developed based on the MMPDE approach 
and finite volume (FV) and finite element (FE) methods \cite{TaTa2003,Tang2005}.  A formulation on smooth tensor 
product meshes enables  the use of finite difference (FD) methods, which are both simpler and more efficient than 
FV and FE methods\footnote{FV schemes are 4 times more expensive than FD schemes in 2$D$, and 9 times more expensive in 3$D$ 
\cite{Wang2008}.}, and have been investigated in several recent papers \cite{Nonomura2010,Jiang2014,LiDuTa2022}. 
In this work, we further develop the $C$-method \cite{RaReSh2019a,RaReSh2019b}, a simplified
WENO-based solver with space-time smooth nonlinear artificial viscosity and 
explicit tracking of material interfaces. 

Special care is given to the so-called \emph{geometric conservation law} (GCL), and  we show that our nonlinear 
WENO reconstruction procedure respects the free-stream preservation property on adaptive meshes. 
The $C$-method dynamically tracks the location and geometry of evolving fronts, and 
is used to add both directionally isotropic and anisotropic artificial viscosity to shocks and contacts. 
Herein, we implement the $C$-method in the ALE context and introduce a new ALE front-tracking algorithm for contact 
discontinuities, which we subsequently use to construct suitable target Jacobian functions for SAM. 
Previous studies have mainly investigated target Jacobian functions constructed based on interpolation 
errors \cite{HuSu2003,Huang2005}, or weighted combinations of solution gradient estimates \cite{DaZe2010}, which sometimes fail 
to capture small scale vortical structures \cite{TaTa2003}. 
Our simple ALE front-tracking algorithm allows us to generate smooth adaptive meshes that capture  
 small scale Kelvin-Helmholtz roll-up  zones in unstable RT problems. 
We apply our coupled SAM-ALE scheme to two challenging test problems, namely the Noh implosion and RT
instability. For the Noh problem, we find that the $50 \times 50$ SAM-ALE solution is more accurate than the $200 \times 200$ uniform 
solution, while running approximately 6 times faster. Moreover, the SAM-ALE solution is completely free of spurious 
numerical anomalies, such as lack of symmetry, unphysical oscillations, and wall-heating.  For the RT problem, we find that the
$64 \times 128$ SAM-ALE solution is comparable to the $256 \times 512$ uniform solution, while running 10 times faster. 

  \subsection{Outline}
  \Cref{sec:preliminaries} introduces notation and definitions that will be used throughout the paper. 
  In \Cref{sec:SAM-static}, we develop the basic SAM algorithm for static mesh generation, upon which we shall build our dynamic 
  scheme. We show that our scheme is high-order accurate and benchmark the algorithm against the 
  MK scheme.  In \Cref{sec:SAM-dyn}, we consider dynamic mesh generation and 
  introduce the perturbation formulation of the MA system. 
  We then perform,  in \Cref{sec:mesh-exp}, a series of challenging mesh generation experiments to demonstrate the capabilities of the 
  scheme. In \Cref{sec:ALE}, we formulate a simple coupled SAM-ALE scheme for the 2$D$ compressible Euler system, and describe 
  some aspects of our numerical method. In \Cref{sec:ALE-sims}, we apply SAM-ALE to the Noh and 
  RT test problems and compare the results  with low-resolution and high-resolution uniform solutions. 
  Finally, in \Cref{sec:conclusion}, we provide some brief concluding remarks. 
Three sections are included in the Appendices: the first concerns the $C$-method regularization for the 2$D$ ALE-Euler system, the 
second describes a simple boundary smoothing technique, and the third provides a machine comparison test for the purposes of 
benchmarking our SAM algorithm. 
  
 \section{Preliminaries}\label{sec:preliminaries}
      
 \subsection{Domains, meshes, and mappings}\label{subsec:meshes}

 The focus of this work is mesh adaptation on 2$D$ rectangles and we provide the mathematical 
 formulation and numerical implementation 
 details of our mesh adaptation strategy in this setting.  However, all of our meshing algorithms can be extended 
 to 3$D$ cuboids\footnote{In fact, our algorithms can also be applied in arbitrary 
 complex geometry (see \Cref{fig:SAM-unstructured} for a preliminary result), though their numerical 
 implementations are more involved.}, and 
 we show in \Cref{subsec:numerics-3D-helical} results from a mesh generation experiment 
 modeling three-dimensional swirling flow.

 Let $\Omegaref \subset \mathbb{R}^2$ be a \emph{reference} domain with coordinates $x = (x^1,x^2) \in \Omegaref$, and 
 given explicitly by the rectangle $\Omegaref = (\xmin^1 \,, \xmax^1) \times  (\xmin^2 \,, \xmax^2)$.  
 The outward pointing unit normal vector to the boundary $\partial \Omegaref$ is defined everywhere on $\partial \Omegaref$, except at the 
 four corners, and is denoted by $\nu$. 
 The domain $\Omegaref$ is also sometimes referred to in the literature as the \emph{logical} or \emph{computational} domain, and in the 
 context of ALE gas dynamics, the \emph{ALE} domain.  
 
 We denote by $\Omega \subset \mathbb{R}^2$ the \emph{physical} or \emph{Eulerian} domain, with coordinates $y = (y^1 \,, y^2) \in \Omega$ 
 and boundary $\partial \Omega$.  We assume that $\Omegaref$ and $\Omega$ represent the same mathematical domain i.e. 
 $\Omegaref = \Omega$.  
 The purpose of using the different notations $\Omegaref$ and $\Omega$ is to clearly distinguish between functions defined on each of these 
 domains, as we shall explain in the next subsection.  We let $\mathrm{id}: \Omegaref \to \Omega$  
 denote the identity map, i.e. $\mathrm{id}(x) = x$. 
 
\begin{figure}[ht]
\centering
\begin{tikzpicture}
\draw[step=0.7cm,black] (0,0) grid (3.5,3.5);
\draw (7.5,0) -- (11,0) -- (11,3.5) -- (7.5,3.5) -- (7.5,0);
\draw[black] (8.25, 0) .. controls(8.875,1.25) .. (8.25, 3.5);
\draw[black] (9, 0) .. controls(9.25,1.25) .. (8.875, 3.5);
\draw[black] (9.75, 0) .. controls(9.55,1.25) .. (9.75, 3.5);
\draw[black] (10.375, 0) .. controls(9.875,1.25) .. (10.25, 3.5);
\draw[black] (7.5, 0.5) .. controls(9.5,1) .. (11, 0.5);
\draw[black] (7.5, 1.25) .. controls(9.5,1.375) .. (11, 1.125);
\draw[black] (7.5, 1.875) .. controls(9.5,1.6) .. (11, 1.875);
\draw[black] (7.5, 2.5) .. controls(9.5,2) .. (11,2.25);
\draw[-latex,line width=0.4mm] (4,2.5) --node[above] {\Large $\psi$} (7,2.5);
\node at (1.75,-0.25) { \footnotesize $x^1$};
\node at (-0.25,1.75) { \footnotesize $x^2$};
\node at (9.25,-0.25) { \footnotesize $y^1$};
\node at (7.25,1.75) { \footnotesize $y^2$};
\node at (1.75,-1) {$\Tref \subset \Omegaref$};
\node at (9.25,-1) {$\mathcal{T} \subset \Omega$};
\end{tikzpicture}
\caption{The uniform $m\times n$ mesh $\Tref$ and the adaptive $m \times n$ mesh $\mathcal{T} = \psi(\Tref)$.} \label{fig:mesh-schematic-static}
\end{figure}
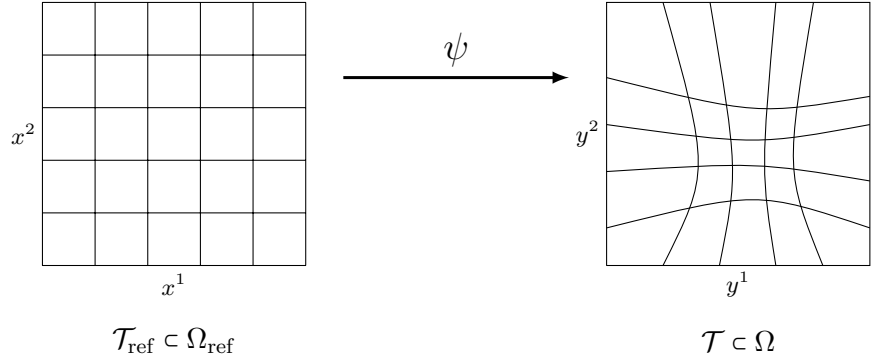

 We discretize $\Omegaref$ and $\Omega$ with $m+1$ nodes in the horizontal direction, and $n+1$ nodes in 
 the vertical direction. and denote by $\Tref$ and 
$\mathcal{T}$ the grids (or meshes) on each of these domains.   Each of these meshes contains $N = m \times n$ cells. 
The domain $\Omegaref$ is discretized uniformly, and we refer to
$\Tref$ as the \emph{reference} or \emph{uniform} mesh.  
The \emph{physical} or \emph{adaptive} 
mesh $\mathcal{T}$ is \emph{a priori} unknown and will be generated through a meshing scheme. The mesh $\mathcal{T}$ 
is not assumed to be uniform, but contains the same number of cells and 
retains the same mesh connectivity structure as the uniform mesh $\Tref$ c.f. \Cref{fig:mesh-schematic-static}.   
The fixed uniform mesh spacing is denoted by $\Delta x = (\Delta x^1 \,, \Delta x^2)$. 

The physical domain $\Omega$ can also be discretized uniformly with a uniform mesh $\mathcal{U}$.  
Since $\Omegaref = \Omega$, the meshes 
$\mathcal{U}$ and $\Tref$ are identical. We stress, however, that functions defined on each 
of these meshes are very different. 

The mesh $\mathcal{T}$ will be the image of $\Tref$ under the action 
of a suitable map $\psi : \Omegaref \to \Omega$.  The map $\psi$ is bijective, continuously differentiable, and has a continuously 
differentiable inverse 
$\psi^{-1} : \Omega \to \Omegaref$ i.e. $\psi$ is a smooth diffeomorphism.  Our SAM scheme solves for the map 
$\psi$ by prescribing its Jacobian determinant, as we shall explain in \Cref{sec:SAM-static,sec:SAM-dyn}. 
Nodes in $\mathcal{T}$ on the boundary $\partial \Omega$ will be allowed to move tangential to the boundary, with the exception of the 
four nodes at the corners of $\Omega$, which must remain fixed.  

In the dynamic setting, we consider $\psi$ to be a time-dependent map $\psi : \Omegaref \times [0,T] \to \Omega$ where, for each 
$t \in [0,T]$, the map $\psi(\cdot,t) : \Omegaref \to \Omega$ is a smooth diffeomorphism with prescribed Jacobian c.f. 
\Cref{fig:mesh-schematic-dynamic}. 
  
\subsection{Eulerian and ALE variables}
A \emph{physical} or \emph{Eulerian} function (scalar, vector-valued, or tensor) is defined on 
$\Omega$ and denoted with the upright mathematical font 
$ \fe : \Omega \to \mathbb{R}^k$.  For a time-dependent function $ \fe : \Omega \times [0,T] \to \mathbb{R}^k$ we shall write 
$\fe(y,t)$.  

Since $\psi$ maps $\Omegaref$ to $\Omega$, we write $y = \psi(x,t)$ for $(x,t) \in \Omegaref \times [0,T]$. Given an Eulerian variable 
$\fe : \Omega \to \mathbb{R}^k$, we define its \emph{computational} or \emph{ALE} counterpart $f : \Omegaref \times [0,T] \to \mathbb{R}^k$ 
by
\begin{equation}\label{composition}
{f}(x,t) = \left[ \fe \circ \psi \right](x,t) = \fe(\psi(x,t),t) \,, \quad \forall (x,t) \in \Omegaref  \times [0,T] \,.
\end{equation}
We shall also denote the function composition in \eqref{composition} by $\fe \circ \psi$. When there is no confusion, we omit 
the function arguments and write $\fe$ or $f$. 

In the discrete setting, computational variables are defined at the nodal points of the uniform reference mesh $\Tref$. 
Physical/Eulerian variables, on the other hand, can be defined on either the adaptive mesh $\mathcal{T}(t)$ or the uniform mesh 
$\mathcal{U}$ on $\Omega$. 
  
\subsection{Derivatives and important geometric quantities}
We denote spatial derivatives on $\Omegaref$ and $\Omega$ by 
$$
\partial_i = \frac{\partial}{\partial x_i}  \quad \text{and} \quad D_{i} = \frac{\partial}{\partial y_i} \,, 
$$
respectively. 
Higher order derivatives are then denoted in the standard fashion, e.g. $\partial_{ij} = \partial_i \partial_j$.  
We use the notation $\nabla = (\partial_1 \,, \partial_2)^\mathcal{T}$ and 
$D = (D_1 \,, D_2 )^\mathcal{T}$ for the 
gradient operators with respect to $x$ and $y$ coordinates, respectively.  The Laplacian operator on 
$\Omegaref$ is  $\Delta = (\partial_1^2 + \partial_2^2)$.   The operator $\Delta$ should not be 
confused with the 
discrete uniform mesh spacing $\Delta x^i$. 

The time derivative of a function $f$ is written as $\partial_t f$, or sometimes with the subscript notation $f_t$.  
Throughout, we shall use Einstein's summation convention wherein a repeated index in the same term indicates summation over all 
values of that index.  We shall also use the standard Kronecker delta symbol $\delta^i_j$.  

We now introduce the following important geometric quantities, all defined on $\Omegaref \times [0,T]$:   
\begin{subequations}
\label{deformation}
\begin{alignat}{2}
\mathcal{A} &= \left[ \nabla \psi \right]^{-1}  \quad && \text{(inverse of the deformation tensor)} \,, \label{deformation-tensor1} \\
\mathcal{J} &= \det \nabla \psi  \quad && \text{(Jacobian determinant)} \,,  \label{Jacobian-determinant1}\\ 
a &= \mathcal{J} \mathcal{A}  \quad && \text{(cofactor matrix of the deformation tensor)} \,. \label{cofactor1}
\end{alignat}
\end{subequations}
We assume that there exists $\varepsilon > 0$ such that 
\begin{equation*}
\mathcal{J}(x,t) \geq \varepsilon > 0\,, \text{ for every } (x,t) \in \Omegaref \times [0,T] \,.
\end{equation*}
Thus, the Jacobian determinant in 2$D$ reads 
$$
\mathcal{J}(x,t) = \partial_1 \psi^1 \, \partial_2 \psi^2 - \partial_1 \psi^2 \, \partial_2 \psi^1 \,.
$$
For a matrix $M = (M^j_i)$, the subscript $i$ indexes the columns of $M$, while the superscript $j$ indexes 
the rows. 

By explicit computation, we can verify the so-called Piola identity, which states that the columns of the cofactor matrix are divergence-free:
\begin{equation}\label{Piola}
\partial_j a_i^j = 0 \,, \text{ for } i=1,2 \,.
\end{equation}

Given an Eulerian variable $\fe(y,t)$ and its ALE counterpart $f(x,t)$, we use the chain rule
to compute 
\begin{equation}\label{spatial-deriv_coordinate}
D_i \fe(y,t) =  \frac{1}{\mathcal{J}(x,t)} a_i^j(x,t) \, \partial_j {f}(x,t) = \frac{1}{\mathcal{J}} \partial_j ( a_i^j {f} ) \,,
\end{equation}
where we have used the Piola identity \eqref{Piola} in the second equality. 
Using \eqref{spatial-deriv_coordinate} and the chain rule again, we have that
\begin{equation}\label{time-deriv_coordinate}
\partial_{t} \fe(y,t) =  \partial_t {f} - \frac{1}{\mathcal{J}}  a_i^j   \psi_t^i \, \partial_j {f} \,,
\end{equation}
where $\psi_t(x,t) \equiv \partial_t \psi(x,t)$ is the \emph{mesh velocity}.

\subsection{Computational platform and code optimization}
All of the algorithms in this work were coded in Fortran90, and 
all of the numerical simulations performed were run on a Macbook Pro laptop with an Apple M1 pro processor 
and 32GB of RAM. The operating system is macOS Ventura 13.1, and the gfortran compiler is used. 
The codes for the numerical methods described in the paper are implemented in the same programming framework, but are not 
otherwise specially optimized, apart from specific calculations described in the paper. 
The same input, output, and timing routines are used in all of the codes. 
This consistency allows for a reliable comparison of the different algorithms
and their associated imposed computational burdens.

% section for static adaptive meshing
\section{Fast static adaptive meshing}\label{sec:SAM-static}

\subsection{Mathematical formulation of static mesh generation}\label{subsec:static}

We construct an adaptive mesh $\mathcal{T}$ as the image of the uniform mesh $\Tref$ under the action 
of a suitable smooth diffeomorphism $\psi : \Omegaref \to \Omega$ c.f. \Cref{fig:mesh-schematic-static}.  
Our objective is to compute the map $\psi$ by prescribing its Jacobian determinant
$\mathcal{J}(x) = \det \nabla \psi(x)$. Specifically, given a strictly positive 
\emph{target Jacobian function}
$\Ge : \Omega \to \mathbb{R}^+$, the map $\psi$ is found as a solution to 
the following nonlinear nonlocal Monge-Amp\`{e}re (MA) equation
\begin{subequations}\label{MA-eqn}
\begin{alignat}{2}[left = \empheqlbrace\,]
\mathrm{det} \nabla \psi (x) &= \Ge \circ \psi (x) \,, \quad && x \in \Omegaref    \label{MA-eqn-a}  \\
\psi (x) \cdot \nu &= x \cdot \nu \,, && x \in \partial \Omegaref    \label{MA-eqn-b}
\end{alignat}
\end{subequations}
with $\nu$ the unit outward normal to the boundary $\partial \Omegaref$. 

The function $\Ge$ is a user prescribed  or constructed function that compresses the mesh 
in regions where $\Ge$ is small, and expands the mesh in regions where $\Ge$ is large. Note 
that $\Ge$ is a physical target Jacobian function
defined on the physical domain $\Omega$.   Assuming that a map $\psi$ 
satisfying \eqref{MA-eqn} is found, the function $\Ge$ then describes 
the size of the cells in $\mathcal{T}$.  Let $\mathcal{V}$ denote a 
cell in $\mathcal{T}$, and $\Vref = \psi^{-1}(\mathcal{V})$ the uniform cell in $\Tref$ 
mapped to $\mathcal{V}$ by $\psi$. 
If $\Ge$ is sufficiently smooth, a Taylor series argument shows that 
\begin{equation*}
|\mathcal{V}| \coloneqq \int_{\mathcal{V}} \,\mathrm{d}y = \int_{\Vref} \det \nabla \psi(x) \,\mathrm{d}x  = |\Vref| \cdot \Ge ( \psi (x_c) ) + \mathcal{O}( |\Delta x|^2 ) \,,
\end{equation*}
where $x_c$ denotes the cell center of $\Vref$.  Thus, the value of $\Ge$ in $\mathcal{V}$ is a scaling factor that 
scales the uniform cell volume $|\Vref| = \Delta x^1 \Delta x^2$ to the volume $|\mathcal{V}|$, up to some spatially fixed constant 
of order $\mathcal{O}(|\Delta x|)$.  

It is convenient to formulate the problem for the inverse map $\phi = \psi^{-1}$, which is found as a solution to
\begin{subequations}\label{MA-inv-eqn}
\begin{alignat}{2}[left = \empheqlbrace\,]
\mathrm{det} D \phi (y) &= \frac{1}{\Ge (y)} \,, \quad && y \in \Omega    \label{MA-inv-eqn-a}  \\
\phi (y) \cdot \nu &= y \cdot \nu \,, && y \in \partial \Omega  \,.   \label{MA-inv-eqn-b}
\end{alignat}
\end{subequations}

For a solution to exist for \eqref{MA-eqn}, the function $\Ge$ is required to 
satisfy the \emph{solvability condition} 
\begin{equation}\label{MA-eqn-solvability}
\int_\Omega \frac{1}{\Ge(y)} \, \mathrm{d}y = \int_{\Omegaref} \frac{ \det \nabla \psi(x) }{\Ge \circ \psi (x) } \,\mathrm{d}x = |\Omegaref| = |\Omega| \,.
\end{equation}

If \eqref{MA-eqn-solvability} holds, then the system 
\eqref{MA-eqn} admits an infinitude of solutions. The question then becomes how to construct
a solution $\psi$ that is in some sense optimal.  Our primary concern in this work is the 
development of a fast-running algorithm that can be 
easily implemented within an ALE framework for 
hydrodynamics simulations.   
We next describe a simple and efficient procedure 
for constructing a solution to \eqref{MA-eqn}.

\subsection{The basic mesh generation procedure}\label{subsec:alg1}

The key to our fast-running algorithm is the reduction of the nonlinear
equation {\eqref{MA-eqn}} to a simple linear Poisson solve and transport equation solve.
Our approach is motivated, as in  
\mbox{\cite{LiaoAnderson1992,LiaoLiuJi1998,Grajewski2009}},  by the deformation method
of \mbox{\citet{DaMo1990}}.
Specifically, a solution to \eqref{MA-eqn} is obtained by the five step construction provided in \Cref{static-alg1}.
We refer to this algorithm as SAM or, in the context of time-dependent meshing, \emph{static} SAM.   
\begin{algorithm}[!htb]
\caption{\textbf{\textsc{: static SAM}}}\label{static-alg1}
\begin{enumerate}[itemsep=0em,leftmargin=2.0cm,label=\textbf{Step \arabic* :}]

\item 
Assume that the physical target Jacobian function $\Ge : \Omega \to \mathbb{R}^+$ is given and 
satisfies the solvability condition
\begin{equation}\label{alg1-solvability}
\int_{\Omega} \frac{1}{\Ge(y)} \,\mathrm{d}y = |\Omega| \,,
\end{equation}
and let $\Fe(y) = 1/\Ge(y)$.  In practice, we are usually given an auxiliary target Jacobian function 
$\bar{\Ge} : \Omega \to \mathbb{R}^+$ that \emph{does not} satisfy \eqref{alg1-solvability}, and we define $\Ge$ and $\Fe$ by 
the following normalization procedure: 
\vspace{-0.5em}
\begin{equation*}
\bar{\Fe}(y) = \frac{1}{\bar{\Ge}(y)} \quad  \longrightarrow  \quad \Fe(y) =  |\Omega| \frac{\bar{\Fe}(y)}{ \int_{\Omega} \bar{\Fe}(y) \,\mathrm{d}y } \quad \longrightarrow \quad
\Ge(y) = \frac{1}{\Fe(y) }\,.
\end{equation*}
%\begin{align*}
%\bar{F}(y) &= 1/\bar{G}(y) \,, \\
%F(y) &=  |\Omega| \frac{\bar{F}(y)}{ \int_{\Omega} F(y) \,\mathrm{d}y } \,, \\
%G(y) &= 1/F(y) \,.
%\end{align*}

\item Solve the following linear Poisson equation with homogeneous Neumann boundary conditions for the \emph{potential} 
$\Phi : \Omegaref \to \mathbb{R}$
\vspace{-0.5em}
\begin{subequations}\label{alg1-Poisson}
\begin{alignat}{2}[left = \empheqlbrace\,]
\Delta \Phi(x) &= \Fe \circ \mathrm{id}(x) -1 \,, \quad && x \in \Omegaref    \label{alg1-Poisson-a}  \\
\nabla \Phi (x) \cdot \nu &= 0 \,, && x \in \partial \Omegaref    \label{alg1-Poisson-b}
\end{alignat}
\end{subequations}

\item Define the velocity $\overline{w} : \overline{\Omegaref} \to \mathbb{R}^2$ as 
\vspace{-0.5em}
\begin{equation}\label{alg1-ubar}
\overline{w}(x) = \nabla \Phi(x) \,.
\end{equation}

\item Solve the following system of transport equations for the \emph{flowmap}  
$\eta : \overline{\Omegaref} \times [0,1] \to \overline{\Omega}$
\vspace{-0.5em} 
\begin{subequations}\label{alg1-flowmap}
\begin{alignat}{2}[left = \empheqlbrace\,]
\partial_{\tau} \eta + w \cdot \nabla \eta &=0 \,, \quad && x \in \overline{\Omegaref} \text{ and } 0 < \tau \leq 1  \label{alg1-flowmap-a}  \\
\eta(x,0) &= x \,, \quad && x \in \overline{\Omegaref} \text{ and }  \tau =0  
\end{alignat}
\end{subequations}
where the \emph{transport velocity} $w : \overline{\Omegaref} \times [0,1] \to \mathbb{R}^2$ is defined as 
\vspace{-0.5em} 
\begin{equation}\label{alg1-flowmap-velocity}
w(x,\tau) = \frac{\overline{w}(x)}{ \tau + (1-\tau) \Fe \circ \mathrm{id}(x) } \,.
\end{equation}

\item Define $\psi(x) \coloneqq \eta(x,1)$. Then $\psi$ solves \eqref{MA-eqn}. 

\end{enumerate}
\end{algorithm}

\subsubsection{Validity of construction}
The proof that the map $\psi$ constructed according to \Cref{static-alg1} satisfies \eqref{MA-eqn} proceeds as follows. 
Define the back-to-labels map $\xi : \Omega \times [0,1] \to \Omegaref$ by $\xi(y,\tau) = \eta^{-1}(y,\tau)$.  The Eulerian transport 
equation for $\eta$ is transformed into a Lagrangian advection equation for $\xi$: 
\begin{subequations}\label{alg1-back-to-labels}
\begin{alignat}{2}[left = \empheqlbrace\,]
\partial_{\tau} \xi(y,\tau) &= w \circ \xi (y,\tau) \,, \quad && y \in \overline{\Omega} \text{ and } 0 < \tau \leq 1  \label{alg1-back-to-labels-a}  \\
\xi(y,0) &= y \,, \quad && y \in \overline{\Omega} \text{ and }  \tau =0   \,.
\end{alignat}
\end{subequations} 
Note that $\xi |_{\tau=1} = \eta^{-1} |_{\tau=1} = \psi^{-1} = \phi$. 

Next, define the quantity 
$$
\mathcal{R}(y,\tau) = J(y,\tau) \left[ \tau + (1-\tau) F  \right] \circ \xi (y,\tau) \,,  
$$
where $J(y,\tau) = \det D \xi(y,\tau)$ and $F = \Fe \circ \mathrm{id}$.  We compute
$$
\partial_\tau \mathcal{R} = \partial_\tau J \left[ \tau  + (1-\tau) F \right]  \circ \xi + J \left[1 - F  \right] \circ \xi + J (1-\tau) \partial_\tau (F \circ \xi) \,.
$$
We recall Euler's lemma, which states that $J(y,\tau)$ evolves according to 
$\partial_\tau J = J \divv w \circ \xi$.  Using \eqref{alg1-Poisson-a} and \eqref{alg1-ubar}, we calculate
\begin{equation*}
\divv w = \frac{\divv \overline{w}}{\tau + (1-\tau) F } -  \frac{(1-\tau) \overline{w} \cdot \nabla F }{ \left[ \tau  + (1-\tau)F \right]^2} =  \frac{F- 1 - (1-\tau) w \cdot \nabla F }{\tau + (1-\tau) F } \,,
\end{equation*}
so that 
$$
\partial_\tau J \left[ \tau  + (1-\tau) F \right]  \circ \xi =J \left[  F -1 - (1-\tau) w \cdot \nabla F  \right] \circ \xi \,.
$$

Next, we have that 
$$
\partial_\tau (F \circ \xi ) = \partial_\tau \xi \cdot \nabla F \circ \xi  = \left[ w \cdot \nabla F  \right] \circ \xi \,,
$$
where we have used equation \eqref{alg1-back-to-labels-a}. 

Using the two formulae above, we find that $\partial_\tau R = 0$, so that $\Fe(y) = R(y,0) = R(y,1) = \mathrm{det} D \phi(y)$ and thus 
$\phi$ satisfies \eqref{MA-inv-eqn-a}, which is in turn equivalent to \eqref{MA-eqn-a}.  The condition $w(x,\tau) \cdot \nu = 0$ for every 
$x \in \partial \Omegaref$ and $0 \leq \tau \leq 1$ ensures that \eqref{MA-eqn-b} is satisfied. \qed

\subsubsection{Discussion} 
The first numerical implementation of the deformation method \cite{LiaoAnderson1992} 
utilized the no slip boundary conditions $\psi(x) = x$,  $\forall x \in \partial \Omegaref$, rather 
than the no penetration boundary conditions \eqref{MA-eqn-b} which permit tangential 
motion of boundary nodes. 
The method of proof in the original paper of \cite{DaMo1990}, which includes an analysis of the 
Poisson problem \eqref{alg1-Poisson}, requires the domain $\Omegaref$ to have smooth boundary 
$\partial \Omegaref$ and so is not valid for the rectangular domains we consider in this work. A modified method, which avoids the use of the 
Poisson problem \eqref{alg1-Poisson} via a direct construction of the deformation velocity field, is provided in 
\cite{LiaoAnderson1992}, but the resulting numerical implementation yields poor quality grids with high levels of distortion \cite{Delzanno2008}. 
On the other hand, as we shall demonstrate in our numerical experiments, the use of the Poisson equation \eqref{alg1-Poisson} together with 
the slip boundary conditions \eqref{MA-eqn-b} produces smooth grids. Moreover, the arguments in   
\cite{DaMo1990} can be modified with the help of elliptic estimates on 
 polygonal domains \cite{Grisvard2011} to show that the procedure outlined in \Cref{static-alg1} yields existence of a solution 
 to \eqref{MA-eqn}.   
 
The basic mesh generation scheme \Cref{static-alg1} differs from other deformation methods in the literature, e.g.   
 \cite{LiaoLiuJi1998,Grajewski2009}, in both its formulation and numerical implementation.  
 Specifically, the use of the transport system 
 \eqref{alg1-flowmap} avoids costly interpolation procedures required for the solution of the Lagrangian advection equations 
 in other deformation methods, which results in an order of magnitude speed-up.  
 Moreover, our numerical algorithm produces solutions that converge with high-order accuracy, in 
 contrast to other methods which only yield second-order accurate solutions, at best.  
 In the following subsection, we describe in detail the two main steps of \Cref{static-alg1}, namely the 
 Poisson solve in Step 2, and the transport equation solve in Step 4. 

\subsection{Numerical implementation details}\label{subsec:numerical-imp}

\subsubsection{FFT-based elliptic solve for $\Phi$}
The Poisson problem \eqref{alg1-Poisson} is solved in frequency space using the Fast Fourier Transform (FFT).  
 The solvability condition \eqref{alg1-solvability} is enforced by 
the normalization procedure described in \Cref{static-alg1} with trapezoidal integration to compute integrals. 
The RHS of \eqref{alg1-Poisson-a} then has zero mean, and a (non-unique) solution to \eqref{alg1-Poisson} exists.  We choose a 
unique solution $\Phi$ with zero mean, enforced in spectral space by zeroing out the first frequency component. 
The use of FFT requires the forcing $\Ge$ to be periodic; we periodize the problem by doubling the size of the 
domain in each direction and extending $\Ge$ symmetrically to the extended 
domain\footnote{An alternative implementation with the discrete cosine transform can also be used.}. 
In Step 3, the velocity $\overline{w}$ is also computed via FFT.

\subsubsection{Boundary conditions and order of convergence}

Solutions to the Poisson problem \eqref{alg1-Poisson} in general have limited regularity due to the presence of 
corner singularities in the domain, unless 
 the function $\Ge$ satisfies certain {compatibility conditions} \cite{HeOs2014}.  In this work, we shall assume the 
 stronger Neumann condition $D \Ge(y) \cdot \nu = 0$ for $y \in \partial \Omega$ to ensure high-order convergence of the numerical 
solution $\psi$ in the limit of zero mesh size.  If  $D \Ge(y) \cdot \nu \neq 0$, then the symmetric extension of 
$\Ge$ is not differentiable on the boundary $\partial \Omega$ and is only Lipschitz continuous.  
In this case, the potential $\Phi$, velocity $\overline{w}$, and solution $\psi$ all converge with 2\textsuperscript{nd} order accuracy, but  
the convergence rate of the Jacobian determinant $\mathcal{J}(x)$ and cofactor matrix $a(x)$ is only 1.5.   

On the other hand, if the  function $\Ge$ does satisfy the 
Neumann condition $D\Ge(y) \cdot \nu = 0$ for $y \in \partial \Omega$, then the symmetric extension of $\Ge$ is at least 
twice continuously differentiable, and the quantities $\psi(x)$, $\mathcal{J}(x)$, and $a(x)$ all converge with (at least) 
4\textsuperscript{th} order accuracy.  
We confirm this high order convergence with a numerical example in \Cref{subsubsec:high-order}. 

For most of the problems we consider in this work, the function $\Ge$ does indeed satisfy the Neumann condition.  However, even if 
the Neumann condition is not satisfied, the errors in the numerical solution are localized to the boundary, and the meshes produced are 
still of high accuracy and quality. Additionally, boundary smoothing techniques \cite{AvIsVo1998,FeZh2020} can be applied to obtain high order 
convergence.  We implement a simplified version of this technique in \Cref{subsec:MK_SAM} and demonstrate that the 
quantity $\mathcal{J}(x)$ converges with 2\textsuperscript{nd} order accuracy.  The details of this boundary smoothing technique 
are provided in Appendix \ref{sec-appendices-a}.

\subsubsection{Numerical solution of the transport equations}

The solution $\eta$ to the transport equations \eqref{alg1-flowmap} is smooth, and we shall therefore utilize the simple 5th order linear 
upwind scheme to compute derivatives, with the upwind direction in the $r$-th coordinate 
determined based on the sign of $w^r$.  For instance, if  $w^1_{i,j} \geq 0$, then we approximate  the derivative $\partial_1 f$ of a function 
$f$ by 
$$
\left[ \partial_1 f \right]_{i,j} =  \frac{ -2 f_{i-3,j} + 15 f_{i-2,j} - 60f_{i-1,j} + 20 f_{i,j} + 30 f_{i+1,j} - 3f_{i+2,j}  }{60 \Delta x^1}  + \mathcal{O}( |\Delta x|^5) \,. 
$$
For time-integration, we utilize the standard explicit RK4 scheme, which has the associated stability condition 
\begin{equation}\label{CFL-pseudo}
\mathrm{CFL}_{\tau} = \Delta \tau \left(  \frac{|| w^1 ||_{\infty}}{\Delta x^1} +\frac{|| w^2 ||_{\infty}}{\Delta x^2}   \right) \leq C \,. 
\end{equation}
Accordingly, the adaptive time-step $\Delta \tau$ is chosen via
\begin{equation}\label{dtau-adaptive}
\mathrm{CFL}_\tau = \frac{\Delta \tau}{\tau + \frac{(1-\tau)}{|| \Ge ||_{\infty}}} \left(  \frac{|| \overline{w}^1 ||_\infty}{\Delta x^1} + \frac{|| \overline{w}^2 ||_\infty}{\Delta x^2}  \right) \,.
\end{equation}
Our numerical experiments have shown that $\mathrm{CFL}_\tau = 2$ is sufficient for a stable scheme.

\subsection{High-order accuracy and a benchmark computation}

\subsubsection{Jacobian error metric}
Assessing the accuracy and convergence behavior of the numerical solutions $\psi$ 
produced with SAM requires an error metric. Since the exact solution 
$\psi_{\mathrm{exact}}$ to the scheme described in \Cref{static-alg1} is not known, we shall instead use the 
$L^2$ Jacobian error, defined as 
\begin{equation}\label{equi-error}
\mathcal{E}_2 \coloneq  || \mathcal{J}(x) - \Ge \circ \psi(x) ||_{L^2} \,.
\end{equation}
We use bicubic interpolation to compute the composition $\Ge \circ \psi$ in \eqref{equi-error} and 
trapezoidal integration to compute the $L^2$ integral norm.

\subsubsection{High order convergence of solutions}\label{subsubsec:high-order}

To demonstrate the high order convergence of numerical solutions computed with SAM,  we 
perform a mesh 
generation experiment on $\Omega=[0,1]^2$ for the circular target Jacobian function 
\begin{equation}\label{static-test-function}
\bar{\Ge}(y) = 1 - \delta \exp \left\{  - \left| \sigma \left( (y^1-0.5)^2 + (y^2 - 0.5)^2 - r^2  \right) \right|^2  \right\} \,,
\end{equation}
which forces the mesh to resolve in an annular region containing the circle of radius $r$ centered at 
$(0.5 \,, 0.5)$.  The parameters $\delta$ and $\sigma$ control the smallest cell-size and 
width of the resolving region, respectively.  We choose $\delta =0.75$, $\sigma = 64$, and 
$r = 0.2$.  See \Cref{fig:dyn-circular-mesh} for the meshes 
associated with a time-dependent version of  \eqref{static-test-function}.  

%The adaptive meshes $\mathcal{T}$ produced using 
%\Cref{static-alg1} are shown in \Cref{fig:static-test-mesh} for $32^2$ and $64^2$ cell resolution. 
%\begin{figure}[ht]
%\centering
%\subfigure[$N = 32 \times 32$.]{\label{fig:static-test-mesh2}\includegraphics[width=55mm]{static/test-mesh2}} 
%\hspace{4em}
%\subfigure[$N = 64 \times 64$.]{\label{fig:static-test-mesh1}\includegraphics[width=55mm]{static/test-mesh1}} 
%\caption{ Adaptive meshes produced with SAM for the circular target Jacobian \eqref{static-test-function}. The SAM solutions 
%are smooth and converge with 4\textsuperscript{th} order accuracy, as confirmed in \Cref{table:high-order}.}
%\label{fig:static-test-mesh}
%\end{figure} 

We generate a sequence of meshes using SAM for cell 
resolutions $N=32^2$ up to $N=1024^2$. 
We compute the Jacobian errors $\mathcal{E}_2$ given by \eqref{equi-error}, with the Jacobian determinant $\mathcal{J}$  approximated 
using 4\textsuperscript{th} order central differencing (CD4).  The errors provided in \Cref{table:high-order} show that SAM solutions
 exhibit the expected 4\textsuperscript{th} order accuracy.  
We note that, to the best of our knowledge, all other grid generation schemes are at best 2\textsuperscript{nd} order accurate.  

\begin{table}[ht]
\centering
\renewcommand{\arraystretch}{1.0}
\scalebox{0.8}{
\begin{tabular}{|lr|cccccc|}
\toprule
\midrule
\multirow{2}{*}{\textbf{Scheme}} &  & \multicolumn{6}{c|}{\textbf{Cells}}\\

{}  &   & $32 \times 32$    & $64 \times 64$   & $128 \times 128 $ & $256 \times 256$ & $512 \times 512$ & $1024 \times 1024$ \\
\midrule
\multirow{2}{*}{SAM} & Error & 
$2.85 \times 10^{-2}$  & $5.10 \times 10^{-3}$  & $5.96 \times 10^{-4}$ & $3.73 \times 10^{-5}$  & $1.87 \times 10^{-6}$ & $9.89 \times 10^{-8}$ \\
				    & Order & -- & 2.5   & 3.1  & 4.0 & 4.3 & 4.2 \\
\midrule
\bottomrule
\end{tabular}}
\caption{Jacobian errors $\mathcal{E}_2$ demonstrating high order convergence 
of SAM solutions for the circular target Jacobian function \eqref{static-test-function}.}
\label{table:high-order}
\end{table}

\subsubsection{Benchmarking against a state-of-the-art mesh generation scheme}\label{subsec:MK_SAM}

Now, we perform a numerical experiment to benchmark SAM against the state-of-the-art 
MK mesh generation scheme \cite{Delzanno2008}, a brief description of which is provided in Appendix \ref{sec:MK}. 
The test problem \cite{Delzanno2008} we consider is as follows: the domain is $\Omega = [0,1]^2$, and the target Jacobian function is 
\begin{equation}\label{radial-forcing}
\bar{\Ge}(y) = 2 + \cos \left( 8\pi r  \right) \,, 
\end{equation}
where $r = \sqrt{ (y^1-0.5)^2 + (y^2-0.5)^2 }$ is the radial coordinate.  

We compute a sequence of meshes for $N=16^2$ up to $N=256^2$ using SAM, and calculate the $L^2$ Jacobian errors $\mathcal{E}_2$. 
For the purposes of consistency with \cite{Delzanno2008}, we use a slightly different formula to compute $\mathcal{E}_2$ (see equations 
(46)-(52) in \cite{Delzanno2008}).  In particular, 2\textsuperscript{nd}-order differencing is used to calculate the Jacobian and, as such, we 
expect only 2\textsuperscript{nd} order convergence of the errors $\mathcal{E}_2$.   Consequently, we 
instead use the 3\textsuperscript{rd} order linear upwind scheme for the transport equation solve.  The 
pseudo-time step is set according to \eqref{dtau-adaptive} with $\mathrm{CFL}_\tau = 8$.  

The errors are listed in \Cref{table:forcing-radial}, along with the errors for the MK scheme obtained from 
\cite{Delzanno2008}.   The function $\bar{\Ge}$ is radially symmetric, and 
consequently does not satisfy the Neumann condition $D \Ge \cdot \nu \neq 0$. 
As such, the  resulting solutions computed with  SAM do not display 2\textsuperscript{nd} order accuracy  
in the limit $N \to \infty$, though, as shown in \Cref{table:forcing-radial}, the order of convergence only degrades to approximately 
1.75 for the resolutions considered.

\begin{table}[ht]
\centering
\renewcommand{\arraystretch}{1.0}
\scalebox{0.8}{
\begin{tabular}{|lr|ccccc|}
\toprule
\midrule
\multirow{2}{*}{\textbf{Scheme}} &  & \multicolumn{5}{c|}{\textbf{Cells}}\\

{}  & & $16 \times 16$   & $32 \times 32$    & $64 \times 64$   & $128 \times 128 $ & $256 \times 256$ \\
\midrule
\multirow{2}{*}{MK} & Error & 
$9.64 \times 10^{-2}$  & $2.80 \times 10^{-2}$  & $5.78 \times 10^{-3}$ & $1.46 \times 10^{-3}$  & $3.67 \times 10^{-4}$ \\
				    & Order & -- & 1.78   & 2.28  & 1.99 & 1.99 \\
\midrule
\multirow{2}{*}{SAM with $\bar{\Ge}$} & Error & 
$6.54 \times 10^{-2}$  & $2.05 \times 10^{-2}$  & $7.82 \times 10^{-3}$ & $2.00 \times 10^{-3}$  & $5.96 \times 10^{-4}$ \\
				    & Order & -- & 1.68   & 1.39  & 1.96 & 1.75 \\
\midrule
\multirow{2}{*}{SAM with $\bar{\Ge}^*$} & Error & 
$2.30 \times 10^{-2}$  & $1.44 \times 10^{-2}$  & $5.46 \times 10^{-3}$ & $1.25 \times 10^{-3}$  & $3.25 \times 10^{-4}$ \\
				    & Order & -- & 0.68   & 1.40  & 2.12 & 1.94 \\
\midrule
\bottomrule
\end{tabular}}
\caption{Comparison of $L^2$ Jacobian errors and convergence rates 
for the MK and SAM schemes applied to \eqref{radial-forcing}. The data for the MK scheme is taken from Table 1 of \cite{Delzanno2008}.}
\label{table:forcing-radial}
\end{table}

\begin{figure}[ht]
\centering
\subfigure[Mesh $\mathcal{T}^*$ for $\bar{\Ge}^*$.]{\label{fig:mesh-radial-modified-a}\includegraphics[width=50mm]{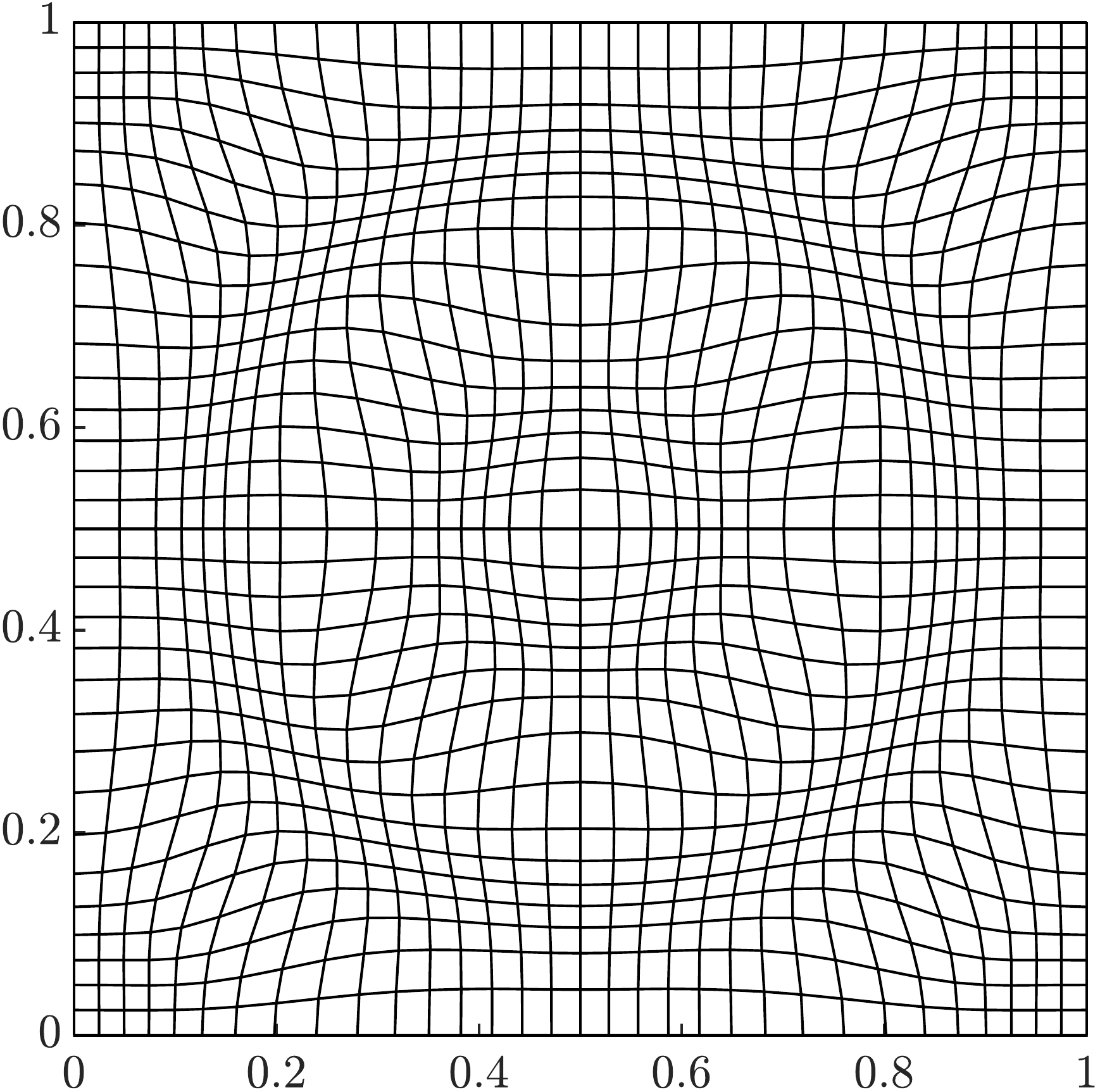}} 
\hspace{4em}
\subfigure[Comparison of $\mathcal{T}$ (black solid) and $\mathcal{T}^*$ (red dashed).]{\label{fig:mesh-radial-modified-b}\includegraphics[width=50mm]{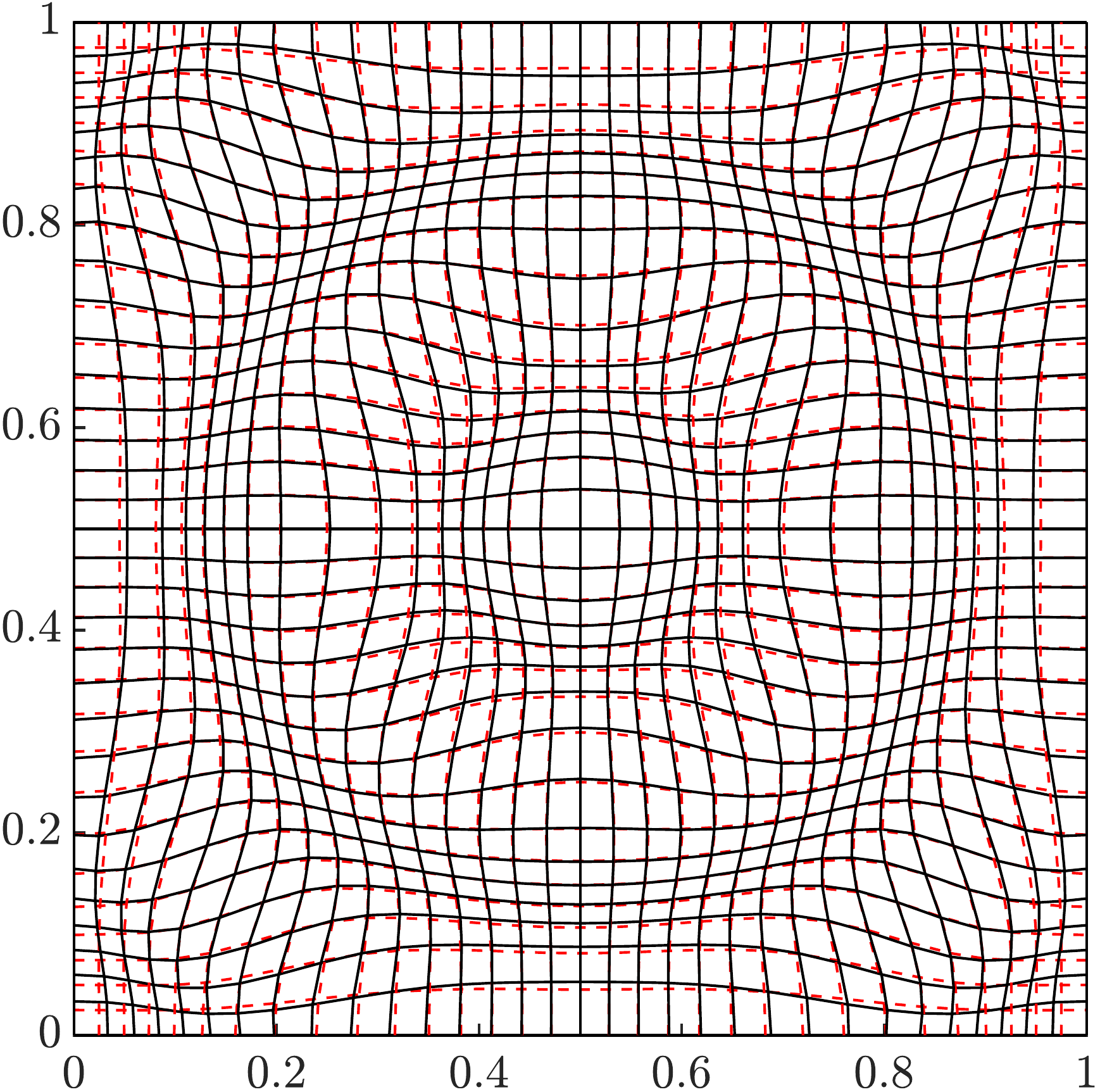}}
\caption{SAM algorithm with boundary smoothing for the radial sinusoidal target function \eqref{radial-forcing}. 
Figure (a) is the $32 \times 32$ cell mesh $\mathcal{T}^*$ produced for the 
modified target Jacobian $\bar{\Ge}^*$, and Figure
(b) is a comparison of the mesh $\mathcal{T}$ without boundary smoothing for the target Jacobian $\bar{\Ge}$ (black solid), and 
the mesh $\mathcal{T}^*$ for $\bar{\Ge}^*$ (red dashed).}
\label{fig:mesh-radial-modified}
\end{figure}

Nonetheless, we shall additionally consider a modified version of this test problem in which 
the function $\bar{\Ge}(y)$ in \eqref{radial-forcing} is replaced by the function $\bar{\Ge}^*(y)$, where $\bar{\Ge}^*(y)$ is such that 
$D \bar{\Ge}^* \cdot \nu = 0$ on $\partial \Omega$.  
The function $\bar{\Ge^*}$ is equal to $\bar{\Ge}$ in the interior of $\Omega$, but is mollified with an 
appropriate cut-off function in a small region near the boundary $\partial \Omega$ to enforce the Neumann condition (see 
Appendix \ref{sec-appendices-a} for further details). 
The mesh $\mathcal{T}^*$ produced using  SAM with $\bar{\Ge}^*$ is 
shown in \Cref{fig:mesh-radial-modified-a}, and a comparison with the mesh $\mathcal{T}$ for $\bar{\Ge}$ is shown in 
\Cref{fig:mesh-radial-modified-b}, from which it can be seen that the two meshes are very similar: they are nearly identical in 
the interior, with small differences near the boundary.    
While the solutions for $\bar{\Ge}$ do not attain the full 2\textsuperscript{nd} order 
accuracy,  the solutions for $\bar{\Ge}^*$ do. 
The SAM solutions for $\bar{\Ge}^*$ have smaller errors than MK across all the resolutions considered.
 Moreover, the SAM solutions for $\bar{\Ge}^*$ display 
2\textsuperscript{nd} order accuracy as $N$ increases\footnote{We have verified this up to $N=2048^2$ but for brevity do not 
show the results here.}.  We also find that the $L^2$ mesh displacement 
$|| \psi(x) - x ||_{L^2} \approx 0.0178$ is comparable to the value of 0.0174 for MK reported in 
Table 1 of \cite{Delzanno2008}.

Next, we benchmark the computational efficiency of our SAM scheme against the MK scheme.
We list in  \Cref{table:mesh-runtime} the CPU runtimes for the MK scheme and the
SAM scheme, where the MK runtimes are taken from Table 1 of \cite{Delzanno2008}. 
To account for the different machines on which the MK and SAM schemes were run on, we divide the MK 
runtimes by $2.2$, where the factor of $2.2$ is determined from a machine comparison experiment, the details of which are 
provided in Appendix \ref{sec:MK}. We then list the  speed-up factor 
of the SAM scheme over the MK scheme in the final row of \Cref{table:mesh-runtime}.  We find that our SAM scheme is almost 
200 times faster than the MK scheme at $N = 256^2$ cell resolution.  We also note that the CPU times reported in \cite{Delzanno2008} 
do not include the cost of an interpolation call, which is non-negligible at high resolutions.

\begin{table}[ht]
\centering
\renewcommand{\arraystretch}{1.0}
\scalebox{0.8}{
\begin{tabular}{|lr|ccccc|}
\toprule
\midrule
\multirow{2}{*}{\textbf{Scheme}} &  & \multicolumn{5}{c|}{\textbf{Cells}}\\

{}  & & $16 \times 16$   & $32 \times 32$    & $64 \times 64$   & $128 \times 128 $ & $256 \times 256$ \\
\midrule
{MK} & $T_{\mathrm{CPU}}$ & 
$0.045$  & $0.182$  & $0.591$ & $2.227$  & $8.636$ \\
\midrule
\multirow{2}{*}{SAM} &  $T_{\mathrm{CPU}}$ & 
$0.0012$  & $0.0017$  & $0.004$ & $0.013$  & $0.042$ \\
				    & speed-up factor & 47 & 131   & 160  & 174 & 190 \\
\midrule
\bottomrule
\end{tabular}}
\caption{CPU runtimes for the MK scheme vs the SAM scheme. The results for the MK scheme are taken from Table 1 of 
\cite{Delzanno2008} then divided by $2.2$ to account for machine difference.}
\label{table:mesh-runtime}
\end{table}

\section{Fast dynamic adaptive meshing}\label{sec:SAM-dyn}

 \subsection{Dynamic formulation}\label{subsec:dynamic}
Given a time interval 
$t \in [0,T]$, we seek to construct a time dependent diffeomorphism $\psi : \Omegaref \times [0,T] \to \Omega$.  
We denote the time-dependent mesh on $\Omega$ by $\mathcal{T}(t)$, which will be 
found as the image of  $\Tref$  under the action of $\psi(\cdot,t)$.  The time-dependent map $\psi : \Omegaref \times [0,T] \to \Omega$ 
is constructed by prescribing, for each $t \in [0,T]$, its 
Jacobian determinant. Let $\Ge : \Omega \times [0,T] \to \mathbb{R}^+$ denote a given time-dependent (physical) target Jacobian function. 
Then $\psi$ satisfies
 \begin{subequations}\label{MA-eqn-dynamic}
\begin{alignat}{2}[left = \empheqlbrace\,]
\mathrm{det} \nabla \psi (x,t) &= \Ge \circ \psi (x,t) \,, \quad && (x,t) \in \Omegaref  \times [0,T]    \label{MA-eqn-dynamic-a}  \\
\psi (x,t) \cdot \nu &= x \cdot \nu \,, && (x,t) \in \partial \Omegaref   \times [0,T] \label{MA-eqn-dynamic-b}
\end{alignat}
\end{subequations}

The target Jacobian function $\Ge$ must satisfy, for each $t \in [0,T]$, the 
following integral constraint to ensure that \eqref{MA-eqn-dynamic} has a solution: 
\begin{equation}\label{MA-eqn-solvability-dynamic}
\int_{\Omega} \frac{1}{\Ge(y,t)} \, \mathrm{d}y = \int_{\Omegaref} \frac{ \det \nabla \psi(x,t) }{\Ge \circ \psi (x,t)} \,\mathrm{d}x =  |\Omega| \,.
\end{equation}

\subsubsection{Temporal discretization}
We uniformly discretize the time domain $[0,T]$ into $K$ intervals of length $\Delta t$ and set $t_k = k \Delta t$ for $k = 0,1,\ldots,K$. 
Denote $\Ge_k \coloneqq \Ge(\cdot,t_k)$,  $\psi_k \coloneqq \psi(\cdot,t_k)$, and $\mathcal{T}_k = \mathcal{T}(t_k)$.  Then each 
$\psi_k : \Omegaref \to \Omega$ is a diffeomorphism satisfying 
 \begin{subequations}\label{MA-eqn-dyn-discrete}
\begin{alignat}{2}[left = \empheqlbrace\,]
\mathrm{det} \nabla \psi_k (x) &= \Ge_k \circ \psi_k (x) \,, \quad && x \in \Omegaref    \label{MA-eqn-dyn-discrete-a}  \\
\psi_k (x) \cdot \nu &= x \cdot \nu \,, && x \in \partial \Omegaref    \label{MA-eqn-dyn-discrete-b}
\end{alignat}
\end{subequations}
with each target Jacobian function $\Ge_k : \Omega \to \mathbb{R}^+$ satisfying the integral constraint 
\begin{equation}\label{MA-eqn-solvability-dyn-discrete}
\int_\Omega \frac{1}{\Ge_k(y)} \, \mathrm{d}y = \int_{\Omegaref} \frac{ \det \nabla \psi_k(x) }{\Ge_k \circ \psi_k(x)} \,\mathrm{d}x =  |\Omega| \,.
\end{equation}

\subsubsection{Algorithmic complexity}
The simplest possible strategy for \eqref{MA-eqn-dyn-discrete} is to compute each map $\psi_k$ using 
static SAM. Our numerical experiments indicate that static SAM is highly efficient 
for a single mesh generation call. However, for unsteady fluids simulations which require dynamic 
meshing at every time-step, the use of static SAM can be expensive at high resolutions due to the 
computational bottleneck in the transport equation solve stage.  

Specifically, suppose that 
the target Jacobian function has large deviation from the identity i.e. 
$||1/\Ge -1||_{L^\infty} \gg 1$. 
Then the associated potential $\Phi$ solving \eqref{alg1-Poisson} has large gradients, and the flowmap velocity 
\eqref{alg1-flowmap-velocity} will therefore be large in magnitude. 
Consequently, many pseudo-time steps will be 
required in the transport equation solve to preserve stability and accuracy of the computed 
numerical solution for $\eta(x,\tau)$.  In particular, the stability condition \eqref{CFL-pseudo} forces the pseudo-time step 
to decay like $\mathcal{O}\left(N^{-1/2}\right)$, so that the overall complexity of the SAM algorithm is $\mathcal{O}\left(N^{3/2}\right)$.  
For large $N$, this can become prohibitively computationally expensive.  
In the next section, we resolve this issue via a novel reformulation of \eqref{MA-eqn-dyn-discrete}. 
   
\subsection{Reformulation in terms of near-identity maps}\label{subsec:perturbation}
  
\subsubsection{The perturbation formulation}
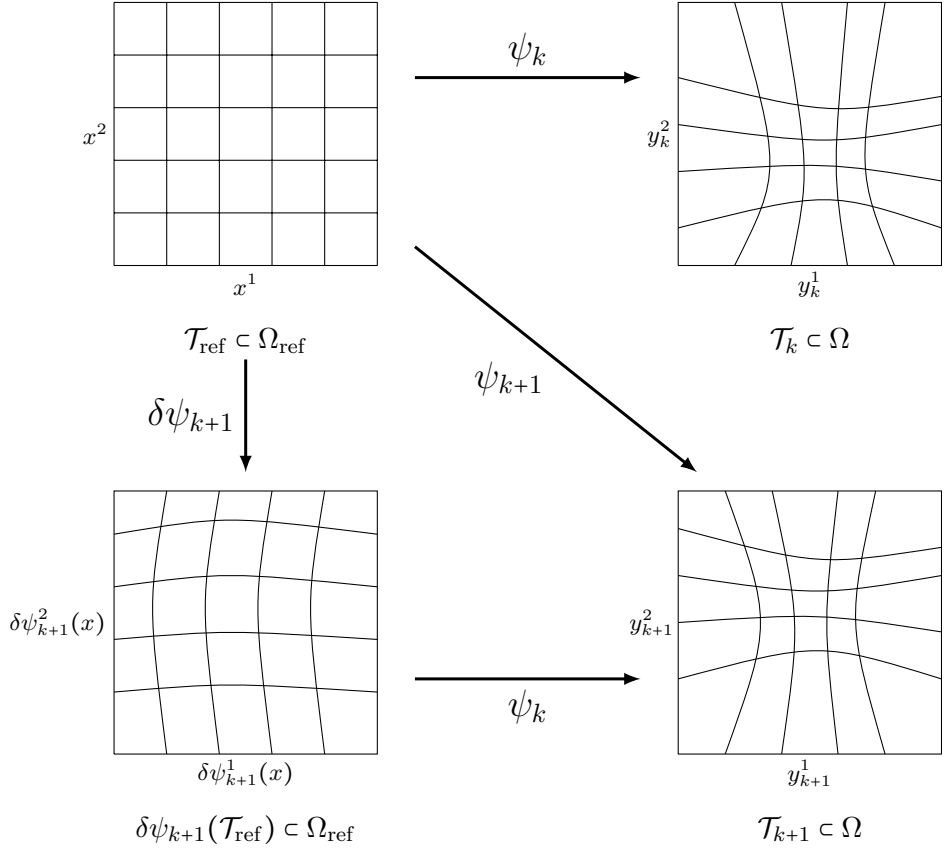
\begin{figure}[ht]
\centering
\begin{tikzpicture}
\draw[step=0.7cm,black] (0,0) grid (3.5,3.5);
\draw (7.5,0) -- (11,0) -- (11,3.5) -- (7.5,3.5) -- (7.5,0);
\draw[black] (8.25, 0) .. controls(8.875,1.25) .. (8.25, 3.5);
\draw[black] (9, 0) .. controls(9.25,1.25) .. (8.875, 3.5);
\draw[black] (9.75, 0) .. controls(9.55,1.25) .. (9.75, 3.5);
\draw[black] (10.375, 0) .. controls(9.875,1.25) .. (10.25, 3.5);
\draw[black] (7.5, 0.5) .. controls(9.5,1) .. (11, 0.5);
\draw[black] (7.5, 1.25) .. controls(9.5,1.375) .. (11, 1.125);
\draw[black] (7.5, 1.875) .. controls(9.5,1.6) .. (11, 1.875);
\draw[black] (7.5, 2.5) .. controls(9.5,2) .. (11,2.25);
\draw[-latex,line width=0.4mm] (4,2.5) --node[above] {\Large $\psi_k$} (7,2.5);
\node at (1.75,-0.25) {\footnotesize $x^1$}; 
\node at (-0.25,1.75) {\footnotesize $x^2$}; 
\node at (9.25,-0.25) {\footnotesize $y_k^1$}; 
\node at (7.25,1.75) {\footnotesize $y_k^2$}; 
\node at (1.75,-1) {$\Tref \subset \Omegaref$};
\node at (9.25,-1) {$\mathcal{T}_k \subset \Omega$};
\draw (7.5,0-6.5) -- (11,0-6.5) -- (11,3.5-6.5) -- (7.5,3.5-6.5) -- (7.5,0-6.5);
\draw[black] (8.25-0.125, 0-6.5) .. controls(8.875-0.125,1.25-6.5+0.5) .. (8.25-0.125, 3.5-6.5);
\draw[black] (9-0.125, 0-6.5) .. controls(9.25-0.125,1.25-6.5+0.5) .. (8.875-0.125, 3.5-6.5);
\draw[black] (9.75-0.125, 0-6.5) .. controls(9.55-0.125,1.25-6.5+0.5) .. (9.75-0.125, 3.5-6.5);
\draw[black] (10.375-0.125, 0-6.5) .. controls(9.875-0.125,1.25-6.5+0.5) .. (10.25-0.125, 3.5-6.5);
\draw[black] (7.5, 0.5-6.5+0.5) .. controls(9.5-0.125,1-6.5+0.5) .. (11, 0.5-6.5+0.5);
\draw[black] (7.5, 1.25-6.5+0.5) .. controls(9.5-0.125,1.375-6.5+0.5) .. (11, 1.125-6.5+0.5);
\draw[black] (7.5, 1.875-6.5+0.5) .. controls(9.5-0.125,1.6-6.5+0.5) .. (11, 1.875-6.5+0.5);
\draw[black] (7.5, 2.5-6.5+0.5) .. controls(9.5-0.125,2-6.5+0.5) .. (11,2.25-6.5+0.5);
\node at (9.25,-0.25-6.5) {\footnotesize $y_{k+1}^1$}; 
\node at (7.15,1.75-6.5) {\footnotesize $y_{k+1}^2$}; 
\node at (9.25,-7.5) {$\mathcal{T}_{k+1} \subset \Omega$};
\draw (0,-6.5) -- (3.5,0-6.5) -- (3.5,3.5-6.5) -- (0,3.5-6.5) -- (0,0-6.5);
\draw[black] (0.7, 0-6.5) .. controls(0.7-0.25,1.75-6.5+0.25) .. (0.7, 3.5-6.5);
\draw[black] (1.4, 0-6.5) .. controls(1.4-0.25,1.75-6.5+0.25) .. (1.4, 3.5-6.5);
\draw[black] (2.1, 0-6.5) .. controls(2.1-0.25,1.75-6.5+0.25) .. (2.1, 3.5-6.5);
\draw[black] (2.8, 0-6.5) .. controls(2.8-0.25,1.75-6.5+0.25) .. (2.8, 3.5-6.5);
\draw[black] (0, 0.7-6.5+0.125) .. controls(1.75-0.25,0.7-6.5+0.25) .. (3.5,0.7-6.5+0.125);
\draw[black] (0, 1.4-6.5+0.125) .. controls(1.75-0.25,1.4-6.5+0.25) .. (3.5,1.4-6.5+0.125);
\draw[black] (0, 2.1-6.5+0.125) .. controls(1.75-0.25,2.1-6.5+0.325) .. (3.5,2.1-6.5+0.125);
\draw[black] (0, 2.8-6.5+0.125) .. controls(1.75-0.25,2.8-6.5+0.375) .. (3.5,2.8-6.5+0.125);
\node at (1.75,-0.25-6.5) {\footnotesize $\delta \psi^1_{k+1}(x)$}; 
\node at (-0.75,1.75-6.5) {\footnotesize $\delta \psi^2_{k+1}(x)$}; 
\node at (1.75,-7.5) {$\delta \psi_{k+1}(\Tref) \subset \Omegaref$};
\draw[-latex,line width=0.4mm] (4,1-6.5) --node[below] {\Large $\psi_k$} (7,1-6.5);
\draw[-latex,line width=0.4mm] (1.75,-1.25) --node[left] {\Large $\delta \psi_{k+1}$} (1.75,-2.75);
\draw[-latex,line width=0.4mm] (4,0.25) -- (7.5+0.25,-2.75);
\node at (5.25,-1.5) {\Large $\psi_{k+1}$};
\end{tikzpicture}
\caption{Schematic of the meshes, and the maps between them, for dynamic mesh generation.} \label{fig:mesh-schematic-dynamic}
\end{figure}

Assume that we are given the map $\psi_k$ and the target Jacobian functions $\Ge_k$ and $\Ge_{k+1}$, and 
suppose that we wish to compute the map $\psi_{k+1}$.  Rather than computing the map 
$\psi_{k+1}$ directly by solving \eqref{MA-eqn-dyn-discrete}, we instead solve for the 
\emph{perturbation map} $\delta \psi_{k+1} : \Omegaref \to \Omegaref$ defined implictly by 
\begin{equation}\label{perturbation-map}
\psi_{k+1} (x) = \psi_k \circ \delta \psi_{k+1} (x) \,.
\end{equation}
That is, we suppose that the map $\psi_{k+1}$ can be found as the image of $\psi_k$ 
acting on a near-identity transformation $\delta \psi_{k+1}$ on the reference domain 
$\Omegaref$ (see \Cref{fig:mesh-schematic-dynamic}). 

By the chain rule and inverse function theorem, we find that $\delta \psi_{k+1}$ satisfies 
 \begin{subequations}\label{MA-eqn-pert}
\begin{alignat}{2}[left = \empheqlbrace\,]
\mathrm{det} \nabla \delta \psi_{k+1} (x) &= P_{k+1} \circ \delta \psi_{k+1} (x) \,, \quad && x \in \Omegaref    \label{MA-eqn-pert-a}  \\
\delta \psi_{k+1} (x) \cdot \nu &= x \cdot \nu \,, && x \in \partial \Omegaref    \label{MA-eqn-pert-b}
\end{alignat}
\end{subequations} 
with the function $P_{k+1} : \Omegaref \to \mathbb{R}^+$ defined as 
\begin{equation}\label{perturbation-target-density}
P_{k+1}(x) = \left( \frac{\Ge_{k+1}}{\Ge_k} \right) \circ \psi_k (x) \,.
\end{equation} 

The system \eqref{MA-eqn-pert} is of exactly the same form as \eqref{MA-eqn}, and the identical
solution procedure described in \Cref{subsec:alg1} for static mesh generation can therefore be used to find the 
solution $\delta \psi_{k+1}$.  Then $\psi_{k+1}$ is computed according to 
\eqref{perturbation-map}. A complete description of the dynamic SAM scheme 
is provided in \Cref{dynamic-alg1}. 

\begin{algorithm}[ht]
\caption{\textbf{\textsc{: dynamic SAM}}}\label{dynamic-alg1}
\begin{enumerate}[itemsep=1.0em,leftmargin=2.0cm,label=\textbf{Step \arabic* :}]

\item Set $t=0$. Given an initial target Jacobian function 
$\Ge_0 : \Omega \to \mathbb{R}^+$, compute the initial diffeomorphism $\psi_0$ 
according to the static solution scheme from \Cref{subsec:alg1}. 

\item For $t = t_{k+1}$, assume that we are given the following: the map $\psi_k$ and 
target Jacobian function $\Ge_k : \Omega \to \mathbb{R}^+$, both from the previous time 
step $t=t_k$, and the target Jacobian function $\Ge_{k+1} : \Omega \to \mathbb{R}^+$ 
at the current time level. 

Define $P_{k+1} : \Omegaref \to \mathbb{R}^+$ by \eqref{perturbation-target-density}, and 
assume that it satisfies the solvability condition 
\begin{equation}\label{alg2-solvability}
\int_{\Omegaref} \frac{1}{P_{k+1} (x)} \,\mathrm{d}x = \int_{\Omegaref} \left( \frac{\Ge_k}{\Ge_{k+1}} \right) \circ \psi_k (x) \,\mathrm{d}x = |\Omegaref| \,,
\end{equation}
In general, we will be given target Jacobian functions $\bar{\Ge}_k$ and $\bar{\Ge}_{k+1}$ such 
that the corresponding $\bar{P}_{k+1} = (\bar{\Ge}_{k+1}/\bar{\Ge}_k) \circ \psi_k$ does not satisfy  \eqref{alg2-solvability}. 
In this case, we define $P_{k+1}$ according to the following normalization procedure:
\begin{equation*}
\bar{Q}_{k+1}(x) = \frac{1}{\bar{P}_{k+1}(x)} \,\,  \longrightarrow  \,\, Q_{k+1}(x) =  |\Omega| \frac{\bar{Q}_{k+1}(x)}{ \int_{\Omegaref} \bar{Q}_{k+1}(x) \,\mathrm{d}x } \,\, \longrightarrow \,\,
P_{k+1}(x) = \frac{1}{Q_{k+1}(x)} \,.
\end{equation*}

\item Solve \eqref{MA-eqn-pert} for the perturbation map 
$\delta \psi_{k+1} : \Omegaref \to \Omegaref$ using the solution procedure in 
\Cref{subsec:alg1}. 

\item Define $\psi_{k+1} : \Omegaref \to \Omega$ by \eqref{perturbation-map}. 
If $t_{k+1} = T$, then stop; otherwise, set $t=t_{k+2}$, and return  to \textbf{Step 2}.

\end{enumerate}
\end{algorithm}

\subsubsection{Discussion}\label{subsubsec:pert-discussion}
The key to the efficiency of dynamic SAM is the reformulation in terms 
of the perturbation map $\delta \psi_{k+1}$ satisfying \eqref{MA-eqn-pert}. 
Specifically, while it may be that both $\Ge_k$ and $\Ge_{k+1}$
 have large deviation from 
1 i.e. $||1/\Ge_k -1||_{L^\infty} \gg 1$ and $||1/\Ge_{k+1} -1||_{L^\infty} \gg 1$, we may nonetheless 
have that $||1/P_{k+1} -1||_{L^\infty} \ll 1$.  Indeed, this is the case in ALE simulations, which are naturally constrained by 
a CFL condition that limits the evolution of the numerical solution over a single time-step.  

More precisely, the usual stability condition for the \emph{physical} time step $\Delta t$ in an (Eulerian) simulation   
forces the time step $\Delta t$ to decay 
like $\Delta t \sim 1/\sqrt{N}$ as $N \to \infty$, which is a constraint of exactly the same form as the condition \eqref{CFL-pseudo}
on  $\Delta \tau$. A Taylor series argument shows that
\begin{align}
1 - \frac{1}{P_{k+1}(x)} &= \Delta t \cdot \frac{\partial_t \Ge(\psi(x,t_k),t_k)}{\Ge(\psi(x,t_k),t_k)}  + \mathcal{O}(\Delta t^2) \nonumber \\
\implies ||1-1/P(\cdot,t) ||_{L^\infty} &= \mathcal{O}(\Delta t) = \mathcal{O}(1/\sqrt{N}) \,, \label{pert-scaling}
\end{align}
since $\partial_t \Ge \sim \mathcal{O}(1)$. Since the stability condition for the pseudo-time step $\Delta \tau$ scales according to \eqref{CFL-pseudo}, and 
$||w||_{L^\infty} = \mathcal{O}(1/\sqrt{N}) $ by \eqref{pert-scaling}, we see that $\Delta \tau = \mathcal{O}(1)$ i.e. the psuedo-time 
step $\Delta \tau$  \emph{can be kept fixed across resolutions} $N$, resulting in a dynamic SAM algorithm with optimal complexity.

%\begin{remark}
%Clearly, we can generalize \eqref{perturbation-map} to the relation 
%$\psi_{k+1}(x) = \psi_{k_0} \circ \delta \psi_{k+1}(x)$, for some $0 \leq k_0 \leq k$, and derive an equation for 
%$\delta \psi_{k+1}$ of the form \eqref{MA-eqn-pert}.  In such a formulation, the map $\psi_{k+1}$ is constructed as the image of 
%$\psi_{k_0}$ acting on a near identity transformation of $\Omegaref$. The choice $k_0 = k$ is natural and is the formulation  adopted 
%in this work, but we note that choosing $k_0 < k$ can sometimes produce better quality grids with less distortion (see also 
%\Cref{subsec:restarted-dyn}). 
%\end{remark}

We emphasize here that our perturbation formulation \eqref{MA-eqn-pert} differs from the methods 
considered in \cite{Liao1995,Grajewski2010} in an important way. 
In particular, our perturbation map $\delta \psi_{k+1}$ is a near identity transformation of 
the \emph{uniform reference domain} $\Omegaref$, whereas the 
schemes in \cite{Liao1995,Grajewski2010} define a perturbation map via 
$\overline{\psi}_{k+1} = \delta \overline{\psi}_{k+1} \circ \overline{\psi}_k$, rather than through \eqref{perturbation-map}. 
This means that the $\overline{\psi}_k$ solutions in \cite{Liao1995,Grajewski2010} are different from our $\psi_k$ SAM solutions. 
Moreover, the equation for $\delta \overline{\psi}_{k+1}$  is posed on 
the deformed mesh $\mathcal{T}_k$; this means that the solvers for the Poisson problem and transport equation must be appropriately 
modified at each time-step $t_k$, requiring, for example, the costly recalculation of the stiffness and mass matrices.
Consequently, SAM is simpler, faster, and more accurate than the schemes in \cite{Liao1995,Grajewski2010}. 
For instance, the scheme in \cite{Grajewski2010} has order of accuracy 1.5, whereas our dynamic 
SAM solutions converge with 4\textsuperscript{th} order accuracy if the data is sufficiently smooth. Additionally, the interpolation routine in 
\cite{Grajewski2010} requires $\mathcal{O}(N^{1.5})$ grid searching on deformed grids, in contrast to our $\mathcal{O}(N)$ 
 SAM algorithm.

\subsection{Restarted dynamic mesh generation}\label{subsec:restarted-dyn}
The perturbation formulation \eqref{MA-eqn-pert}, by design, follows the time history of $\psi(x,t)$.  That is to say, the solution 
$\psi(x,t)$ at time $t=t_k$ depends upon the solution for all $t < t_k$. As such, numerical solutions 
to \eqref{MA-eqn-pert} are susceptible to increasing grid distortion and mesh tangling, a common ailment of Lagrangian-type methods. 
To mitigate this issue, we can periodically \emph{restart} the dynamic mesh generation by computing at time $t=t_k$ the map 
$\psi_k$ directly with static SAM, rather than with dynamic SAM. 
In this way, the greater 
efficiency of dynamic SAM is utilized, while grid distortion errors 
are controlled with the use of static SAM, thereby preventing mesh tangling. 
The restarting criterion is chosen as $\lambda_k > \Lambda \lambda_{\mathrm{ref}}$, where 
$\lambda_k$ is the $L^1$ grid distortion at time step $t_k$, $ \lambda_{\mathrm{ref}}$ is a ``reference'' 
grid distortion (defined in \Cref{dynamic-alg2}), and 
$\Lambda$ is a user prescribed parameter.  
A description of our restarted dynamic SAM scheme
is provided in \Cref{dynamic-alg2}.

\begin{algorithm}[ht]
\caption{\textbf{\textsc{: restarted dynamic SAM}}}\label{dynamic-alg2}
\begin{enumerate}[itemsep=1.0em,leftmargin=2.0cm,label=\textbf{Step \arabic* :}]
\setcounter{enumi}{-1}

\item Choose the maximum grid distortion parameter $\Lambda > 1$. 

\item Set $t=0$. Given an initial target Jacobian function 
$\Ge_0 : \Omega \to \mathbb{R}^+$, compute the initial diffeomorphism $\psi_0$ 
according to the static solution scheme from \Cref{subsec:alg1}. Let $\lambda_{\mathrm{ref}}$ be the (reference) $L^1$ grid distortion 
of the adaptive mesh $\mathcal{T}_0$, computed according to \eqref{L1-distortion}.

\item For $t = t_{k+1} > 0$, compute the average grid distortion of the map $\psi_k$
\begin{equation}\label{L1-distortion}
\lambda_k \coloneqq \left\Vert \frac{1}{2} \mathrm{Tr}\left( \nabla \psi_k \nabla \psi_k^\mathcal{T} \right) \right\Vert_{L^1} \,.
\end{equation}

\item If $\lambda_k > \Lambda \lambda_{\mathrm{ref}}$, then compute the map $\psi_{k+1}$ using static SAM 
\Cref{static-alg1} and recalculate $\lambda_{\mathrm{ref}}$  according to \eqref{L1-distortion}. Otherwise, compute 
$\psi_{k+1}$ using dynamic SAM \Cref{dynamic-alg1}. If $t_{k+1} = T$, then stop; otherwise, set $t=t_{k+2}$, and return to \textbf{Step 2}. 

\end{enumerate}
\end{algorithm}

\section{Dynamic mesh generation experiments}\label{sec:mesh-exp}

In this section, we present and discuss the results of several dynamic mesh generation 
experiments conducted with the static, dynamic, and restarted SAM algorithms. 
Unless otherwise stated, all experiments are conducted on the unit square $\Omega = [0,1]^2$ with 
an equal number of cells in 
the horizontal and vertical directions $m = n =\sqrt{N}$.

\subsection{Static mesh with large zoom-in factor}\label{subsec:large-zoom-in}

This static test problem demonstrates the ability of dynamic SAM to generate smooth meshes with large zoom-in factors which, in 
practical applications, can be used to track very small scale structures with only a few total number of cells. 
On the other hand, when 
 when the target function $\Ge$ has large gradients (as is the case for large zoom-in meshes), numerical errors in the Poisson solve 
often lead to poor quality grids containing non-convex elements \cite{Delzanno2008,Grajewski2010}. 

As an example, consider the circular target Jacobian function $\Ge_{\delta}(y)$ given by  
\eqref{static-test-function} with $\sigma=64$, $r=0.2$, and $\delta \in [0,1)$. 
More generally, we have a family of target functions
$ \{ \Ge_{\delta}(y) \}_{0 \leq \delta < 1}$ parametrized by $\delta$, with each such $\Ge_\delta$ forcing 
the mesh to resolve around some given curve (see equation \eqref{target-jacobian-z}).  
The zoom-in parameter $\delta$ determines the 
zoom-in factor $\Upsilon = 1 / \min \mathcal{J}$ of the adaptive mesh $\mathcal{T}$, and thus the smallest scales that can be 
represented on $\mathcal{T}$.  When $\delta = 0$ (uniform mesh) we have $\Upsilon = 1$. 
As $\delta$ increases, so does $\Upsilon$, with smaller and smaller scales captured 
on $\mathcal{T}$.  As $\delta \to 1$, $\Upsilon \to \infty$ and the mapping is degenerate at $\delta =1$. 
 From the point of view of efficiency, we would like to have $\Upsilon$ large since we then  
require few total number of grid points.  Moreover, for unstable RT problems that have evolving 
interfaces with large curvature, it is essential that we construct adaptive meshes with large enough
$\Upsilon$ such that they can capture the small-scale vortical structures of the flow. 
As such, we often want to choose 
$\delta \approx 1$, and refer to the associated meshes as ``large zoom-in'' meshes. 

\begin{figure}[ht]
\centering
\subfigure[Static SAM with $\delta = 0.97$]{\label{fig:mesh-zoom-in-static-a}\includegraphics[width=55mm]{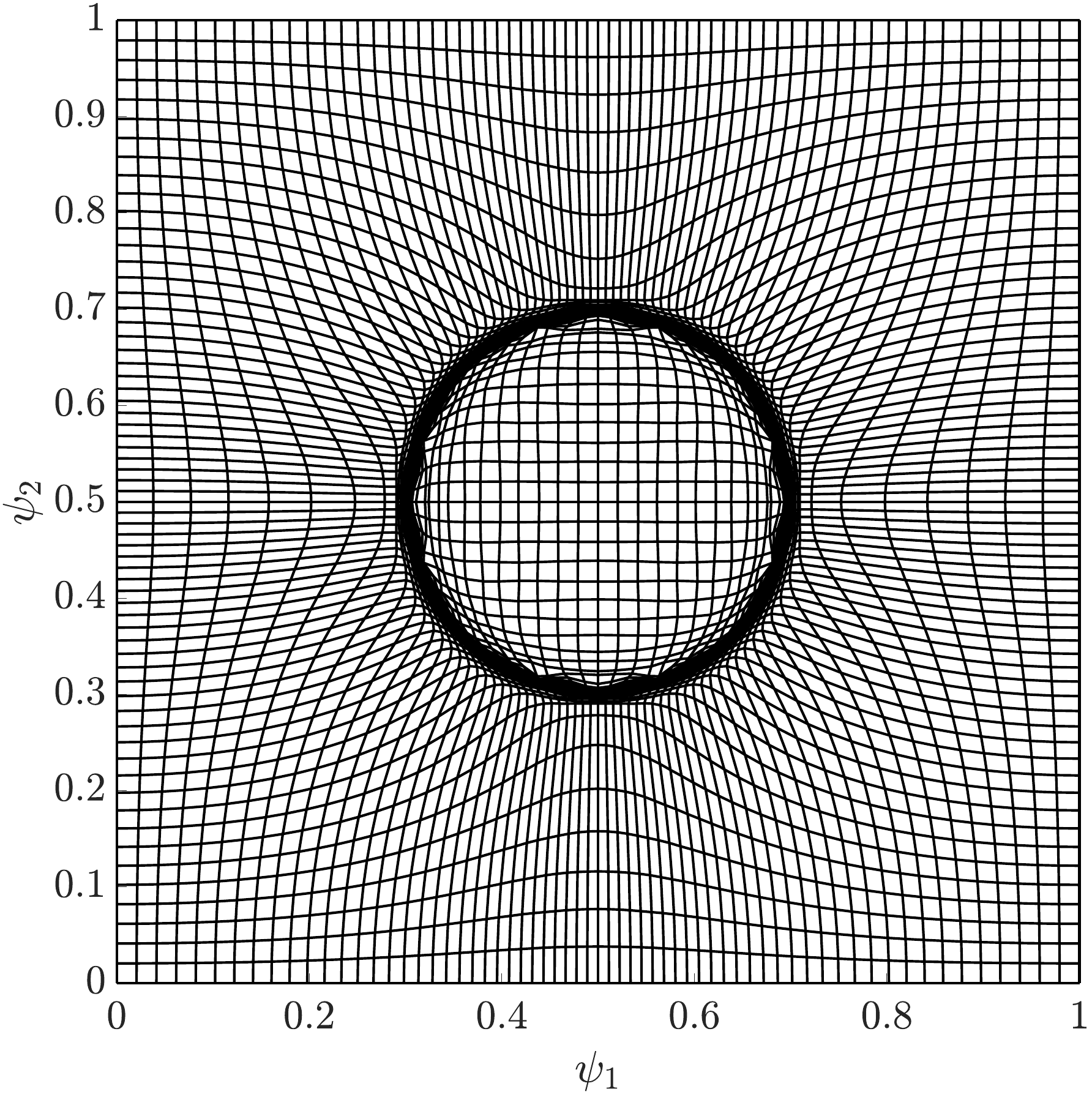}} 
\hspace{4em}
\subfigure[Zoom-in of (a)]{\label{fig:mesh-zoom-in-static-b}\includegraphics[width=55mm]{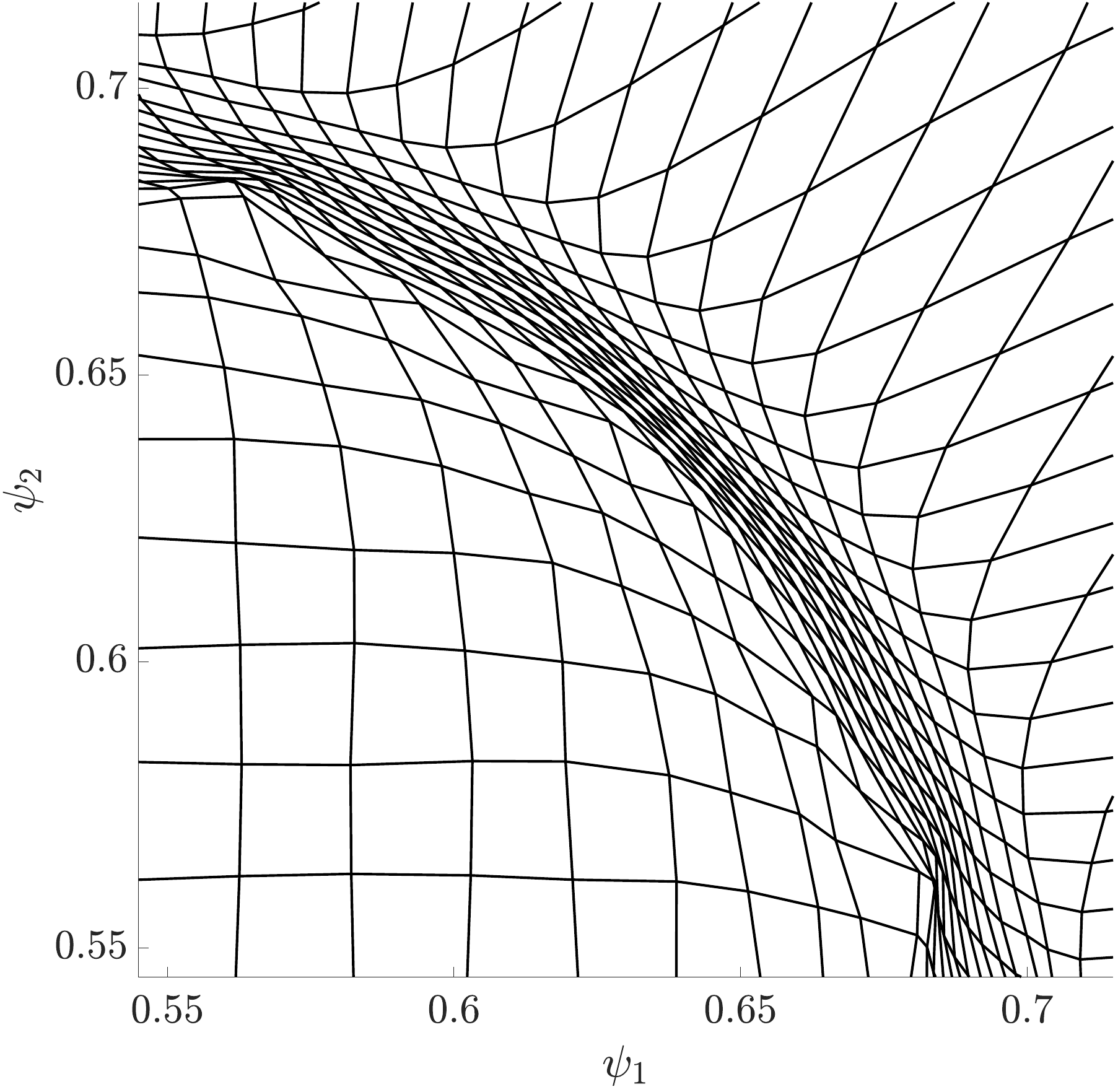}}
\hspace{4em}
\subfigure[Dynamic SAM with $\delta = 0.996$]{\label{fig:mesh-zoom-in-dyn-a}\includegraphics[width=55mm]{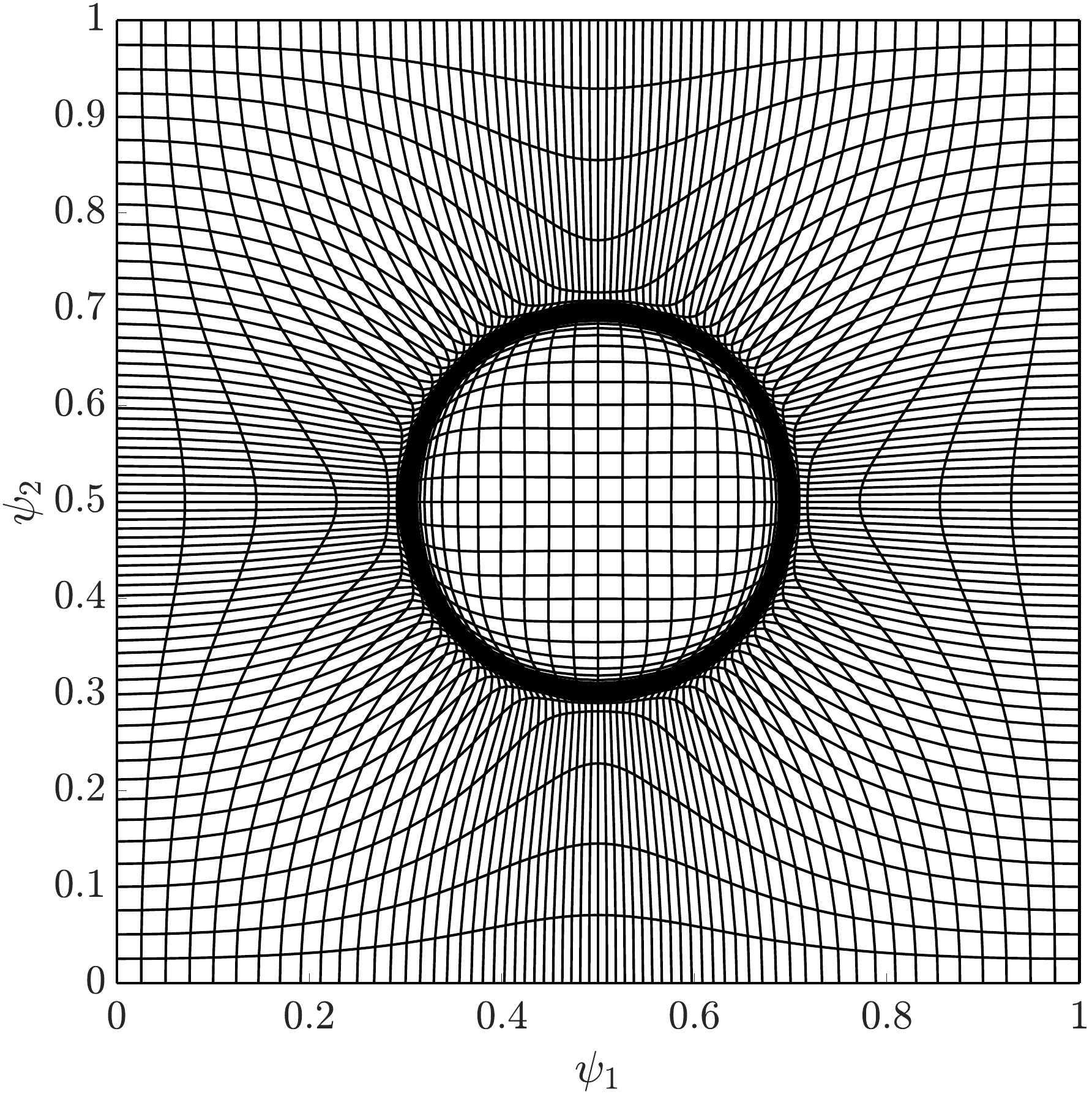}}
\hspace{4em}
\subfigure[Zoom-in of (c)]{\label{fig:mesh-zoom-in-dyn-b}\includegraphics[width=55mm]{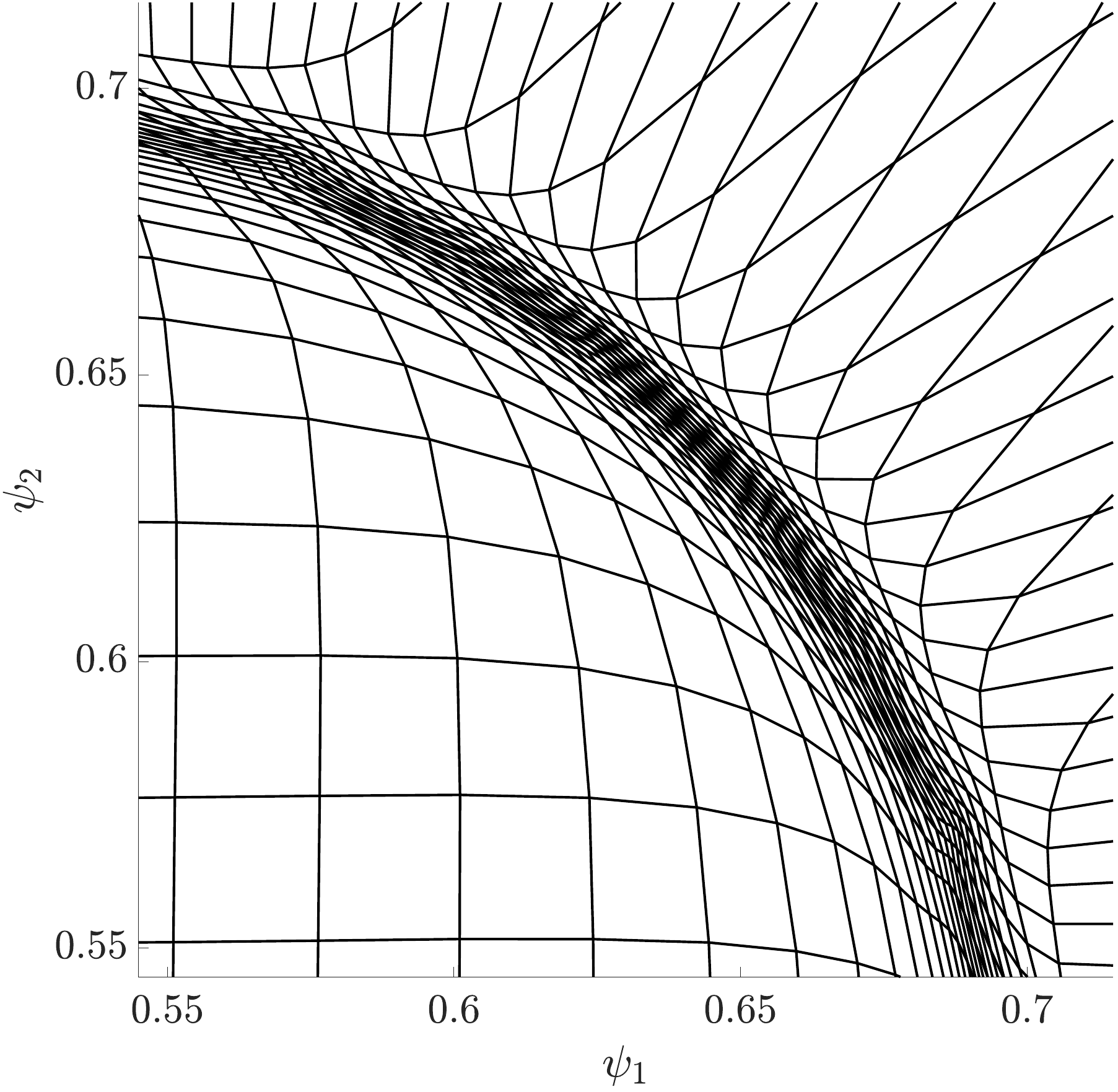}}
\caption{
Test problem \ref{subsec:large-zoom-in} demonstrating smooth large zoom-in meshing using dynamic SAM.  Shown are the
$64^2$ cell meshes with large zoom-in parameter $\delta$ for the circular target Jacobian function \eqref{static-test-function}. 
Figure (a) is the poor quality mesh containing non-convex elements produced with static SAM with $\delta =0.97$, and (b) is a 
zoom-in of (a) near the refining region. Figure (c) is the smooth large zoom-in mesh produced 
with dynamic SAM with $\delta=0.996$ and with smallest cell 100 times smaller than a uniform cell, and (d) is a zoom-in.}
\label{fig:mesh-zoom-in}
\end{figure}

While the continuous mapping $\psi$ is non-degenerate for all $0 \leq \delta < 1$, in practice 
numerical errors in static SAM will produce folded grids if $\delta$ is sufficiently close to 1.  
That is, for each $N$, there exists a corresponding $\delta_{\max}$ such that the grids produced with 
static SAM for $\delta > \delta_{\max}$ contain non-convex elements.
An example of such a grid is shown for 
$N =64^2$ and $\delta =0.97$ in \Cref{fig:mesh-zoom-in-static-a,fig:mesh-zoom-in-static-b}. 
The function  $\Ge_\delta$ is such that $||1- 1/\Ge_\delta||_{L^\infty} \approx \frac{1}{1-\delta} \to \infty$ as 
$\delta \to 1$. When $\delta \approx 1$, large errors in the numerical solution of the Poisson 
problem lead to grids with non-convex 
elements.

Dynamic SAM provides a simple method for producing smooth grids with $\delta \approx 1$. 
We define the time-dependent function
\begin{equation}\label{zoom-in-interp}
\Ge(y,s) = (1-s) + s \Ge_{\delta}(y)\,.
\end{equation}
Then \eqref{zoom-in-interp} linearly interpolates between 1 at $s=0$ and $\Ge_{\delta}$ at $s=1$, and applying dynamic SAM with 
sufficiently many time steps $\Delta s$ yields smooth grids with no  non-convex elements. As an example, we set 
$\Delta s = 0.05$ and construct a $64^2$ cell mesh using \Cref{dynamic-alg1} 
with $\delta = 0.996$ in \eqref{static-test-function}. The resulting grid, shown in 
\Cref{fig:mesh-zoom-in-dyn-a,fig:mesh-zoom-in-dyn-b}, is smooth with $\Upsilon \approx 100$.  
In \Cref{sec:ALE-sims}, we consider large zoom-in meshing for the 
more complicated Rayleigh-Taylor test.

\subsection{Propagating circular front}\label{subsec:numerics-circular-front}
\subsubsection{Problem description}
Our first dynamic mesh generation experiment tracks a circular front propagating radially outwards with radial velocity 1. 
The time-dependent target Jacobian function is defined as 
\begin{equation}\label{dyn-circular-fn}
\bar{\Ge}(y,t) = 1 - \delta \exp \left\{  - \left| \sigma \left[ (y^1-0.5)^2 + (y^2 - 0.5)^2 - r(t)^2  \right] \right|^2  \right\} \,.
\end{equation}
The parameters are chosen as $\delta =0.75$, $\sigma = 64$, and the radius is $r(t) = 0.2 + t$. 
We generate a sequence of meshes for $0 \leq t \leq 0.1$. 

The choice of time step $\Delta t$ depends upon $N$ as
\begin{equation}\label{CFL-simple}
\Delta t = \frac{0.64}{2\sqrt{N}} \,.
\end{equation}
This choice of scaling for $\Delta t$ is motivated by the CFL condition.  Since the radial velocity of the propagating front is 1, we can 
estimate that the CFL number associated with \eqref{CFL-simple} is 0.64.

\subsubsection{Results}

\begin{figure}[ht]
\centering
\subfigure[$t=0.025$]{\label{fig:dyn-circular-mesh1}\includegraphics[width=40mm]{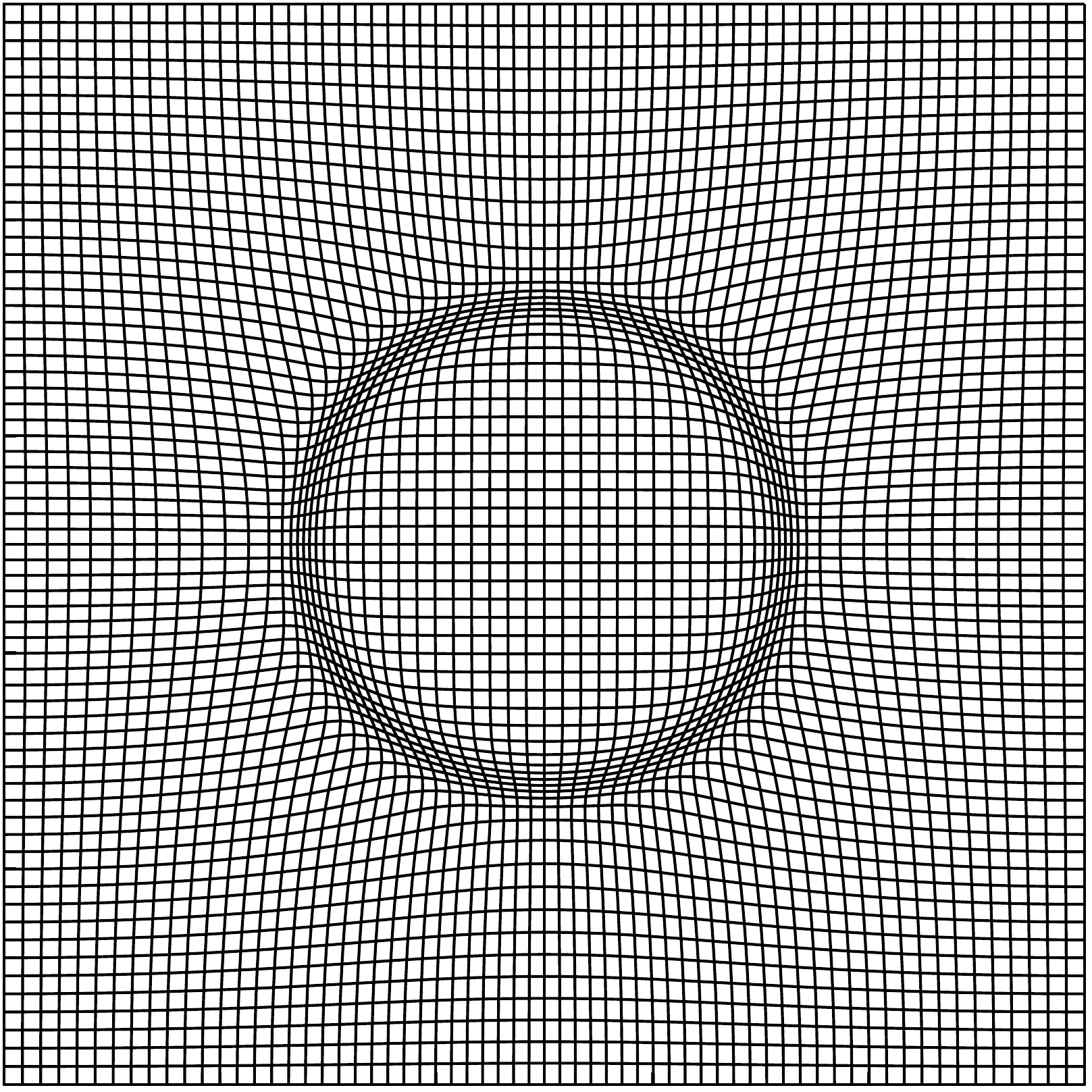}} 
\hspace{2em}
\subfigure[$t=0.05$]{\label{fig:dyn-circular-mesh2}\includegraphics[width=40mm]{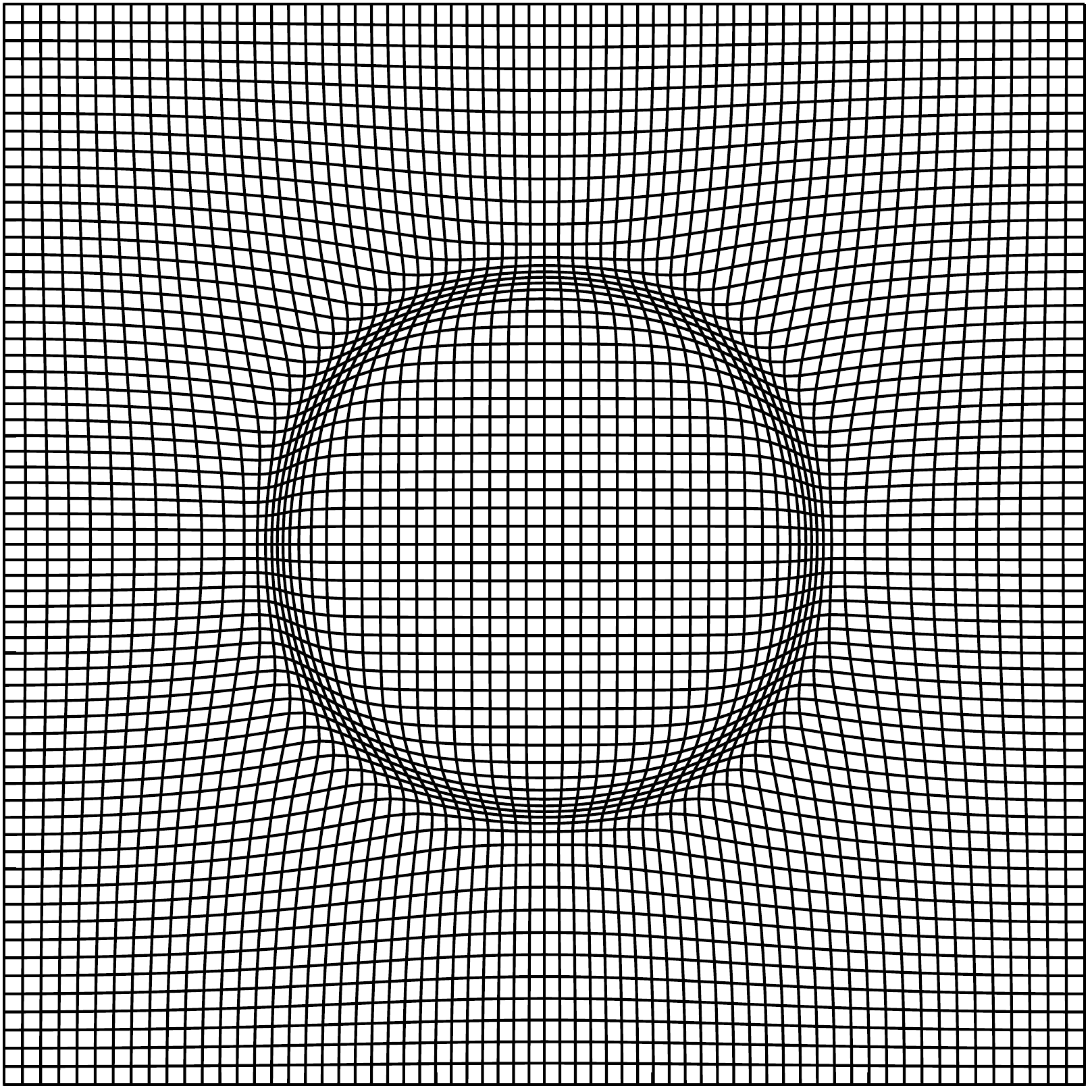}}
\hspace{2em}
\subfigure[$t=0.1$]{\label{fig:dyn-circular-mesh3}\includegraphics[width=40mm]{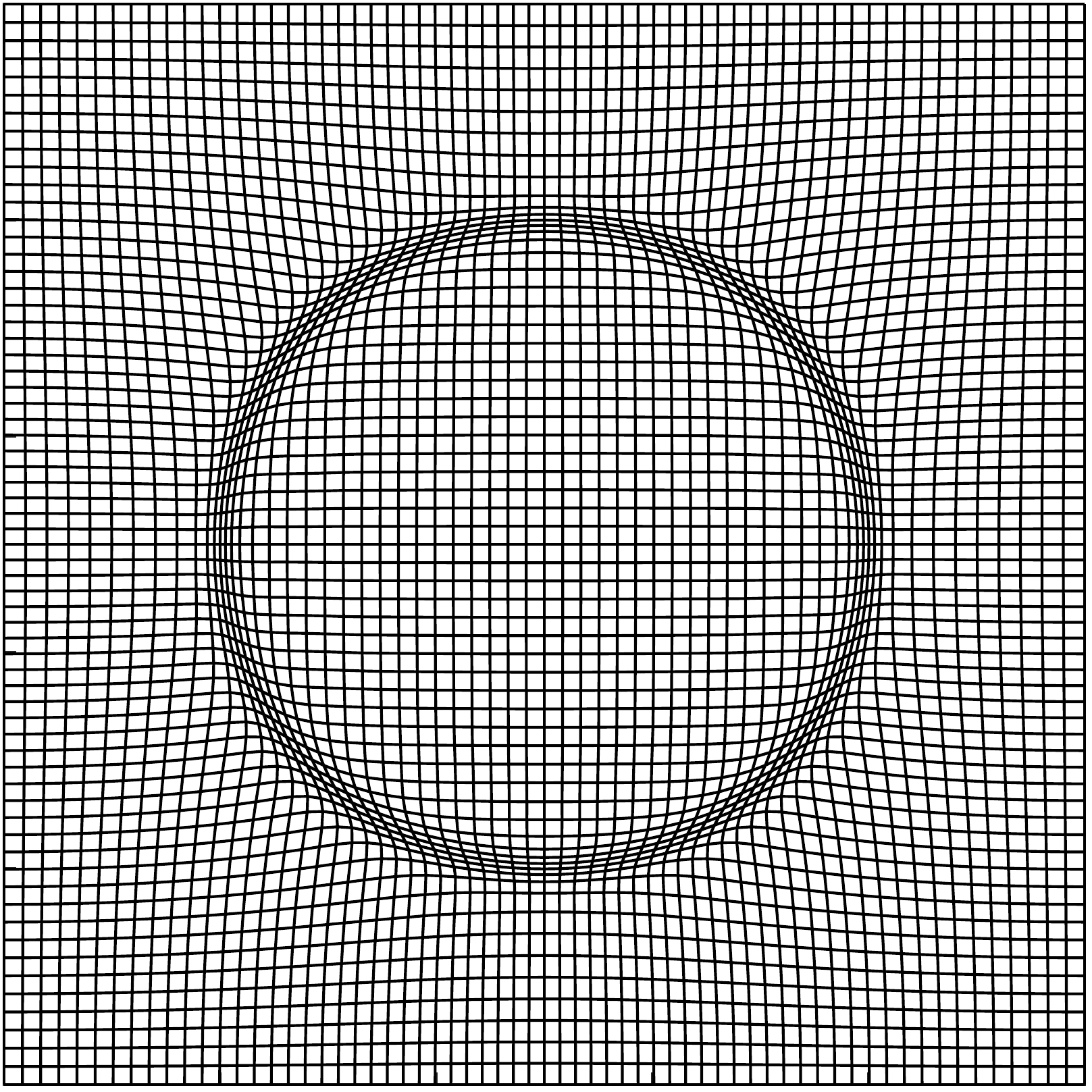}}
\hspace{2em}
\subfigure[$t=0.025$]{\label{fig:dyn-circular-mesh4}\includegraphics[width=40mm]{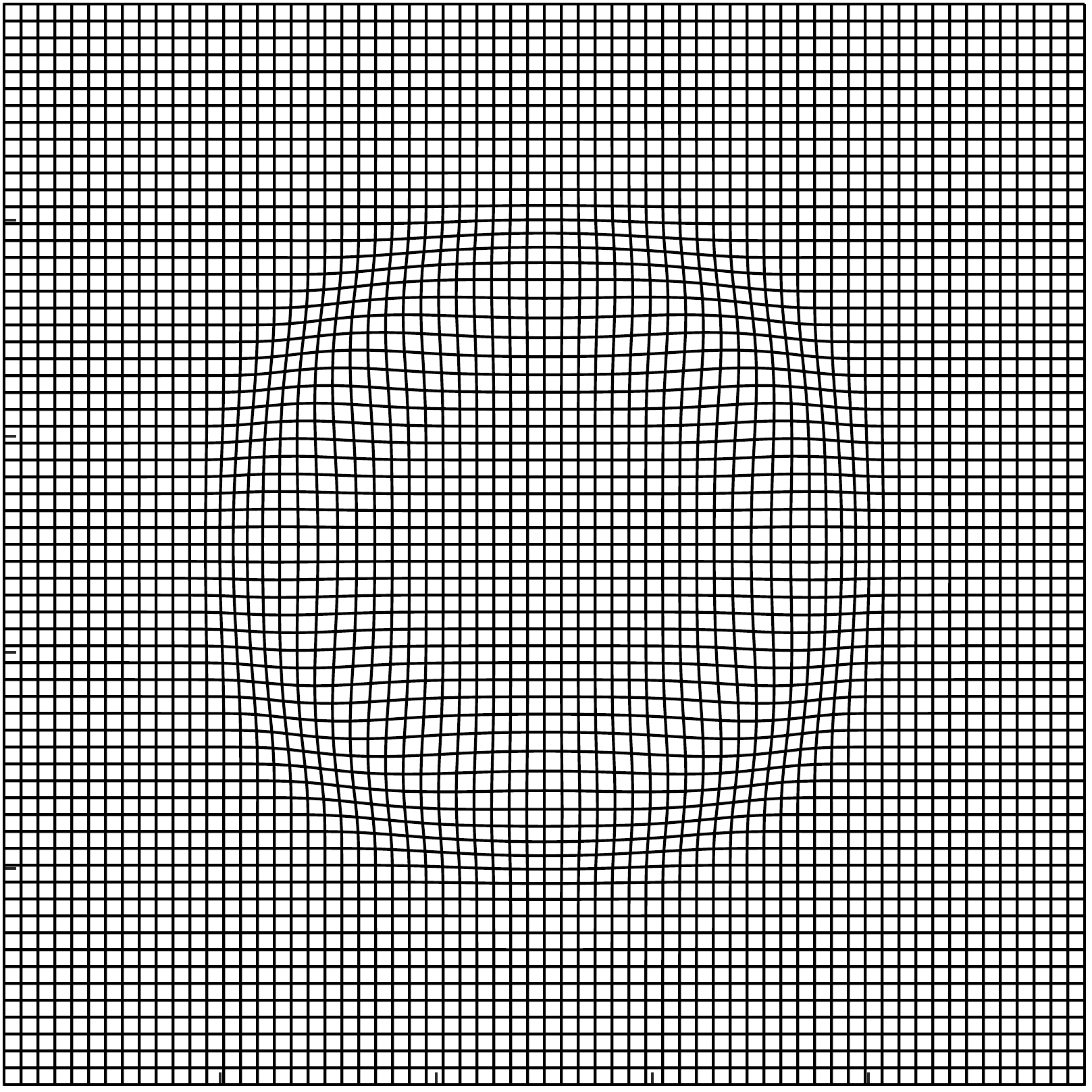}}
\hspace{2em}
\subfigure[$t=0.05$]{\label{fig:dyn-circular-mesh5}\includegraphics[width=40mm]{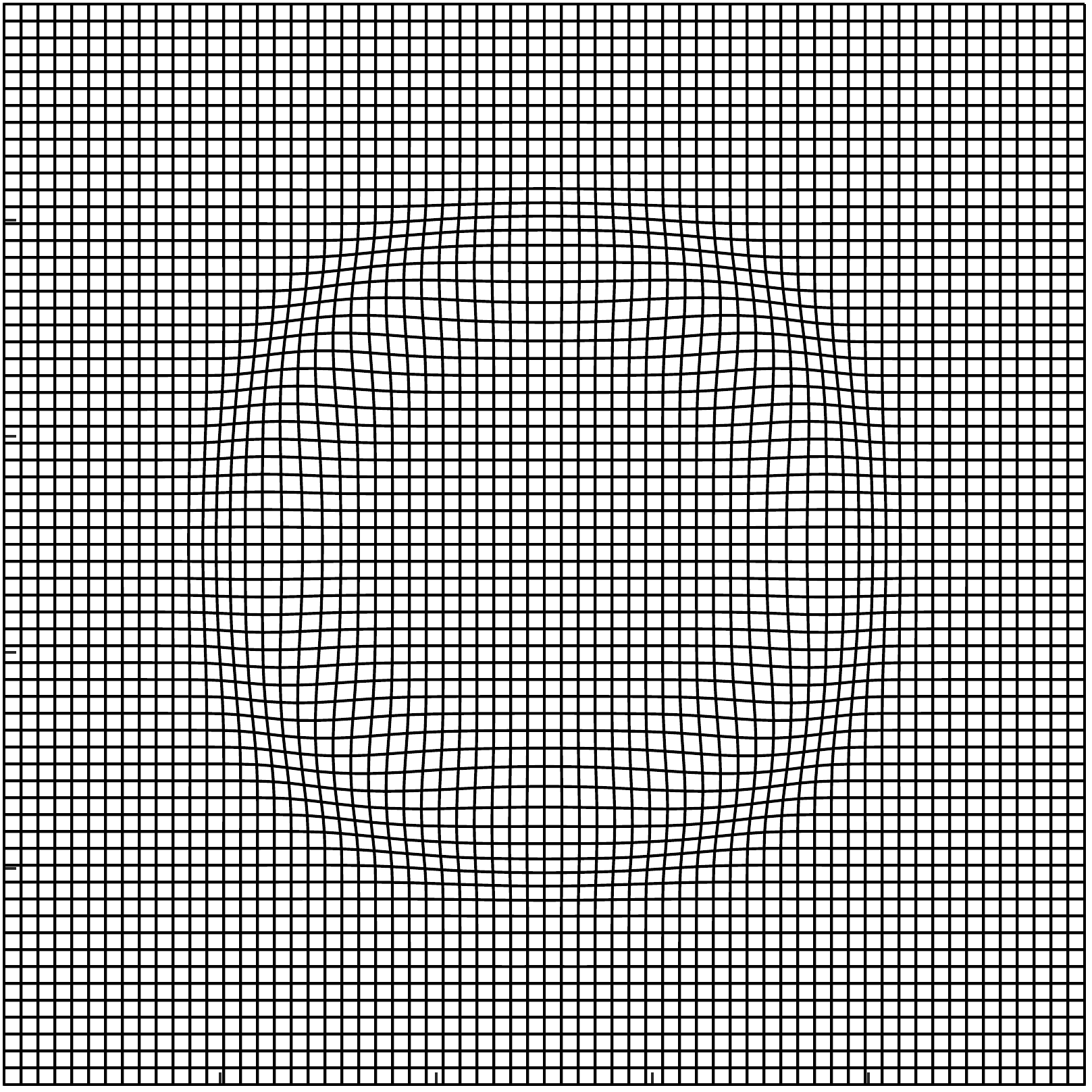}}
\hspace{2em}
\subfigure[$t=0.1$]{\label{fig:dyn-circular-mesh6}\includegraphics[width=40mm]{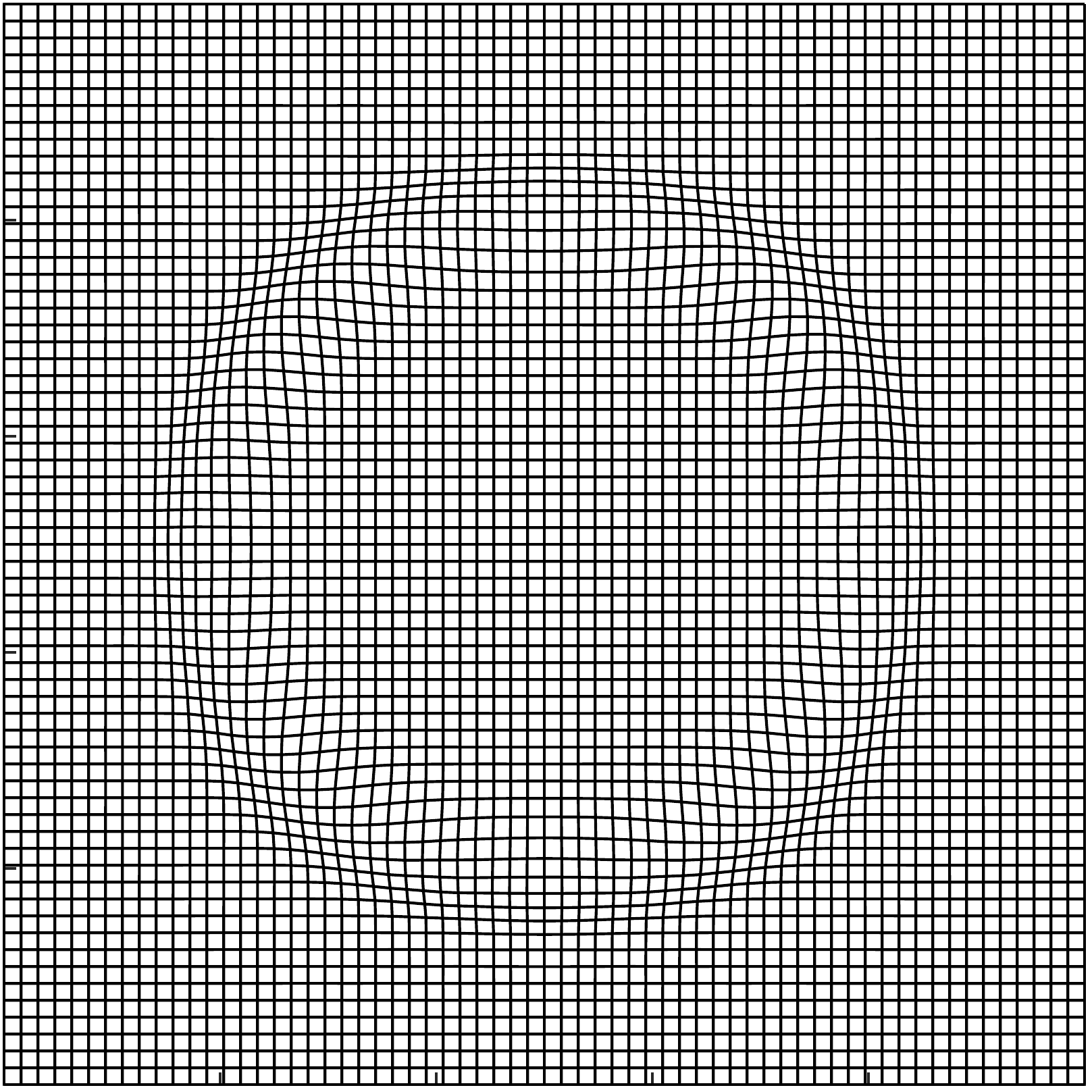}}
\caption{
Test problem \ref{subsec:numerics-circular-front}: tracking a propagating circular front modeling a shock wave.  Shown are the
$64^2$ cell meshes produced using dynamic SAM for the circular target Jacobian function \eqref{dyn-circular-fn}. 
The top row shows the adaptive meshes $\mathcal{T}(t)$, while the bottom row shows the corresponding 
``perturbation meshes'' $\delta \psi_k(\Tref)$.}
\label{fig:dyn-circular-mesh}
\end{figure}

The $64^2$ cell adaptive meshes $\mathcal{T}(t)$ for \eqref{dyn-circular-fn} are shown in the top row of 
\Cref{fig:dyn-circular-mesh} at various times $t$. The computed meshes $\mathcal{T}_k$ are smooth and are correctly resolved around 
the evolving circular front.  
The meshes $\delta \psi_{k}(\Tref)$ are shown at the same times in the bottom row of 
\Cref{fig:dyn-circular-mesh}; from these figures, it is clear that $\delta \psi_{k}(\Tref)$ is a near-identity transformation of the uniform 
 mesh $\Tref$. For this problem, the function $\Ge$ is such that 
 $||1 - 1/\Ge(\cdot,t) ||_{L^\infty} \approx 2.43$, whereas the perturbation density $P$ is such that 
$||1 - 1/P(\cdot,t) ||_{L^\infty} \approx 0.3$. 

\subsubsection{Comparison with static SAM}
Next, we conduct a grid resolution study with $N$ ranging from $N=32^2$ to $N=512^2$, and compare the results of dynamic SAM with 
those of static SAM. The Jacobian errors $\mathcal{E}_2$ at the final time $t=0.1$ are shown 
in \Cref{table:dyn-circular-static}. Both schemes exhibit 4\textsuperscript{th} order accuracy, as expected, but the dynamic 
SAM solutions have smaller errors. This is due to the higher accuracy of the Poisson solve in the dynamic 
method vs the static method.

\begin{table}[ht]
\centering
\renewcommand{\arraystretch}{1.0}
\scalebox{0.8}{
\begin{tabular}{|lr|cccccc|}
\toprule
\midrule
\multirow{2}{*}{\textbf{Scheme}} &  & \multicolumn{6}{c|}{\textbf{Cells}}\\

{}  &    & $32 \times 32$    & $64 \times 64$   & $128 \times 128 $ & $256 \times 256$ & $512 \times 512$ & $1024 \times 1024$ \\
\midrule
\multirow{3}{*}{Static SAM} & $\mathcal{E}_2$ & 
$4.23 \times 10^{-2}$  & $1.15 \times 10^{-2}$  & $1.22 \times 10^{-3}$ & $9.29 \times 10^{-5}$  & $4.98 \times 10^{-6}$ & $2.65 \times 10^{-7}$ \\
				    & Order & -- & 1.9   & 3.2  & 3.7 & 4.2 & 4.2 \\[0.5em]
				    & $T_{\mathrm{CPU}}$ (sec) & 0.006 & 0.049   & 0.59  & 7.26 & 112.8 & 1663 \\
\midrule
\multirow{4}{*}{Dynamic SAM} & $\mathcal{E}_2$ & 
$4.05 \times 10^{-2}$  & $6.66 \times 10^{-3}$  & $6.18 \times 10^{-4}$ & $4.00 \times 10^{-5}$  & $2.23 \times 10^{-6}$ & $1.33 \times 10^{-7}$ \\
				    & Order & -- & 2.6   & 3.4  & 3.9 & 4.2 & 4.1 \\[0.5em]
				    & $T_{\mathrm{CPU}}$ (sec) & 0.017 & 0.101   & 0.869  & 7.31 & 65.2 & 582 \\
				    & speed-up factor & 0.33 & 0.48   & 0.68  & 0.99 & 1.73 & 2.86 \\
\midrule
\bottomrule
\end{tabular}}
\caption{Test problem \ref{subsec:numerics-circular-front}: tracking a propagating circular front. We list the
$L^2$ Jacobian errors $\mathcal{E}_2$ at $t=0.1$, convergence rates, and total CPU runtimes
for static and dynamic SAM. The results confirm that dynamic SAM produces high order 
accurate solutions and is of optimal complexity.}
\label{table:dyn-circular-static}
\end{table}

At low resolutions, the dynamic SAM runtimes are greater than those for static SAM.  This is due to the interpolation required in 
the dynamic SAM algorithm.  On the other hand, static SAM is of complexity 
$\mathcal{O}\left(N^{3/2}/\Delta t \right) = \mathcal{O}\left( N^2 \right)$, whereas dynamic 
SAM is of optimal complexity 
$\mathcal{O}\left( N/\Delta t \right) = \mathcal{O}\left(N^{3/2} \right)$.  
For this test, dynamic SAM becomes more efficient than 
static SAM at $N = 256^2$.

\subsection{Uniformly rotating patch}\label{subsec:numerics-rotating-patch}

\subsubsection{Problem description}
Our next mesh generation experiment assesses the performance of SAM for target Jacobian functions of the form 
\begin{equation}\label{patch-fn}
\bar{\Ge}(y,t) = \frac{1}{1 + M  \exp \left\{  - \left( \sigma \Bigl\lvert \left[ (y^1-0.5-r\cos(2 \pi t))^2 + (y^2 - 0.5-r\sin(2\pi t))^2 - R^2  \right] \Bigr\rvert \right)^2  \right\} } \,.
\end{equation}
\Cref{patch-fn} forces the mesh to concentrate nodes within a uniformly rotating (with angular velocity $\omega=2\pi$) 
circular patch of radius $R > 0$, whose 
center is a distance $r \geq 0$ from $(0.5,0.5)$.  The constant $M \geq 0$ determines the zoom-in factor, and $\sigma$ controls the 
width of the transition region from fine to coarse scale of the mesh. 

\subsubsection{Comparison with the schemes in \cite{Sulman2011}}

The case $M=5$, $\sigma=50$, $r=0.25$, and $R=0.1$ in \eqref{patch-fn} corresponds to a test problem from
\cite{Sulman2011}.  Therein, the authors compare four different mesh generation methods and conclude 
that the so-called Parabolic Monge-Kantorovich method (PMKP) is the best 
method among the four for \eqref{patch-fn}, both in terms of accuracy as well as efficiency. 
The PMKP method is similar to the MK scheme, but replaces the nonlinear Newton-Krylov solver in MK with 
a parabolization (in pseudo-time $\tau$) and time-stepping until a steady state is reached. 
The solution in PMKP is only found in the asymptotic limit $\tau \to \infty$, whereas the SAM solution is 
computed at pseudo-time $\tau=1$. Moreover, the explicit integration of the parabolic PDE requires that the 
pseudo-time step scales like $\Delta \tau \sim \frac{1}{N}$ to ensure 2\textsuperscript{nd} order convergence.  
In contrast, static SAM 
requires only that $\Delta \tau \sim \frac{1}{\sqrt{N}}$, while for dynamic 
SAM we can keep  $\Delta \tau = \mathcal{O}(1)$.

\begin{figure}[!htb]
\centering
\subfigure[$t=0.25$]{\label{fig:patch-static1}\includegraphics[width=38mm]{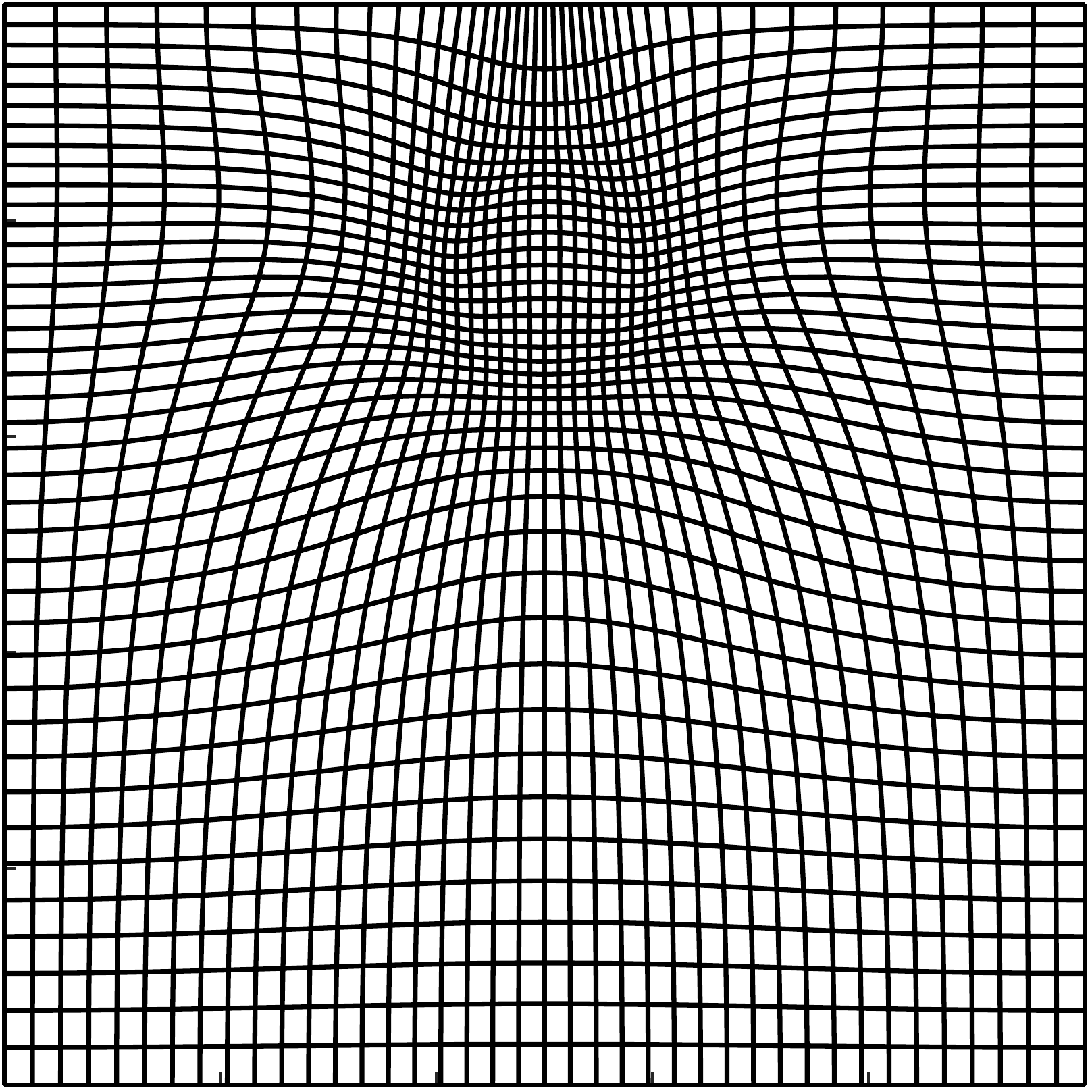}} 
\hspace{.25em}
\subfigure[$t=0.5$]{\label{fig:patch-static2}\includegraphics[width=38mm]{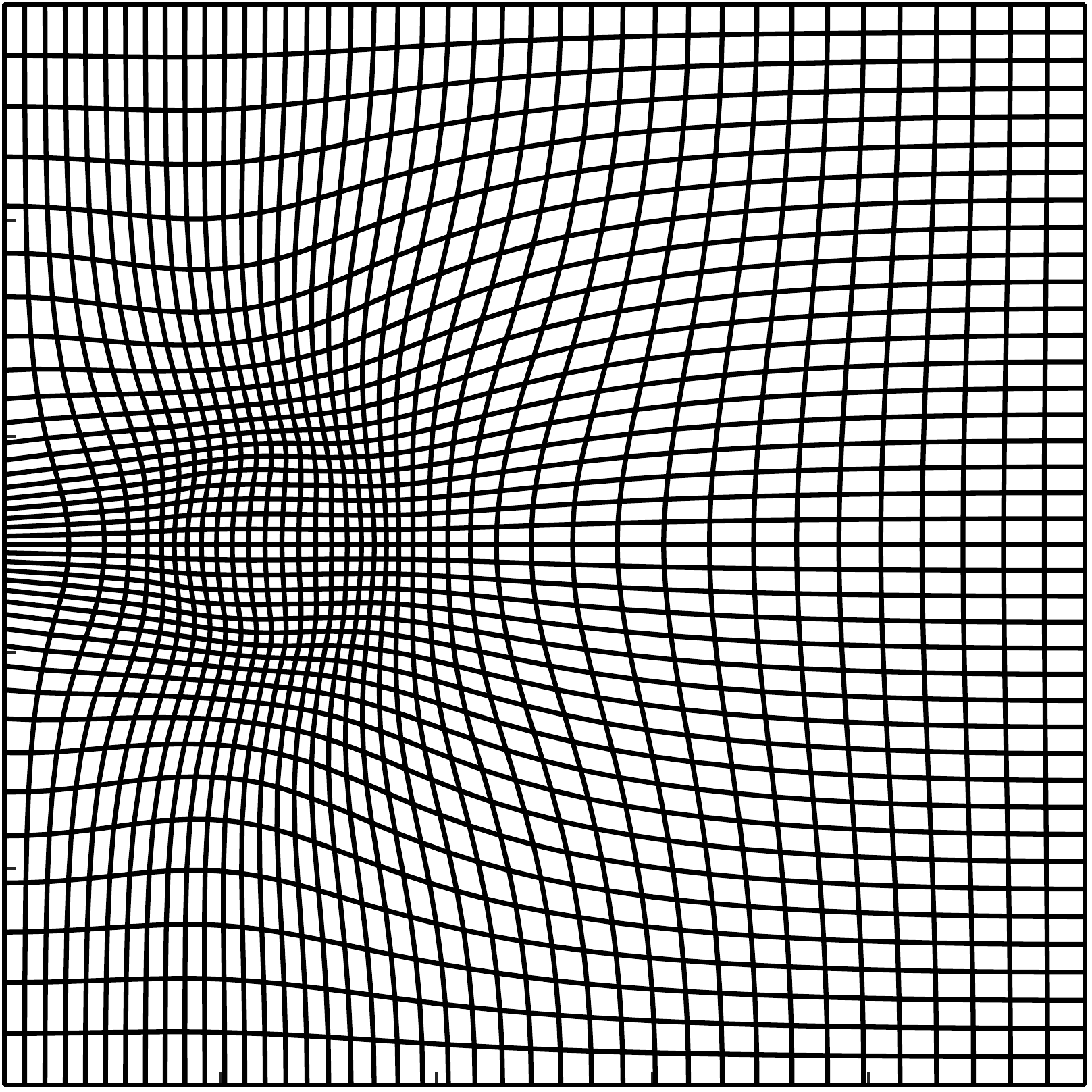}}
\hspace{.25em}
\subfigure[$t=0.75$]{\label{fig:patch-static3}\includegraphics[width=38mm]{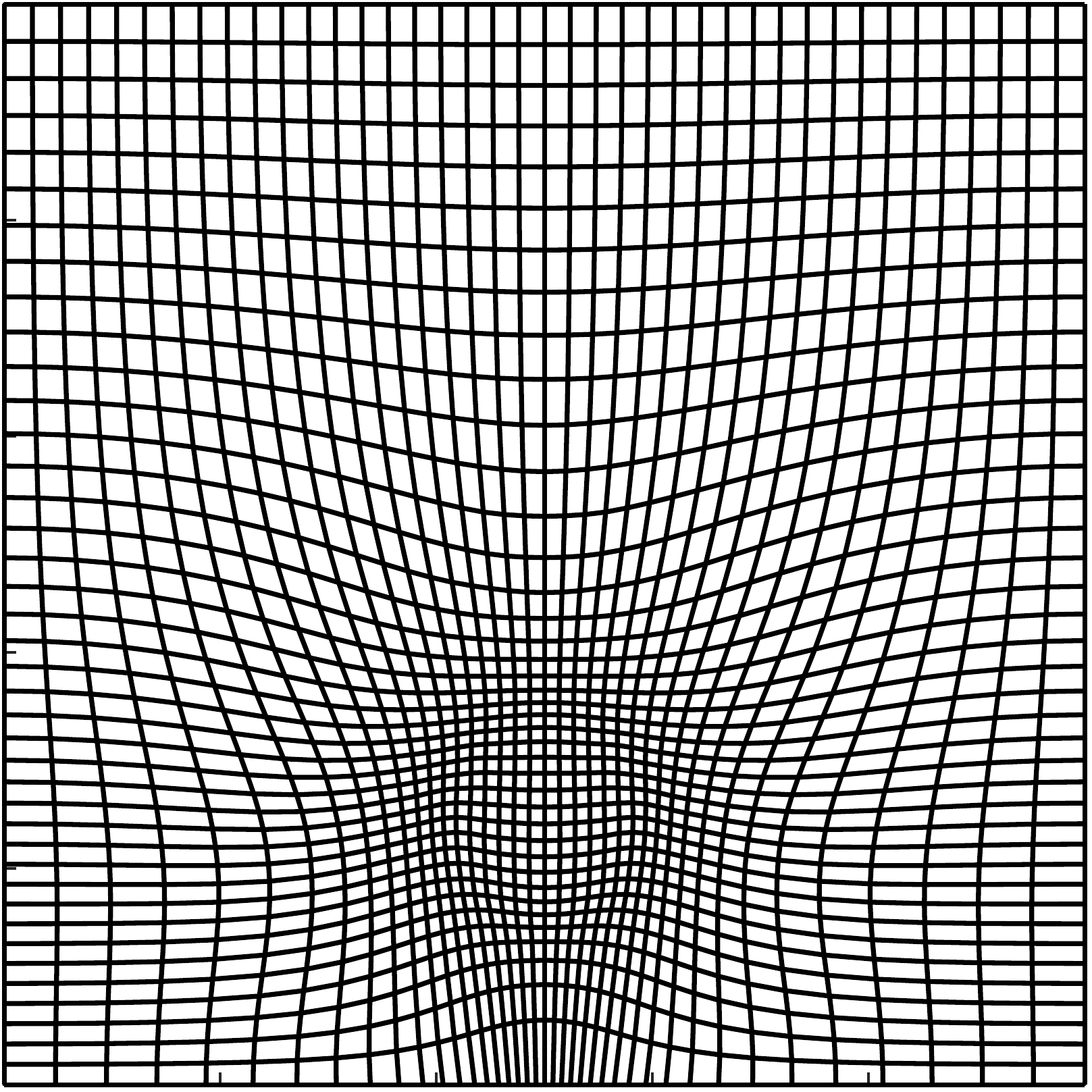}}
\hspace{.25em}
\subfigure[$t=1.0$]{\label{fig:patch-static4}\includegraphics[width=38mm]{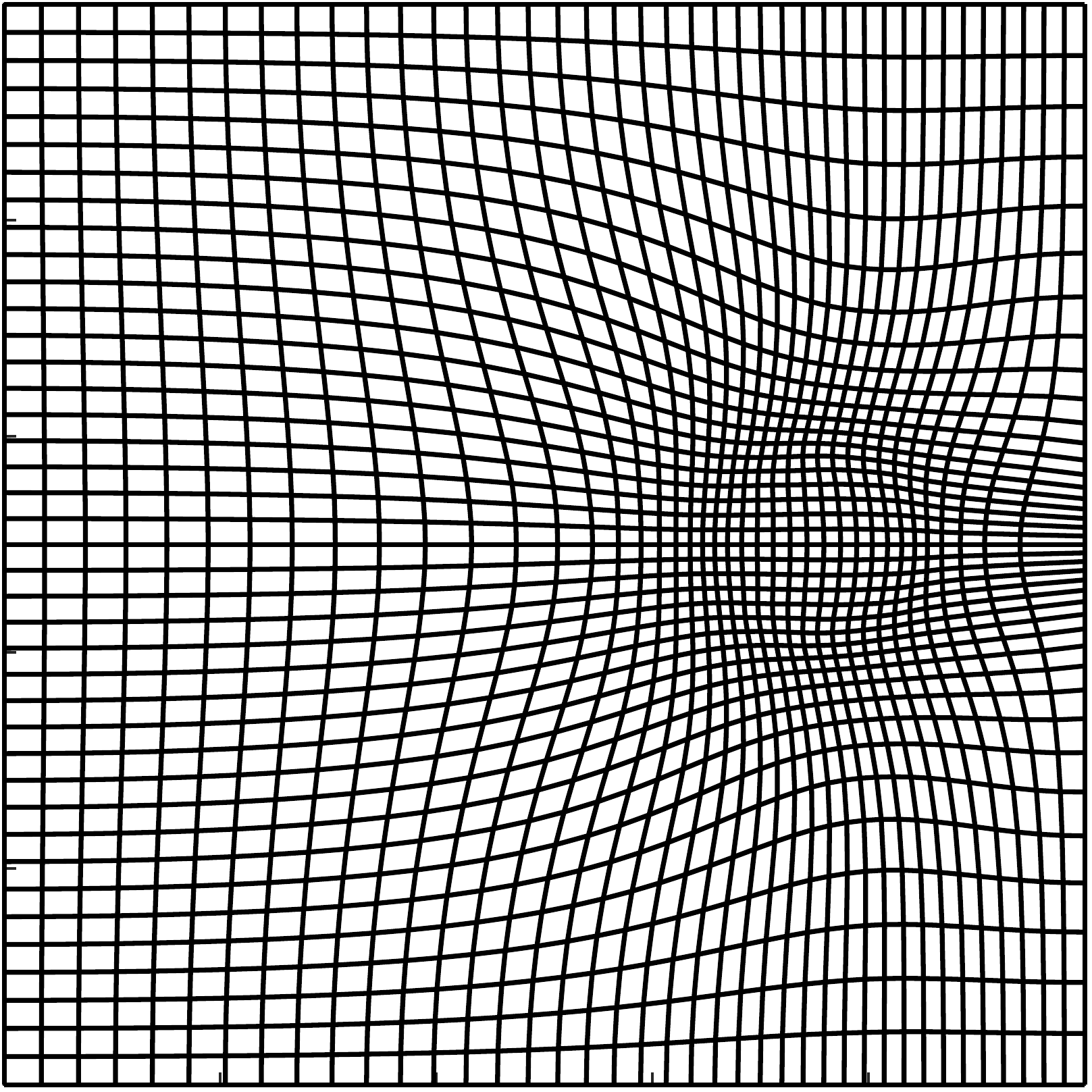}}
\hspace{.25em}
\subfigure[$t=0.25$]{\label{fig:patch-dyn1}\includegraphics[width=38mm]{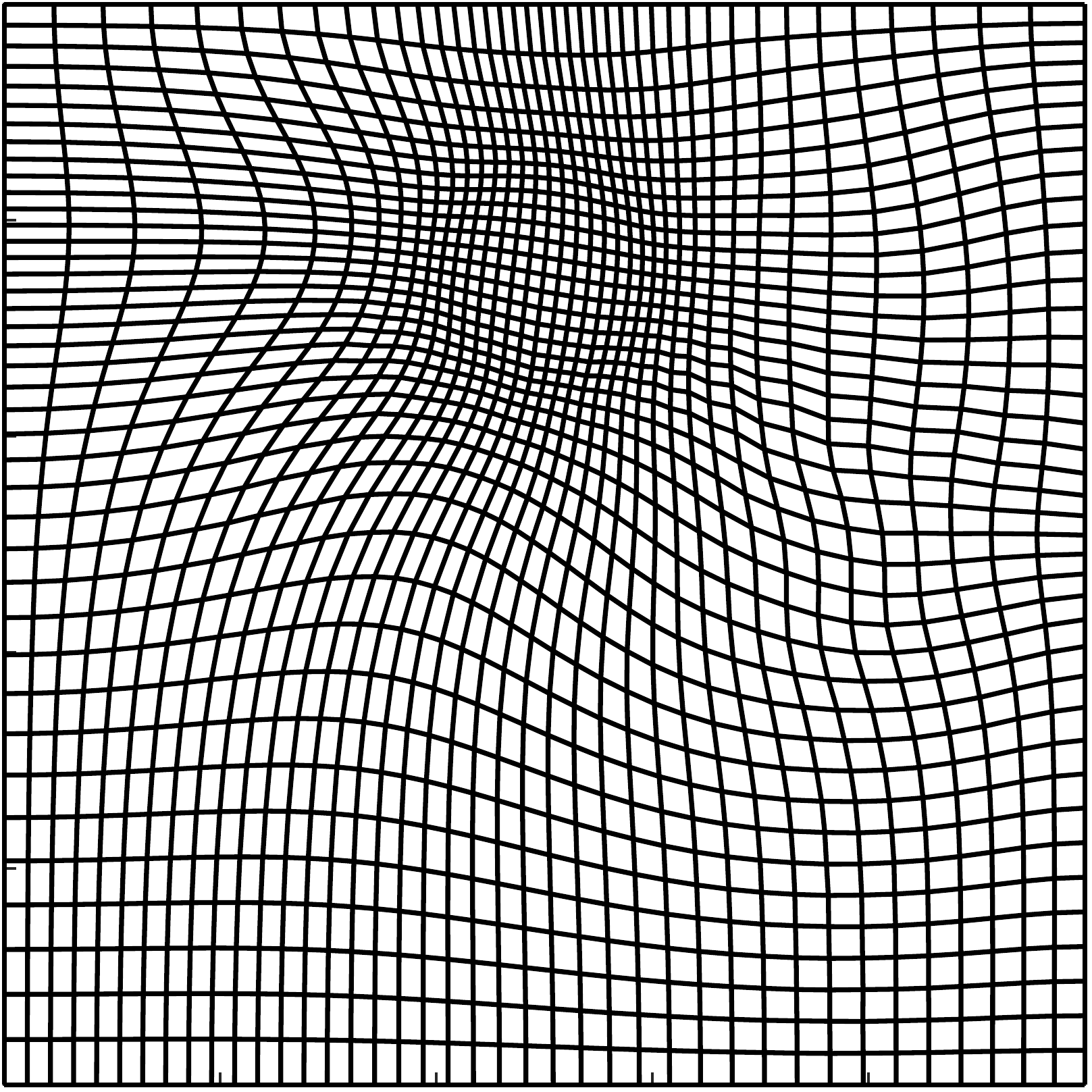}}
\hspace{.25em}
\subfigure[$t=0.5$]{\label{fig:patch-dyn2}\includegraphics[width=38mm]{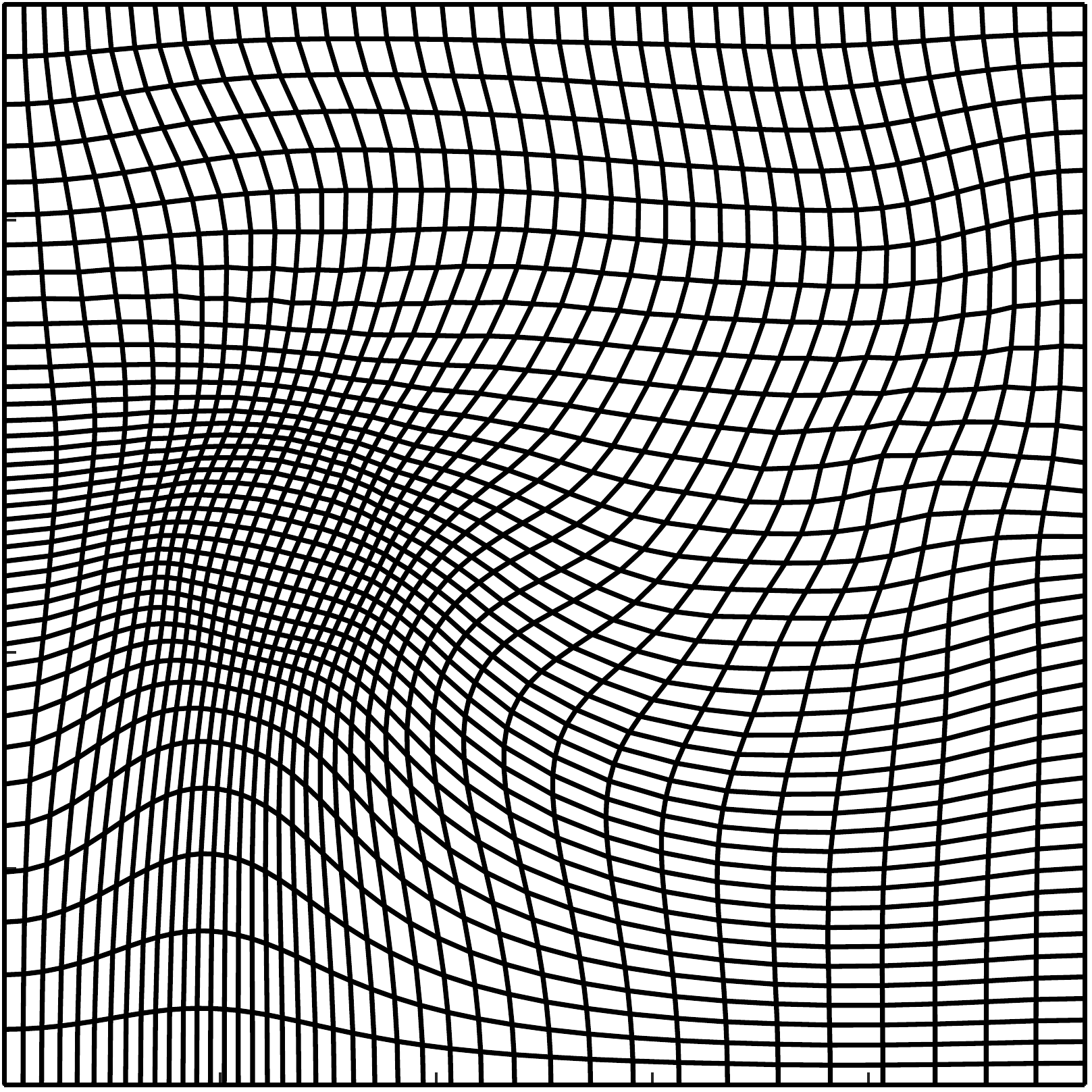}}
\hspace{.25em}
\subfigure[$t=0.75$]{\label{fig:patch-dyn3}\includegraphics[width=38mm]{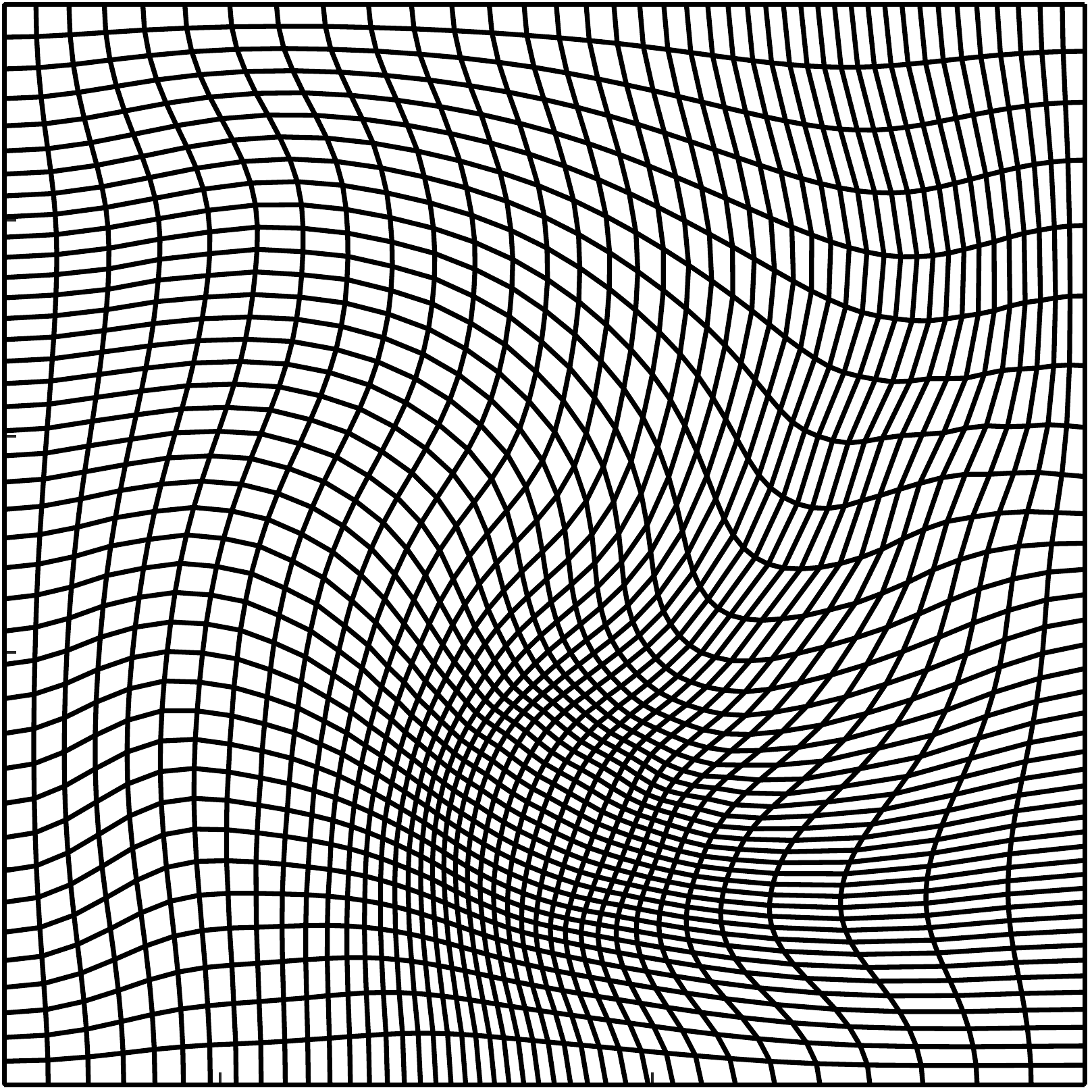}}
\hspace{.25em}
\subfigure[$t=1.0$]{\label{fig:patch-dyn4}\includegraphics[width=38mm]{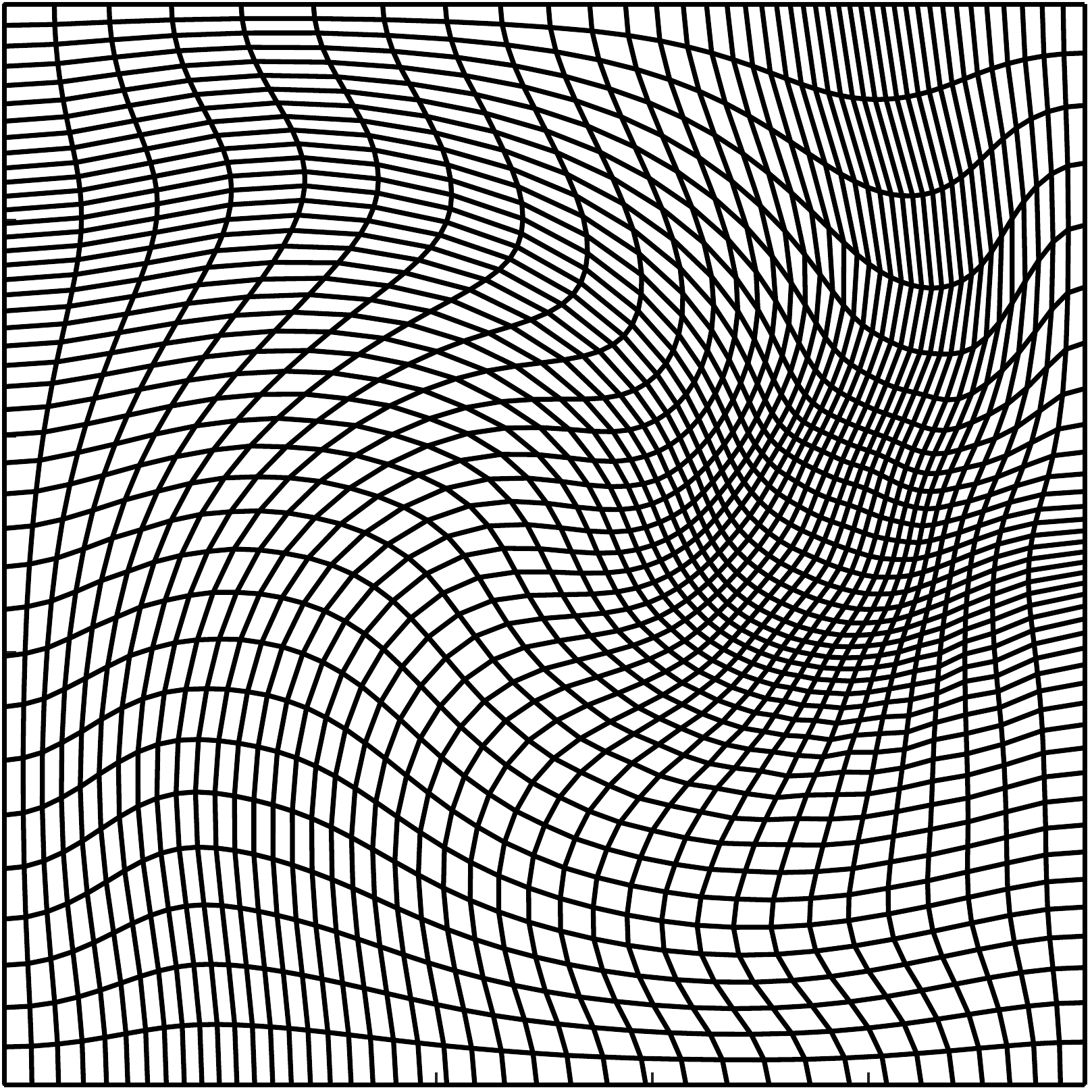}}
\hspace{.25em}
\subfigure[$t=0.25$]{\label{fig:patch-rest1}\includegraphics[width=38mm]{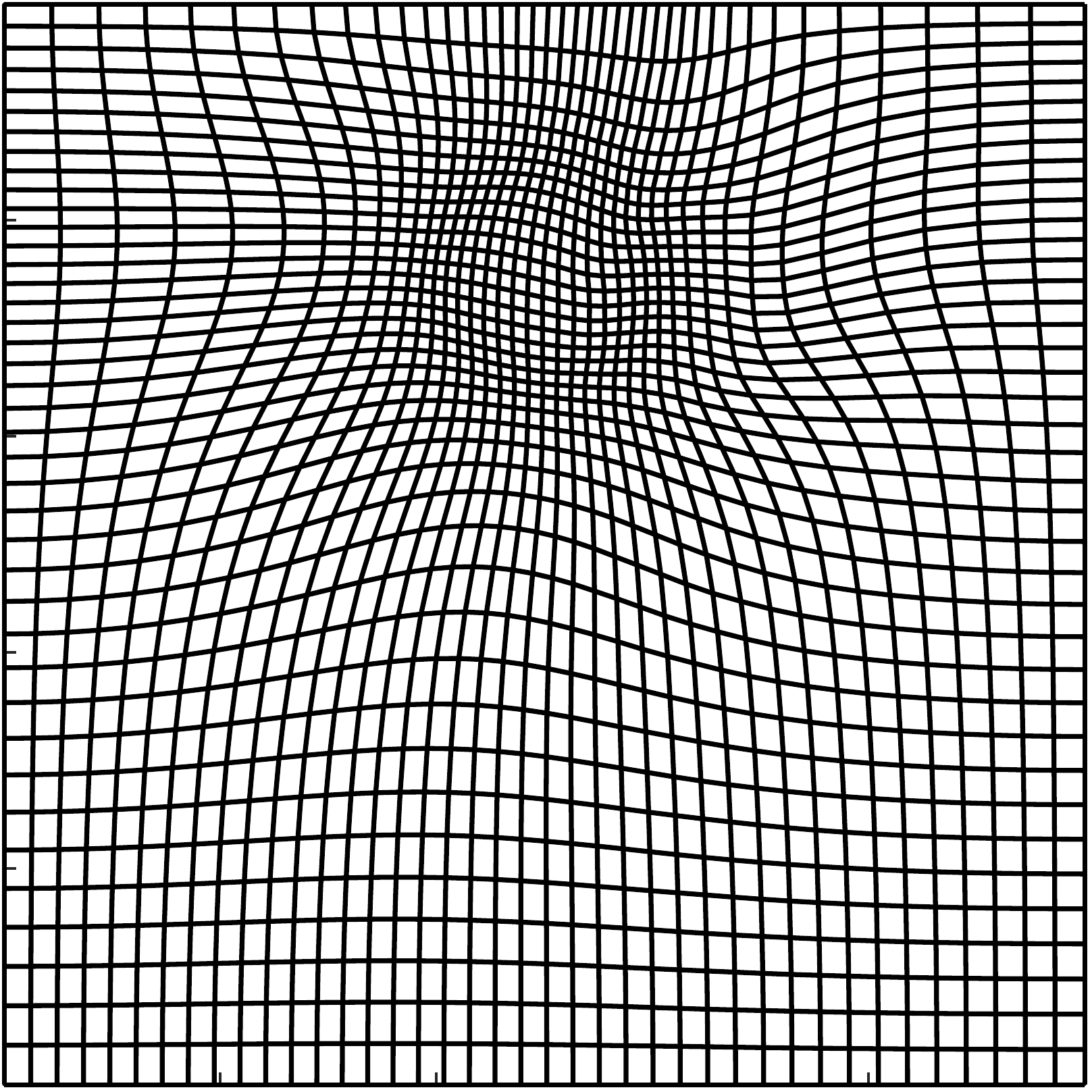}}
\hspace{.25em}
\subfigure[$t=0.5$]{\label{fig:patch-rest2}\includegraphics[width=38mm]{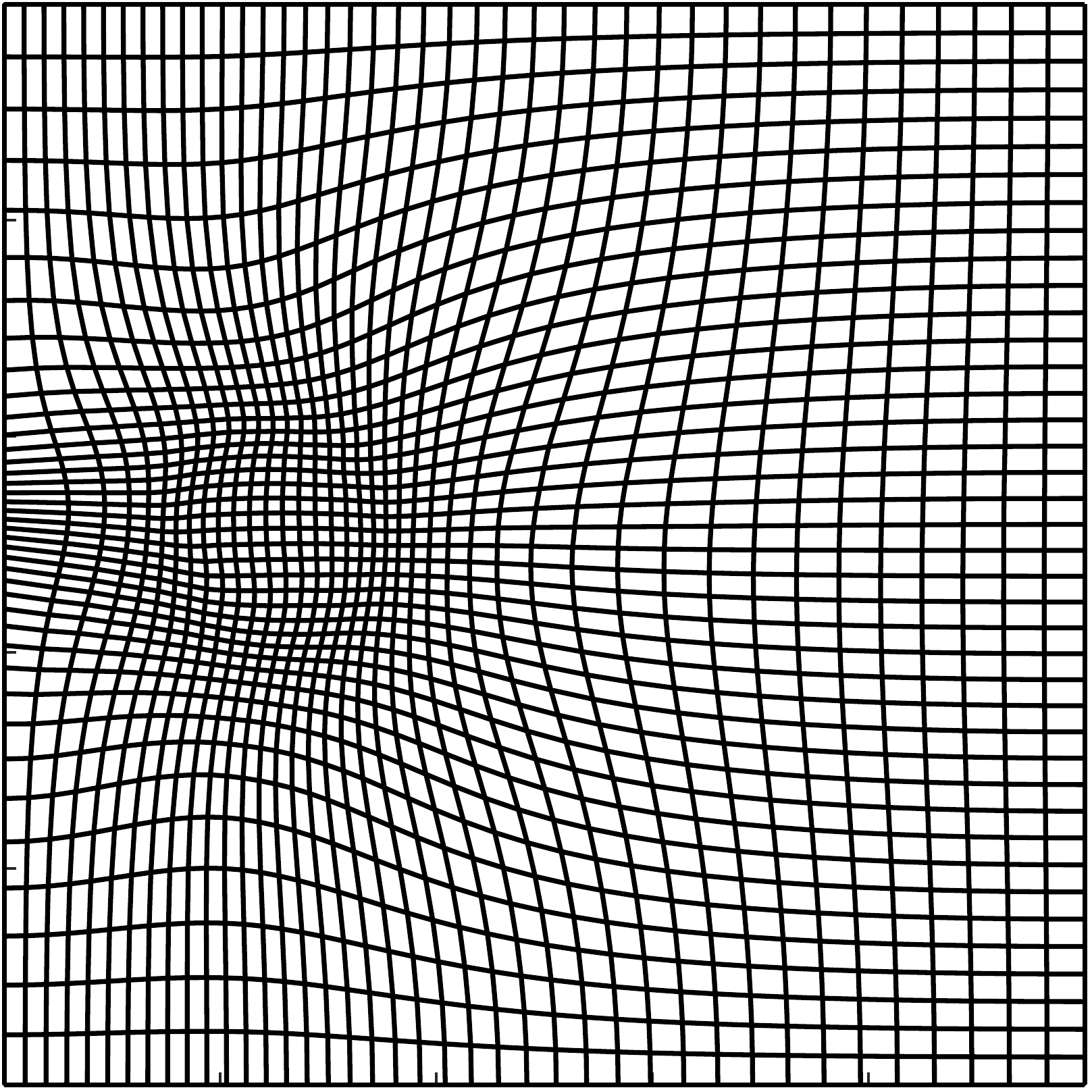}}
\hspace{.25em}
\subfigure[$t=0.75$]{\label{fig:patch-rest3}\includegraphics[width=38mm]{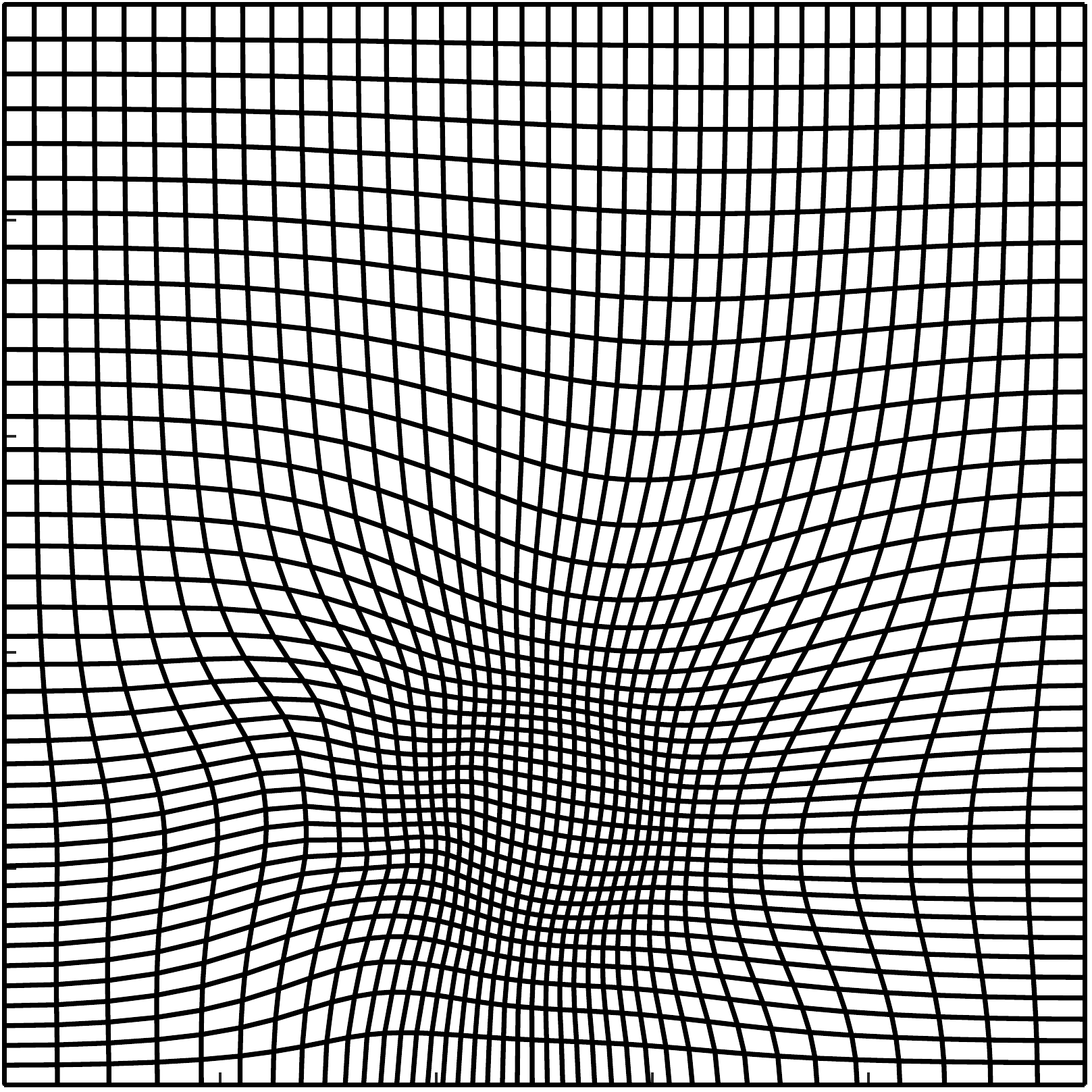}}
\hspace{.25em}
\subfigure[$t=1.0$]{\label{fig:patch-rest4}\includegraphics[width=38mm]{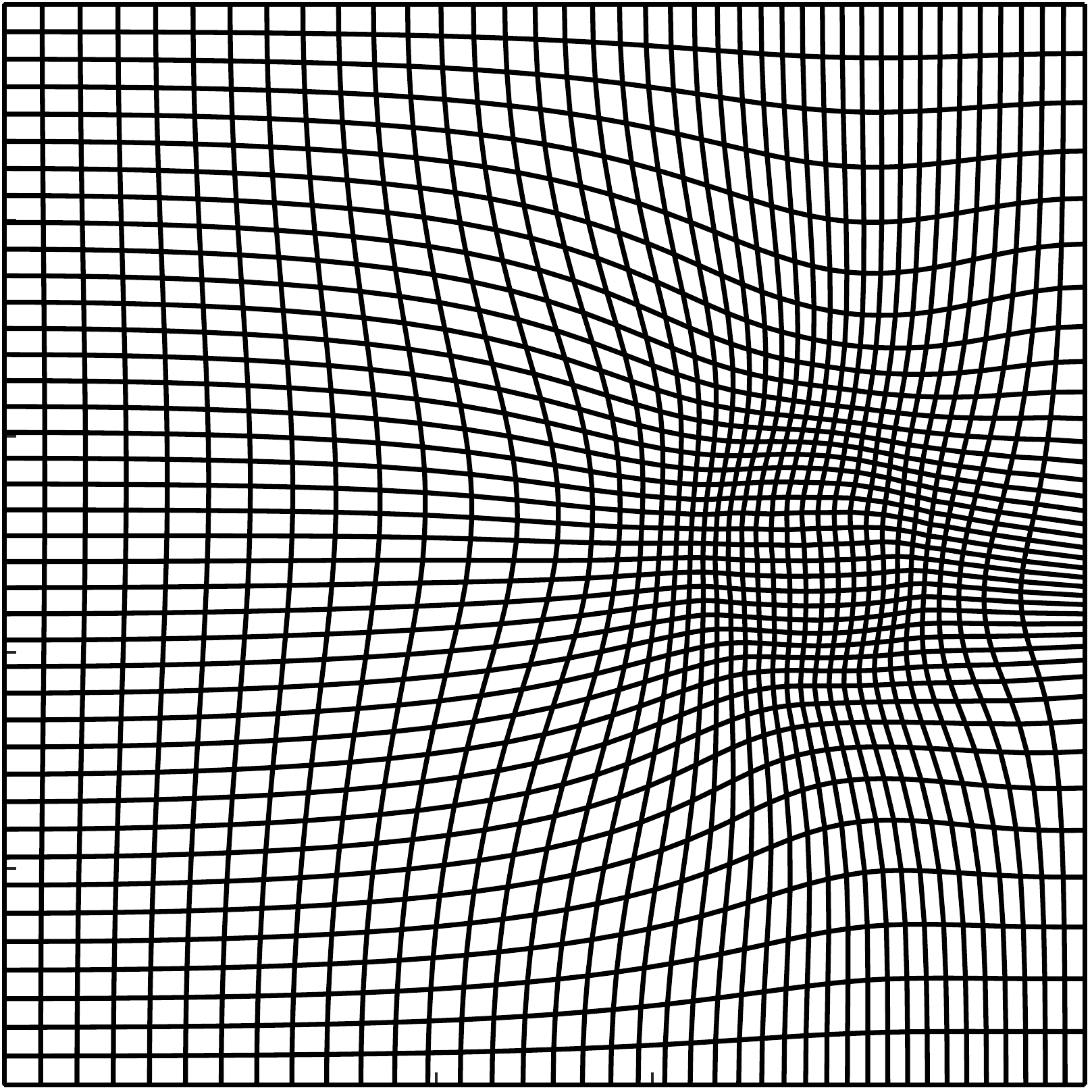}}
\caption{
Test problem \ref{subsec:numerics-rotating-patch}: tracking a uniformly rotating patch with target function \eqref{patch-fn}. 
Shown are plots of the $40^2$ cell adaptive meshes at various times $t$. 
The meshes are produced using static SAM (top), dynamic SAM (middle), and restarted SAM (bottom). 
Restarted SAM removes the grid distortion errors associated with Lagrangian methods.}
\label{fig:patch}
\end{figure}

We set $N=40^2$, $\Delta t = 0.01$, and generate a sequence 
of meshes for $0 \leq t \leq 1$.  The adaptive meshes generated with static, dynamic, and restarted SAM  are shown in \Cref{fig:patch} 
at various times $t$. The Jacobian errors, mean grid distortion, and cumulative simulation runtimes
are provided in \Cref{table:patch}. For the purposes of comparison with \cite{Sulman2011}, we also provide the 
\emph{mesh fidelity measure} $\hat{\mathcal{E}}_2$, defined by 
\begin{equation}\label{mesh-fidelity}
\hat{\mathcal{E}}_2 \coloneqq \Bigl\lvert  ||  \mathcal{J}(\cdot,t)/\Ge \circ \psi(\cdot,t) ||_{L^2} - 1 \Bigr\rvert \,. 
\end{equation}
The superior accuracy of SAM produces fidelity measures $\hat{\mathcal{E}}_2$ that are an {order of magnitude smaller}
than those produced with the PMKP method (see Tables 6 and 7 in \cite{Sulman2011}). 
Moreover, the SAM runtimes are more than two orders of magnitude smaller than the PMKP runtimes provided in \cite{Sulman2011} e.g. 
0.179 sec vs 75 sec for static SAM vs PMKP.

\begin{table}[ht]
\centering
\renewcommand{\arraystretch}{1.0}
\scalebox{0.8}{
\begin{tabular}{|lr|ccccc|}
\toprule
\midrule
\multirow{2}{*}{\textbf{Scheme}} &  & \multicolumn{5}{c|}{\textbf{Time}}\\

{}  &    & $t=0$    & $t=0.25$   & $t=0.5$ & $t=0.75$ & $t=1.0$ \\
\midrule
\multirow{4}{*}{\vspace{-1.5em}Static SAM} 
& $\hat{\mathcal{E}}_2$ & $3.86 \times 10^{-3}$  & $3.86 \times 10^{-3}$  & $3.86 \times 10^{-3}$ & $3.86 \times 10^{-3}$  & $3.86 \times 10^{-3}$ \\[0.5em]
& $\mathcal{E}_2$ & $3.79 \times 10^{-2}$ & $3.79 \times 10^{-2}$   & $3.79 \times 10^{-2}$  & $3.79 \times 10^{-2}$ & $3.79 \times 10^{-2}$ \\[0.5em]
& $\lambda$ & 1.251 & 1.251   & 1.251  & 1.251 & 1.251 \\[0.5em]
& $T_{\mathrm{CPU}}$ (sec) & 0.003 & 0.048   & 0.091  & 0.134 & 0.179 \\
\midrule
\multirow{4}{*}{\vspace{-1.5em}Dynamic SAM} 
& $\hat{\mathcal{E}}_2$ & $3.86 \times 10^{-3}$  & $1.13 \times 10^{-3}$  & $1.13 \times 10^{-3}$ & $1.49 \times 10^{-3}$  & $6.65 \times 10^{-3}$ \\[0.5em]
& $\mathcal{E}_2$ & $3.79 \times 10^{-2}$ & $3.28 \times 10^{-2}$   & $3.49 \times 10^{-2}$  & $4.79 \times 10^{-2}$ & $6.31 \times 10^{-2}$ \\[0.5em]
& $\lambda$ & 1.251 & 1.328   & 1.544  & 1.877 & 2.311 \\[0.5em]
& $T_{\mathrm{CPU}}$ (sec) & 0.003 & 0.049   & 0.094  & 0.139 & 0.185 \\
\midrule
\multirow{4}{*}{\vspace{-1.5em}Restarted SAM} 
& $\hat{\mathcal{E}}_2$ & $3.86 \times 10^{-3}$  & $2.85 \times 10^{-3}$  & $1.18 \times 10^{-3}$ & $3.86 \times 10^{-3}$  & $2.85 \times 10^{-3}$ \\[0.5em]
& $\mathcal{E}_2$ & $3.79 \times 10^{-2}$ & $3.05 \times 10^{-2}$   & $2.15 \times 10^{-2}$  & $3.79 \times 10^{-2}$ & $3.05 \times 10^{-2}$ \\[0.5em]
& $\lambda$ & 1.251 & 1.252   & 1.259  & 1.251 & 1.252 \\[0.5em]
& $T_{\mathrm{CPU}}$ (sec) & 0.003 & 0.048   & 0.093  & 0.138 & 0.183 \\
\midrule
\bottomrule
\end{tabular}}
\caption{Test problem \ref{subsec:numerics-rotating-patch}: tracking a uniformly rotating patch. We provide the
mesh fidelity measure $\hat{\mathcal{E}}_2$, $L^2$ Jacobian error $\mathcal{E}_2$, $L^1$ distortion $\lambda$, and 
cumulative CPU runtime
 $T_{\mathrm{CPU}}$ at various times $t$ for the static, dynamic, and restarted SAM schemes. The mesh fidelity 
 measures $\hat{\mathcal{E}}_2$ of SAM solutions are an order of magnitude smaller than those provided in 
 \cite{Sulman2011}. Additionally, static SAM is more than 400 times faster than the schemes in \cite{Sulman2011}.}
\label{table:patch}
\end{table}

Since static SAM constructs the map $\psi$ directly from the uniform mesh, the meshes at 
$t=0.25, 0.5, 0.75, 1.0$ are simply rotated versions of the initial grid. This is confirmed in \Cref{table:patch}, which shows that 
static SAM produces grids with identical grid quality metrics at these times.  Dynamic SAM, on the other hand, necessarily tracks 
the history of the simulation, and the rotating target Jacobian produces  grids with increasing levels of 
distortion.  The Jacobian errors of dynamic SAM are smaller than 
static SAM for $t=0.25$ and $t=0.5$ due to the high accuracy with which the Poisson problem is solved for in the perturbation 
formulation.  For $t > 0.6$, however, grid distortion errors outweigh the improved accuracy for the Poisson solve,  and dynamic 
SAM errors become larger than static SAM errors.  The restart criterion parameter in restarted SAM is set as $\Lambda = 1.01$. 
The mesh restarting controls the grid distortion errors, which in turn 
prevents the Jacobian errors from growing.  As shown in the bottom row of \Cref{fig:patch}, and confirmed in 
\Cref{table:patch}, restarted SAM  grids are of comparable accuracy and smoothness to static SAM grids

At this low resolution, dynamic SAM is actually slower than the static 
algorithm.  For higher resolutions, however, dynamic SAM is much more efficient than static SAM.  To demonstrate this, we repeat
the above experiment with $N = 400^2$ cells. For brevity, we do not report the Jacobian errors or $L^1$ distortion, since the conclusions 
are similar to the $N=40^2$ case.  The computational runtimes, however, are very different:  1414 sec for static SAM vs 184 sec for 
dynamic SAM, and 194 sec for restarted SAM.  We thus see that restarted SAM combines the best aspects of static and 
dynamic SAM i.e. smoothness and efficiency, respectively.

\subsection{Differential rotation with small scales} \label{subsec:numerics-diff-rot}
\subsubsection{Problem description}
This dynamic mesh generation test models a Gaussian ``blob'' deforming under a rotating flow in which the angular velocity 
is dependent upon the distance from the center of the blob \cite{Chacon2011}. The time-dependent 
target Jacobian function is defined as 
\begin{equation}\label{Chacon-fn}
\bar{\Ge}(y,t) = \frac{1}{1 + 4\exp \left[  -r(y)^2 \left(  \frac{\cos^2 \theta_0(y,t)}{\sigma_1} +  \frac{\sin^2 \theta_0(y,t)}{\sigma_2}  \right)  \right]} \,,
\end{equation} 
where $r(y) = |y-0.5|$ is the radial coordinate,  $\theta_0(y,t) = \theta(y) + \omega(r) t$, and 
$\theta(y) = \arctan \left( \frac{y^2-0.5}{y^1-0.5} \right)$. The parameters $\sigma_1$ and $\sigma_2$ control the aspect ratio of the blob, while 
$\omega(r)$ is the angular velocity. As in \cite{Chacon2011}, we set $\sigma_1 = 0.05$, $\sigma_2=0.001$, and 
$$
\omega(r) = 1.6 \max \left[  (0.5-r)r, 0  \right] \,.
$$ 
The function \eqref{Chacon-fn} describes the evolution of an initially smooth Gaussian blob $\bar{\Ge}(y,0)$ advected by an incompressible 
velocity field $V = (V_r \,,V_\theta) = (0,r \omega(r))$, where $V_r$ and $V_\theta$ are the velocity components in the $r$ and $\theta$ 
directions, respectively.  The initial blob is smooth but will develop arbitrarily small scales for $t > 0$ due to the radial dependence of 
the angular velocity.  As in \cite{Chacon2011}, we set the grid resolution at $N=128^2$, the time-step as $\Delta t =1$, and 
generate a sequence of meshes for $0 \leq t \leq 90$.  

\subsubsection{Comparison of static, dynamic, and restarted SAM}
\begin{figure}[!hbt]
\centering
\subfigure[static SAM]{\label{fig:chacon-static3}\includegraphics[width=40mm]{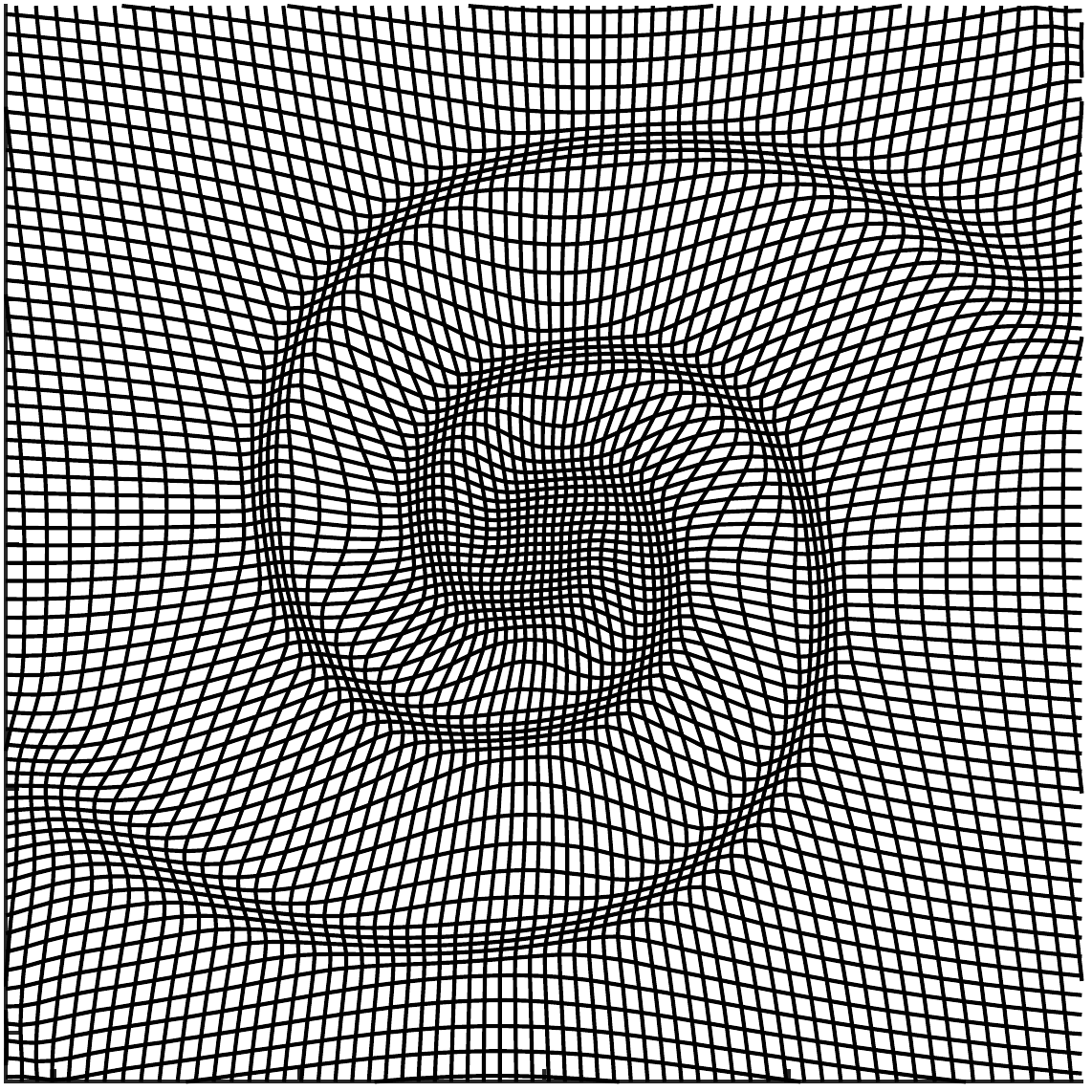}}
\hspace{2em}
\subfigure[dynamic SAM]{\label{fig:chacon-dyn3}\includegraphics[width=40mm]{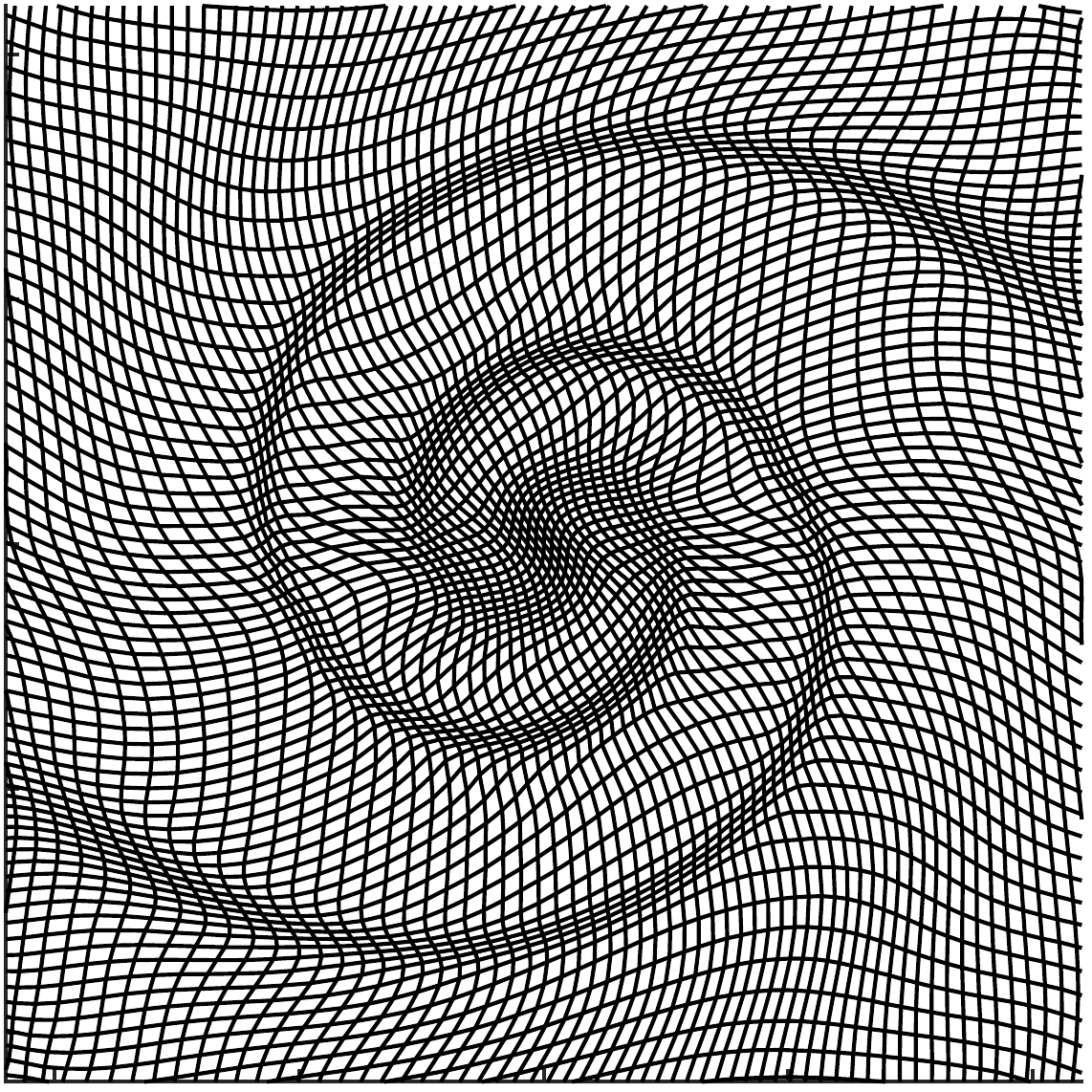}}
\hspace{2em}
\subfigure[restarted SAM]{\label{fig:chacon-rest3}\includegraphics[width=40mm]{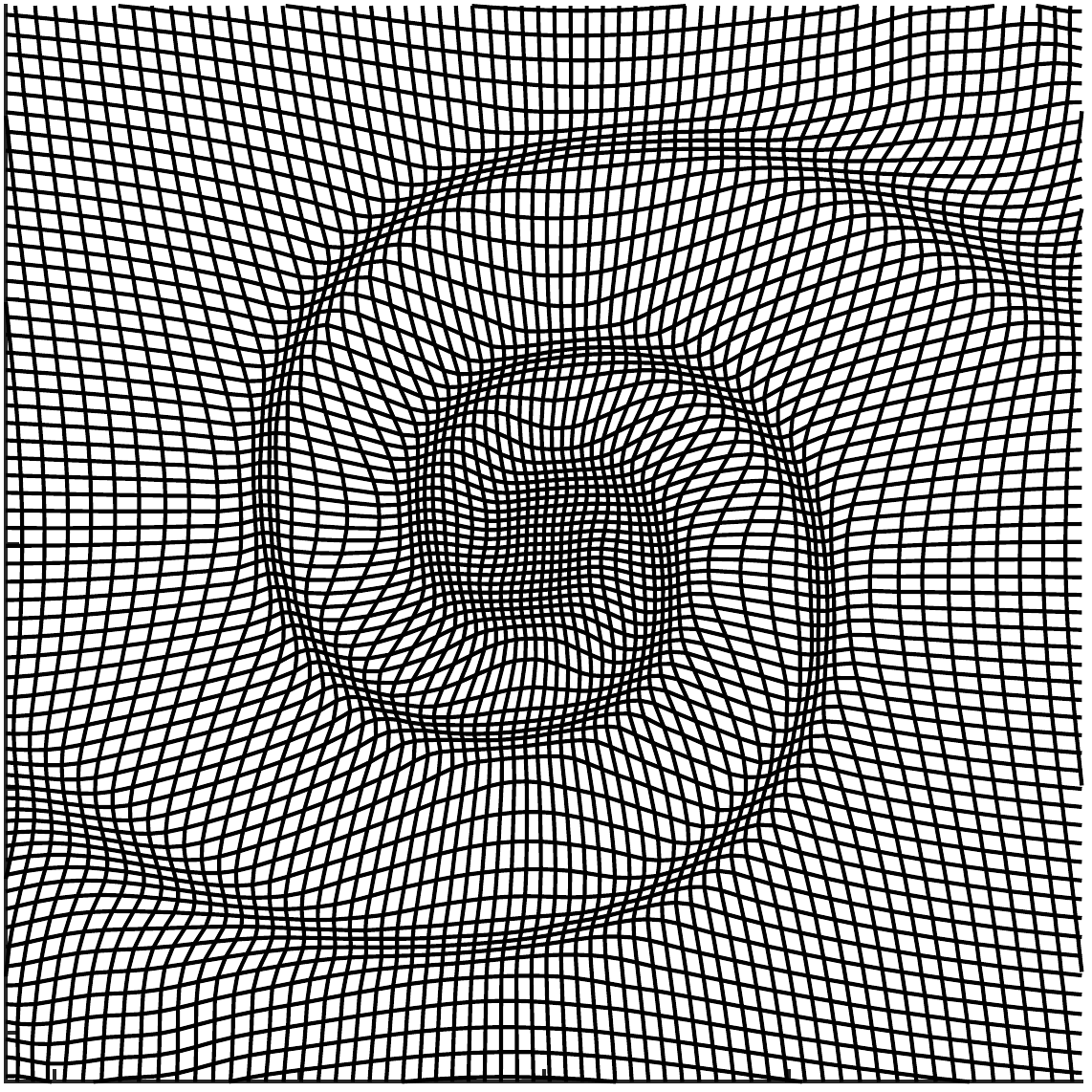}}
\caption{Test problem \ref{subsec:numerics-diff-rot}: tracking small scale vortical structures in flows 
with differential rotation using \eqref{Chacon-fn}. 
Shown are zoomed in plots of the $128^2$ cell adaptive meshes at $t=90$. 
SAM produces smooth meshes without the grid distortion errors associated with 
Lagrangian-type schemes.}
\label{fig:chacon-zoom}
\end{figure}

\begin{figure}[ht]
\centering
\subfigure[$\mathcal{E}_2$ vs $t$]{\label{fig:chacon-compare1}\includegraphics[width=66mm]{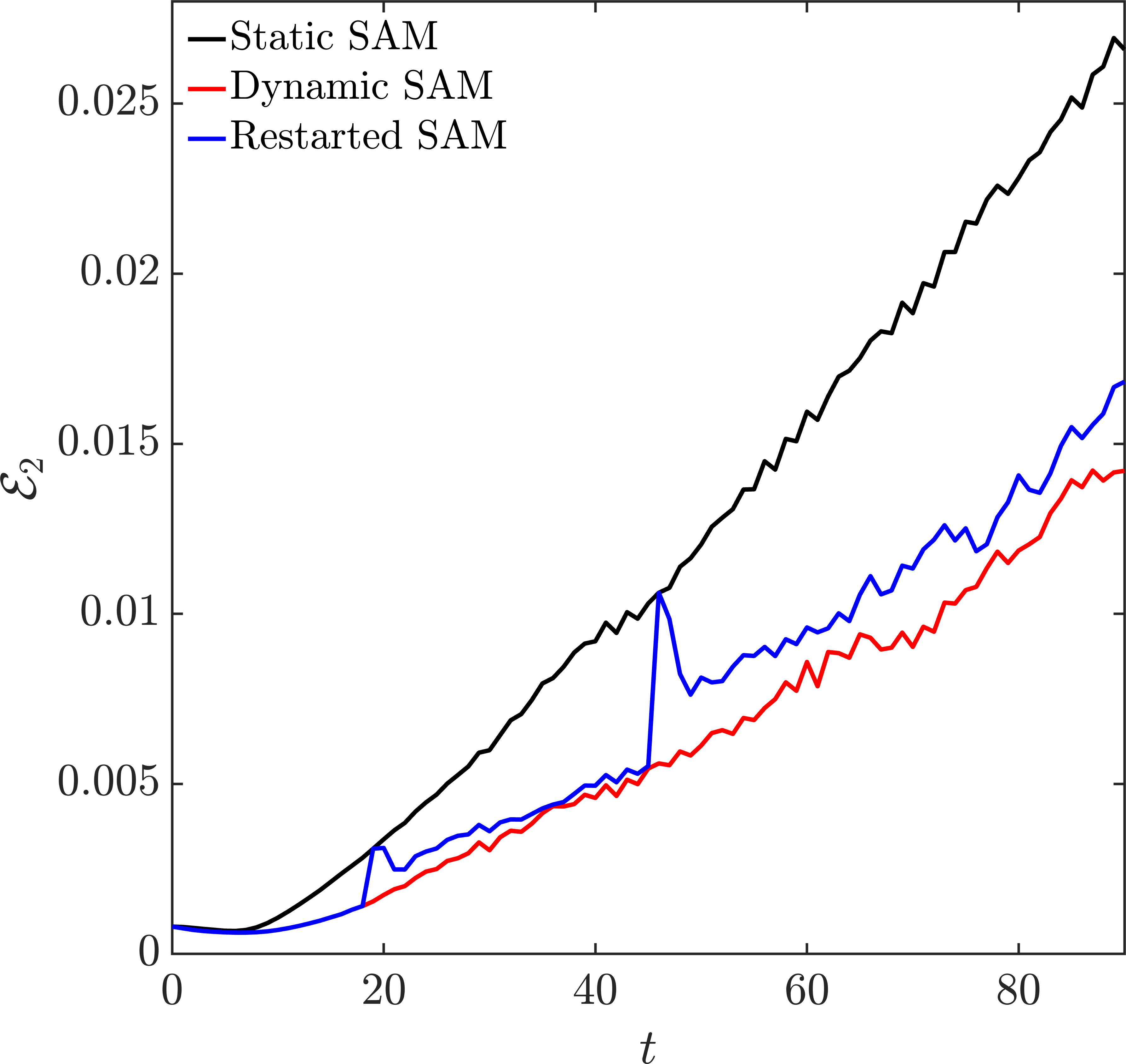}} 
\hspace{4em}
\subfigure[$L^1$ grid distortion vs $t$]{\label{fig:chacon-compare2}\includegraphics[width=65mm]{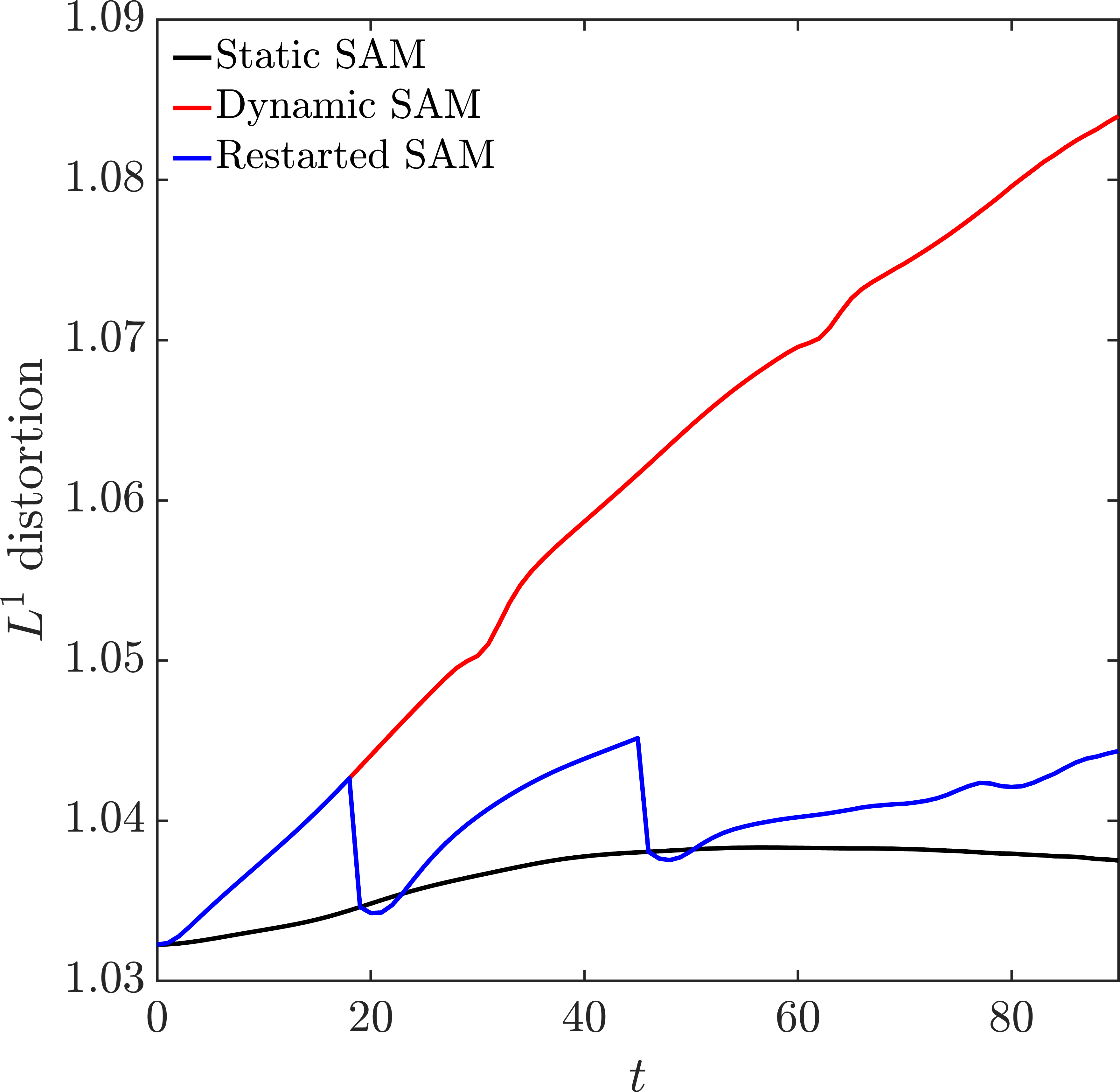}}
\caption{Test problem \ref{subsec:numerics-diff-rot}: tracking small scale vortical structures in flows 
with differential rotation using \eqref{Chacon-fn}. Shown are (a) $L^2$ Jacobian error $\mathcal{E}_2$ and (b) $L^1$ grid 
distortion history of the grids produced 
using static, dynamic, and restarted SAM. The errors are comparable to those provided in \cite{Chacon2011}, and
restarted SAM controls the grid distortion associated with Lagrangian-type schemes.}
\label{fig:chacon-compare}
\end{figure}

The results of static, dynamic, and restarted SAM simulations are provided in 
\Cref{fig:chacon-zoom}, which shows zoomed-in plots of the meshes near $(0.5,0.5)$ at 
the final time $t=90$. All three schemes produce grids that are untangled, but 
while static SAM grids are smooth, the dynamic SAM grids contain more distorted 
elements. This is confirmed in \Cref{fig:chacon-compare}, which provides plots of the time history of the 
$L^2$ Jacobian error \eqref{equi-error} and $L^1$ distortion \eqref{L1-distortion}.  
As with the rotating patch problem, the distorted dynamic SAM grids are still more accurate than the static SAM grids, though 
the Jacobian errors are roughly comparable.   

% \begin{figure}[!hbt]
%\centering
%\subfigure[$t=50$]{\label{fig:chacon-static1}\includegraphics[width=40mm]{dynamic/chacon-static1}} 
%\hspace{2em}
%\subfigure[$t=70$]{\label{fig:chacon-static2}\includegraphics[width=40mm]{dynamic/chacon-static2}}
%\hspace{2em}
%\subfigure[$t=90$]{\label{fig:chacon-static3}\includegraphics[width=40mm]{dynamic/chacon-static3}}
%\hspace{2em}
%\subfigure[$t=50$]{\label{fig:chacon-dyn1}\includegraphics[width=40mm]{dynamic/chacon-dyn1}}
%\hspace{2em}
%\subfigure[$t=70$]{\label{fig:chacon-dyn2}\includegraphics[width=40mm]{dynamic/chacon-dyn2}}
%\hspace{2em}
%\subfigure[$t=90$]{\label{fig:chacon-dyn3}\includegraphics[width=40mm]{dynamic/chacon-dyn3}}
%\hspace{2em}
%\subfigure[$t=50$]{\label{fig:chacon-rest1}\includegraphics[width=40mm]{dynamic/chacon-rest1}}
%\hspace{2em}
%\subfigure[$t=70$]{\label{fig:chacon-rest2}\includegraphics[width=40mm]{dynamic/chacon-rest2}}
%\hspace{2em}
%\subfigure[$t=90$]{\label{fig:chacon-rest3}\includegraphics[width=40mm]{dynamic/chacon-rest3}}
%\caption{Test problem \ref{subsec:numerics-diff-rot}: tracking small scale vortical structures in flows 
%with differential rotation using \eqref{Chacon-fn}. 
%Shown are zoomed in plots of the $128^2$ cell adaptive meshes at various times $t$. 
%The meshes are produced using static SAM (top), dynamic SAM (middle), and restarted dynamic SAM (bottom). 
%SAM produces smooth meshes without the grid distortion errors associated with 
%Lagrangian-type schemes.}
%\label{fig:chacon-zoom}
%\end{figure}  

For restarted SAM, the restart criterion $\lambda_k > \Lambda \lambda_{\mathrm{ref}}$, with $\Lambda = 1.01$, 
 forces the grid to reset 2 times during the simulation. As shown in \Cref{fig:chacon-compare} and in the final row 
of \Cref{fig:chacon-zoom}, the restarted SAM grids are smooth and comparable to 
static SAM grids, and are almost as accurate as the dynamic SAM grids. 
A comparison of \Cref{fig:chacon-compare} with Figure 10 in \cite{Chacon2011} shows that 
static and restarted SAM grids are of similar quality to the MK grids.

\subsection{3$D$ swirling flow}\label{subsec:numerics-3D-helical}

\subsubsection{Problem description}
\begin{figure}[ht]
\centering
\subfigure[$\Ge \circ \psi (x,t)$ at $t = 1$]{\label{fig:3D-helical-target1}\includegraphics[width=80mm]{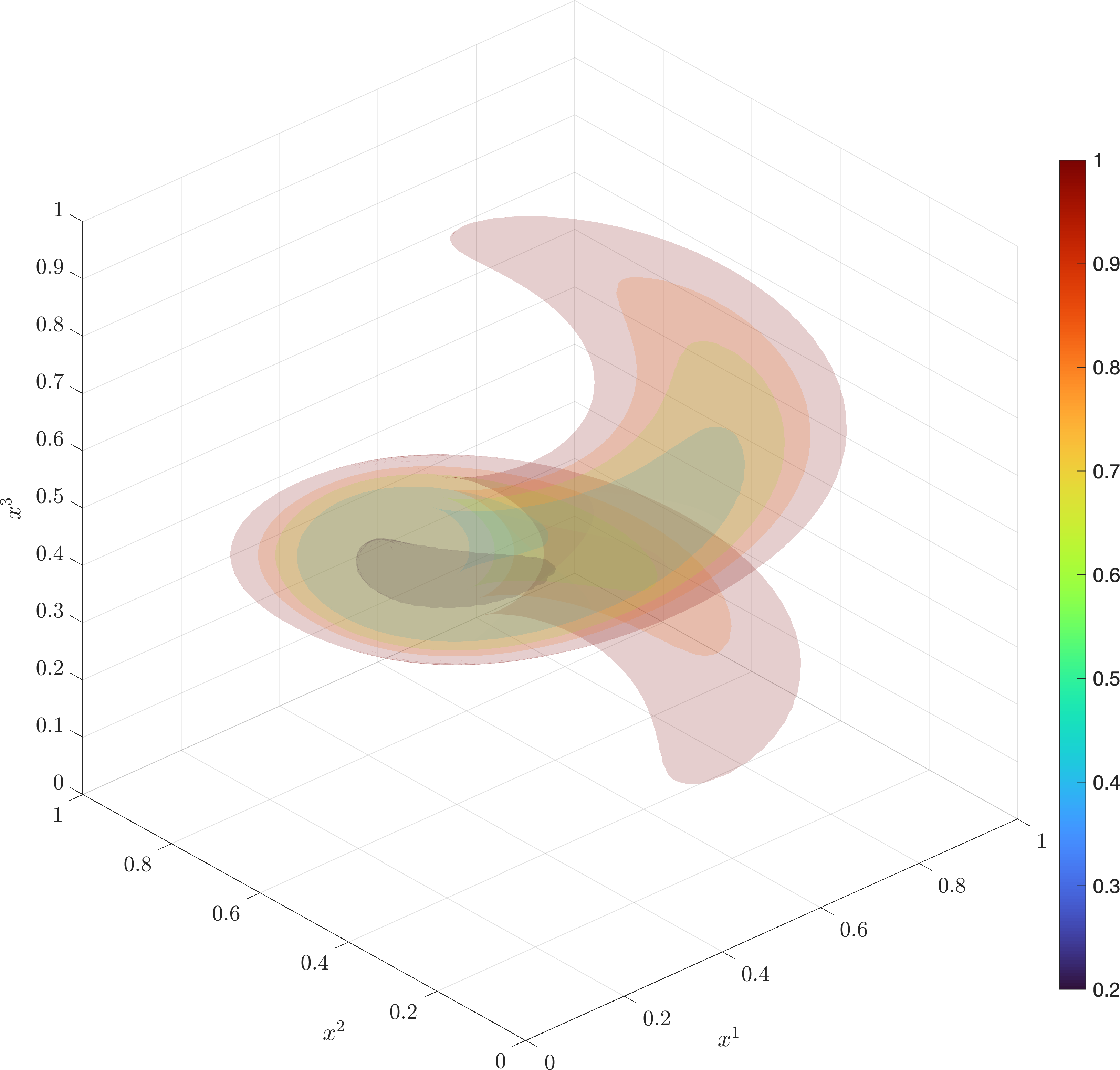}} 
\hspace{2em}
\subfigure[$\mathcal{T}(t)$ at $t=1$]{\label{fig:3D-helical-mesh}\includegraphics[width=74mm]{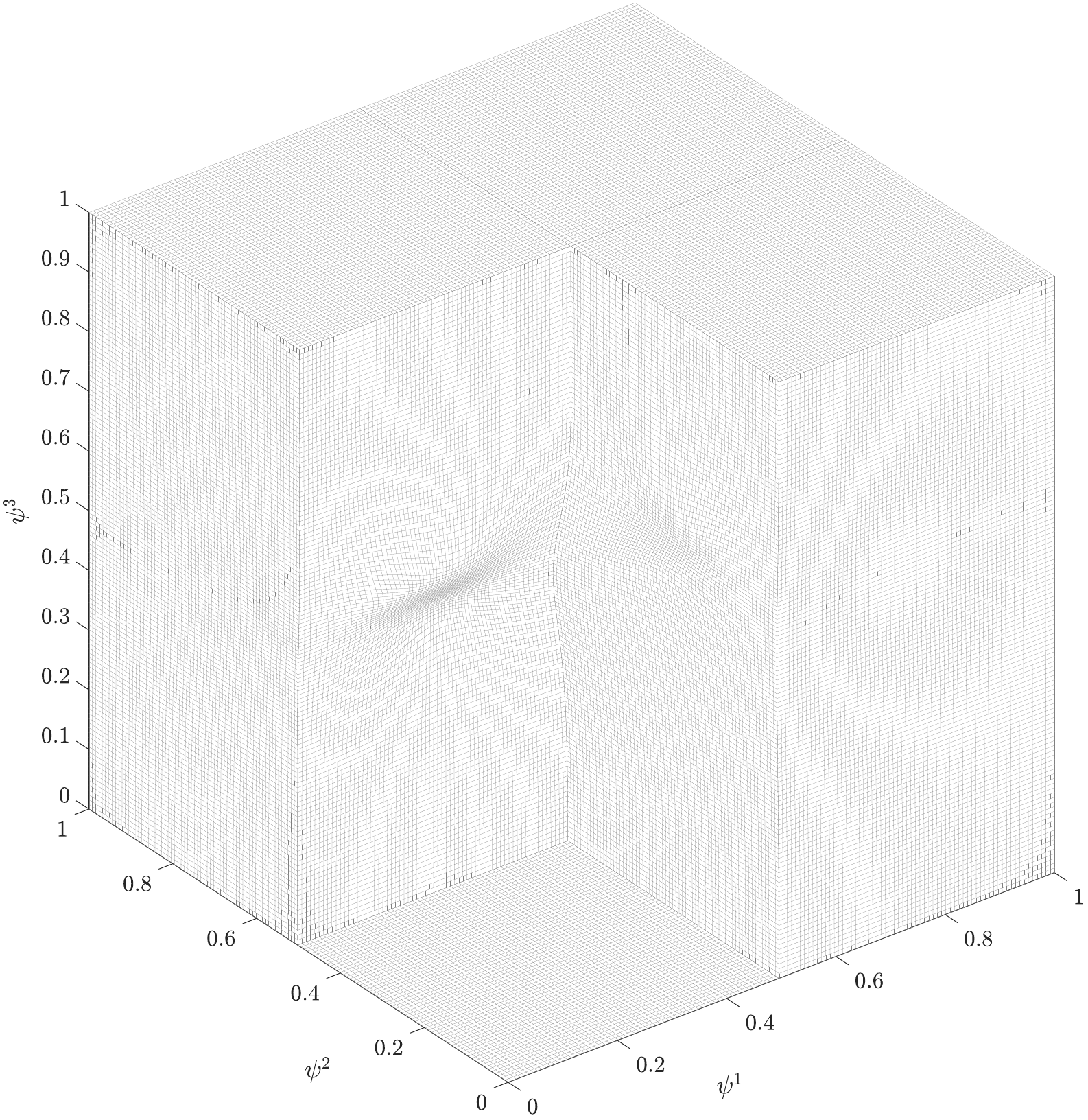}}
\hspace{0em}
\caption{
Test problem \ref{subsec:numerics-3D-helical}: 3$D$ swirling flow with the helical target function \eqref{3D-helical}.
\Cref{fig:3D-helical-target1} shows isosurfaces of the target function $\Ge$ and \Cref{fig:3D-helical-mesh} 
shows a portion of the corresponding mesh. }
\label{fig:3D-helical}
\end{figure}

\begin{figure}[ht]
\centering
\subfigure[Planes in $\Tref$ showing the pre-image of the meshes in 
\Cref{fig:3D-helical-mesh-x,fig:3D-helical-mesh-z,fig:3D-helical-mesh-y}.]{\label{fig:3D-helical-mesh-comp}\includegraphics[width=60mm]{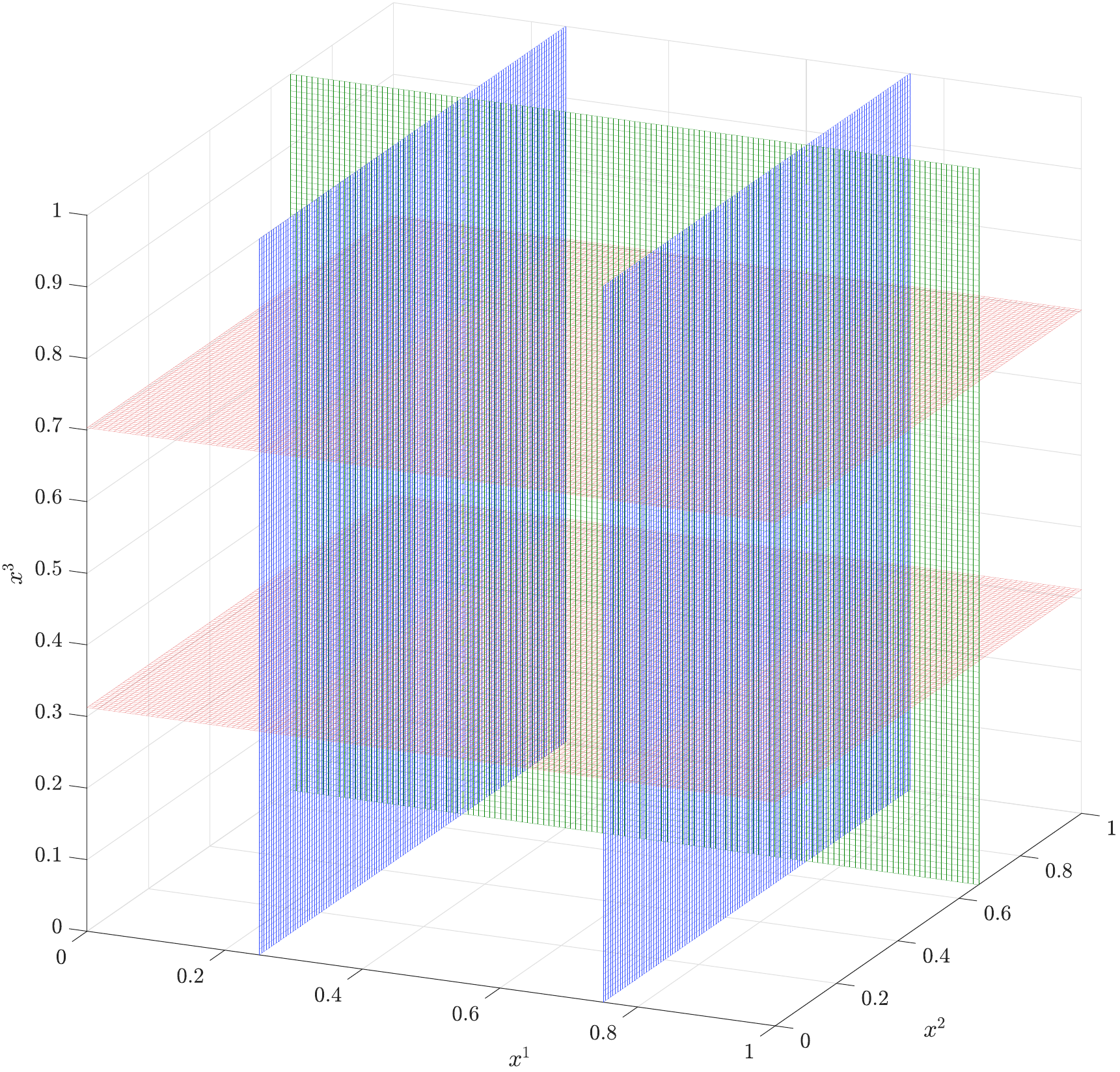}}
\hspace{3em}
\subfigure[Image of the planes $\{x^1=32/128\}$ and $\{x^1 = 96/128\}$ under the map $\psi$.]{\label{fig:3D-helical-mesh-x}\includegraphics[width=60mm]{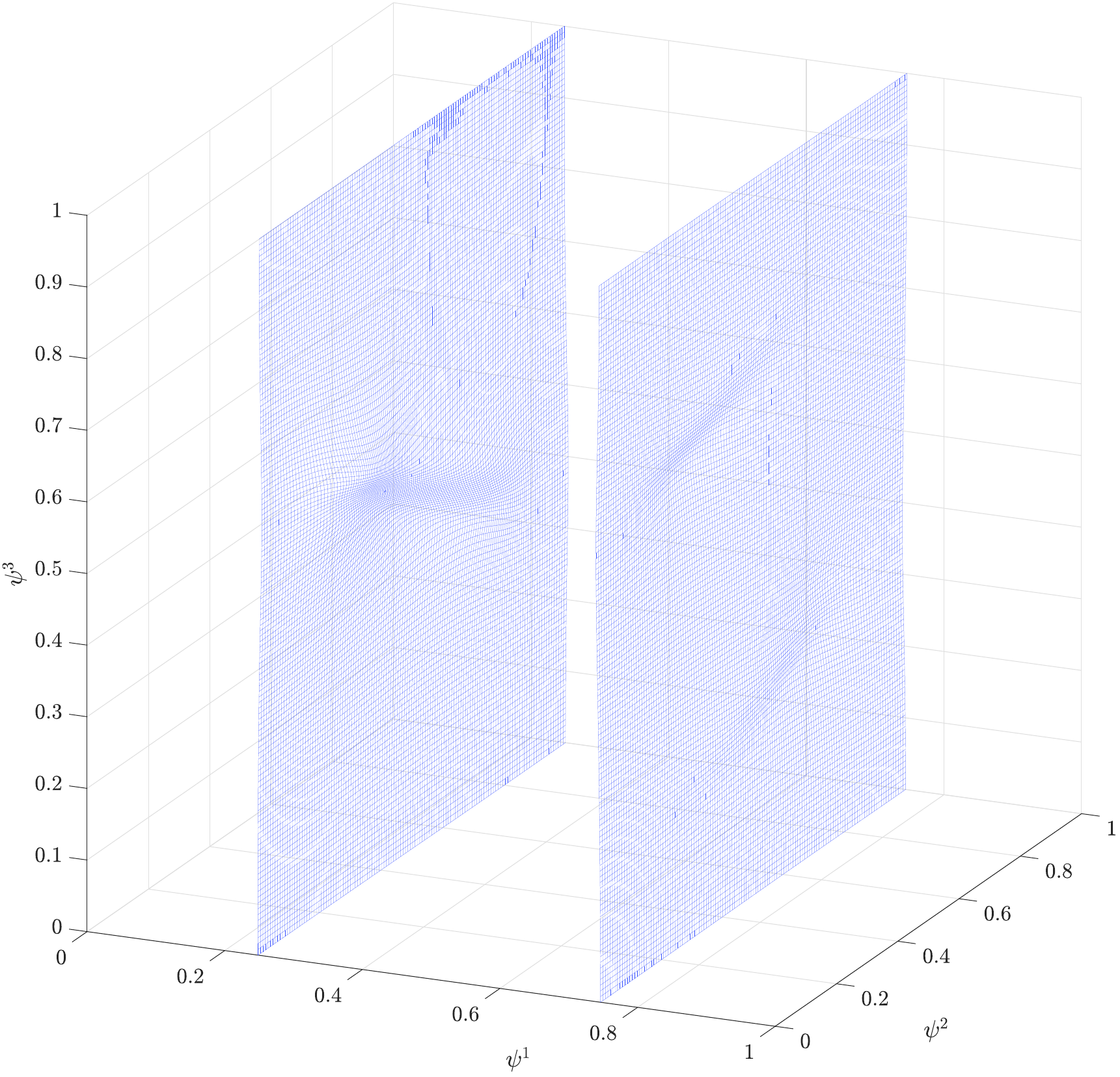}}
\hspace{0em}
\subfigure[Image of the planes $\{x^3=40/128\}$ and $\{x^3 = 90/128\}$ under the map $\psi$.]{\label{fig:3D-helical-mesh-z}\includegraphics[width=60mm]{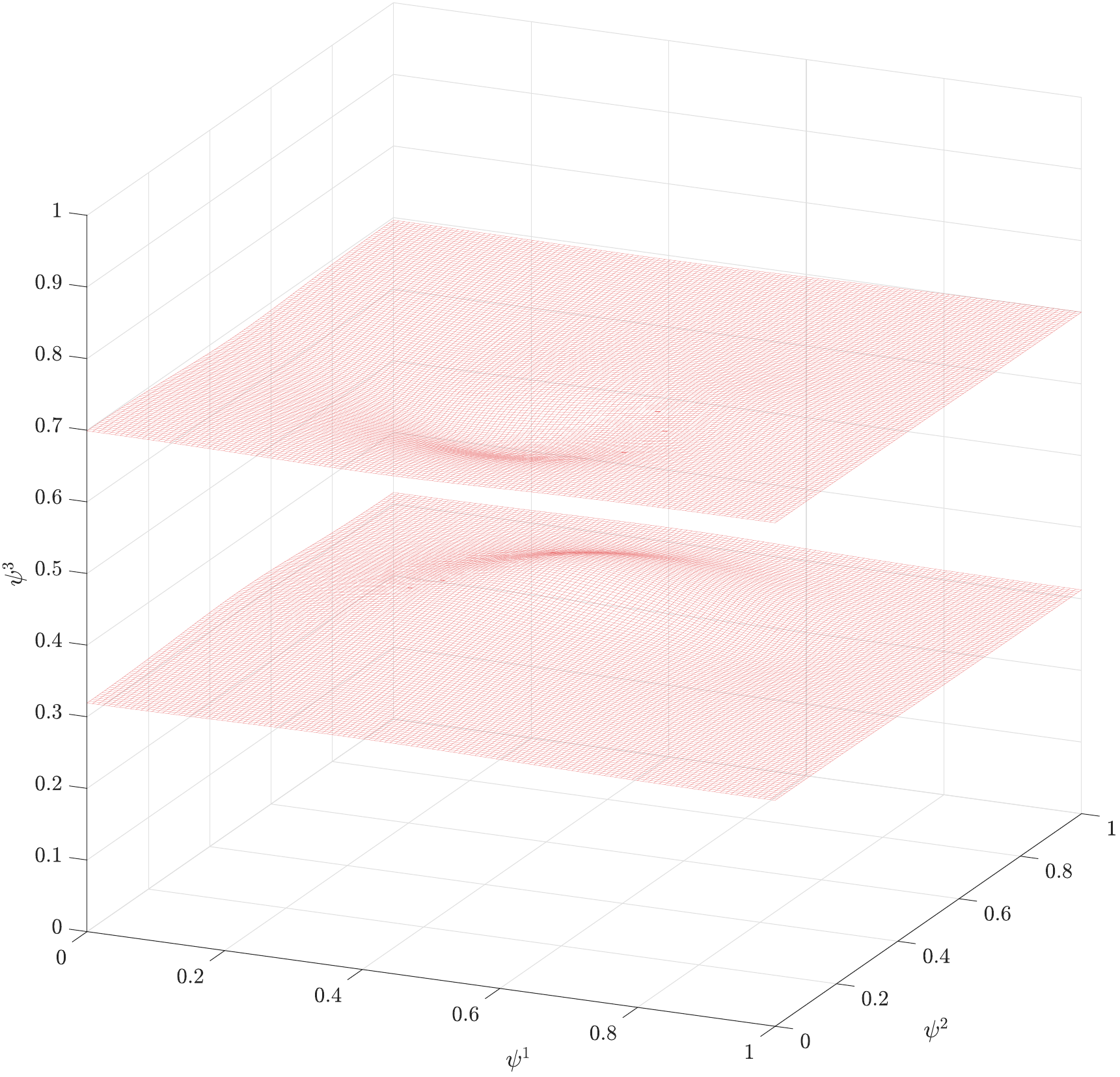}}
\hspace{3em}
\subfigure[Image of the planes $\{x^2=85/128\}$ under the map $\psi$.]{\label{fig:3D-helical-mesh-y}\includegraphics[width=60mm]{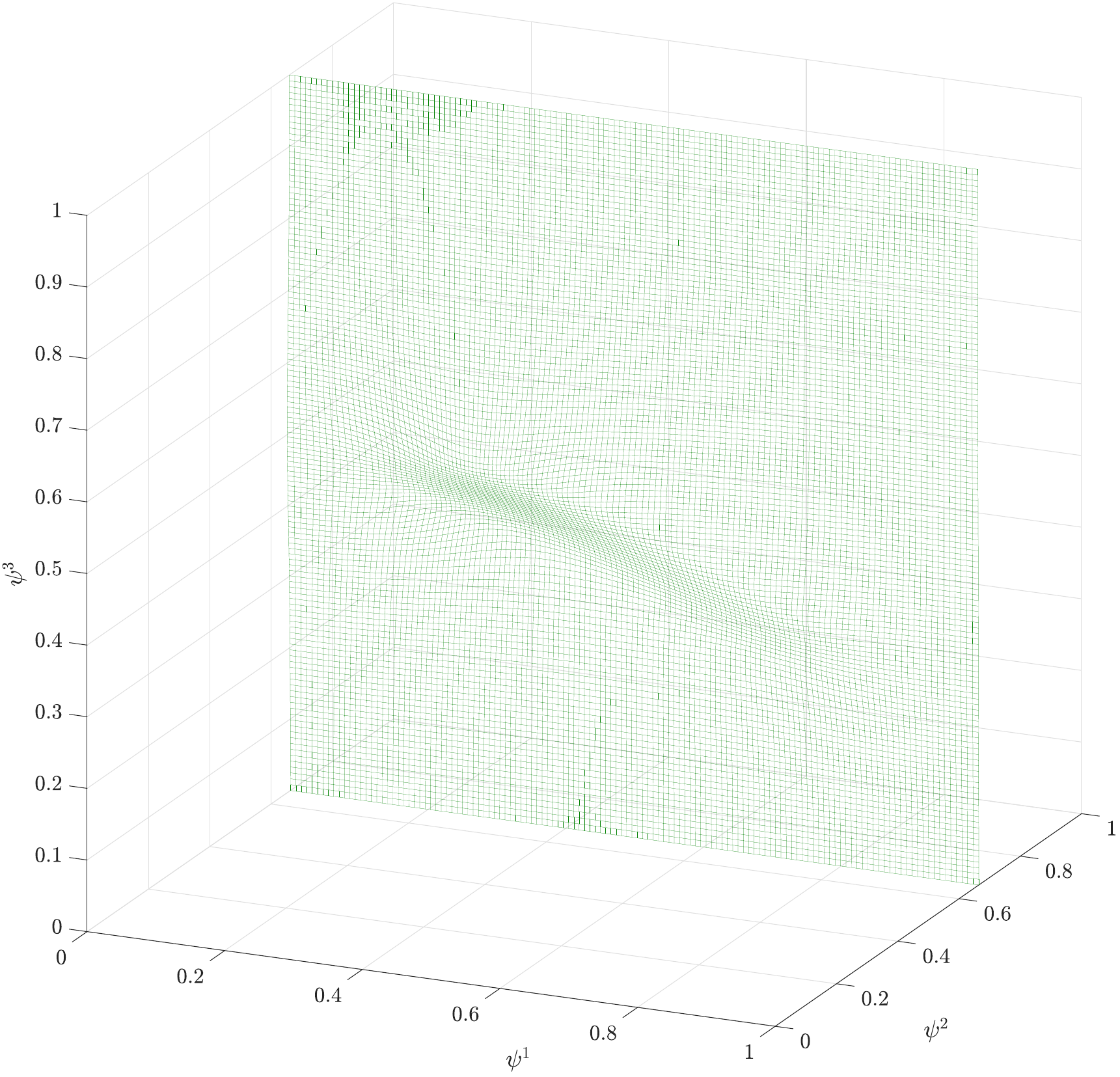}}
\caption{
Test problem \ref{subsec:numerics-3D-helical}: 3$D$ swirling flow with the helical target function \eqref{3D-helical}. Shown 
are plots of the images of various planes $P \subset \Tref$ under the map $\psi$.}
\label{fig:3D-helical-cs}
\end{figure}

Our final experiment is a 3$D$ dynamic version of the test in \cite{BrBuPiCu2014}. 
The domain is $\Omega = [0,1]^3$, the 
time interval is $0 \leq t \leq 1$,  and the 
target Jacobian function  is given by
\begin{equation}\label{3D-helical}
\bar{\Ge}(y^1,y^2,y^3,t) =  \frac{1}{1 + 5 e^{-36(y^3-\frac{1}{2})^2} \exp \left( - \omega_1 R(y^1,y^2,y^3,t)   \right)} \,,
\end{equation} 
with 
$$
R(y^1,y^2,y^3,t) = \left(y^1-\frac{1}{2}-\omega_2 \cos(4 \pi (y^3-t/4) \right)^2 + \left( y^2-\frac{1}{2}-\omega_2 \sin(4 \pi (y^3-t/4) \right)^2 \,,
$$
and $\omega_1 = 100$ and $\omega_2 = 0.25$.  As discussed in \cite{BrBuPiCu2014}, the target function \eqref{3D-helical} describes a 
complex 3$D$ helical surface and poses a major challenge for mesh generation algorithms since it leads to highly non-uniform and twisted 
meshes.  See \Cref{fig:3D-helical} for plots of the target function and (a portion of) the associated mesh at 
$t=1$ and at $N=128^3$ cell resolution.  In \Cref{fig:3D-helical-cs}, we provide plots of $\psi(P)$, where 
$P \subset \Tref$ is some planar subset (lying in ether the $x^1x^2$-, $x^2x^3$-, or $x^1x^3$-planes) 
of the reference mesh.

We generate a sequence of meshes starting with $N = 32^3$ resolution and doubling in each direction until $N=256^3$.  The time-step 
$\Delta t$ depends on $N$ according to the CFL scaling and is set as 
$\Delta t = \frac{2}{\sqrt[3]{N}}$.  
It is straightforward to adapt the 2$D$ numerical scheme described in 
\Cref{subsec:numerical-imp} to the 3$D$ setting, and for 
brevity we omit the details.  

\subsubsection{$N=128^3$ simulations using static, dynamic, and restarted SAM}

\begin{figure}[ht]
\centering
\subfigure[$\mathcal{E}_2$ vs $t$]{\label{fig:helix-t-compare1}\includegraphics[width=62.3mm]{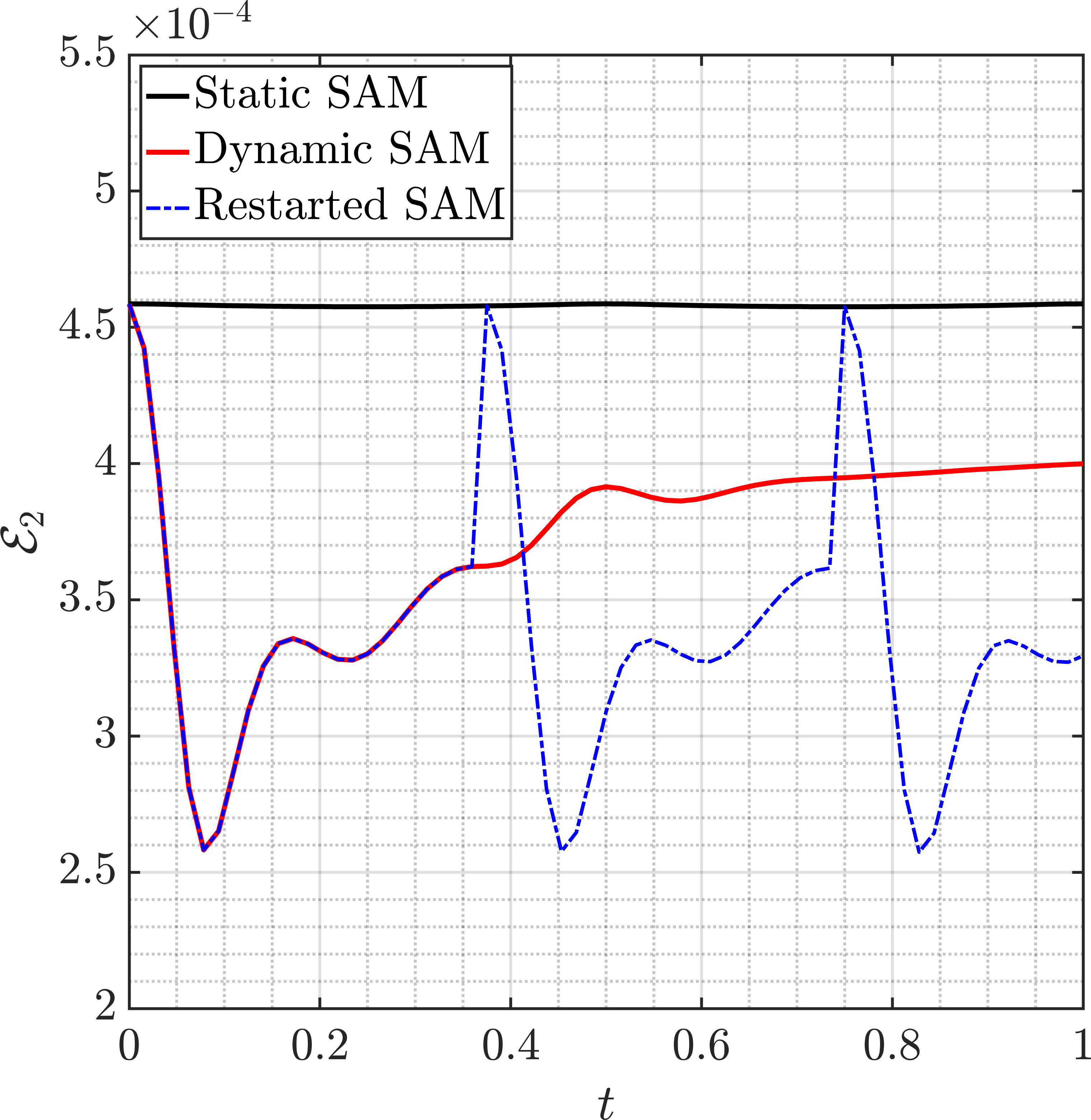}} 
\hspace{4em}
\subfigure[$L^1$ grid distortion vs $t$]{\label{fig:helix-t-compare2}\includegraphics[width=65mm]{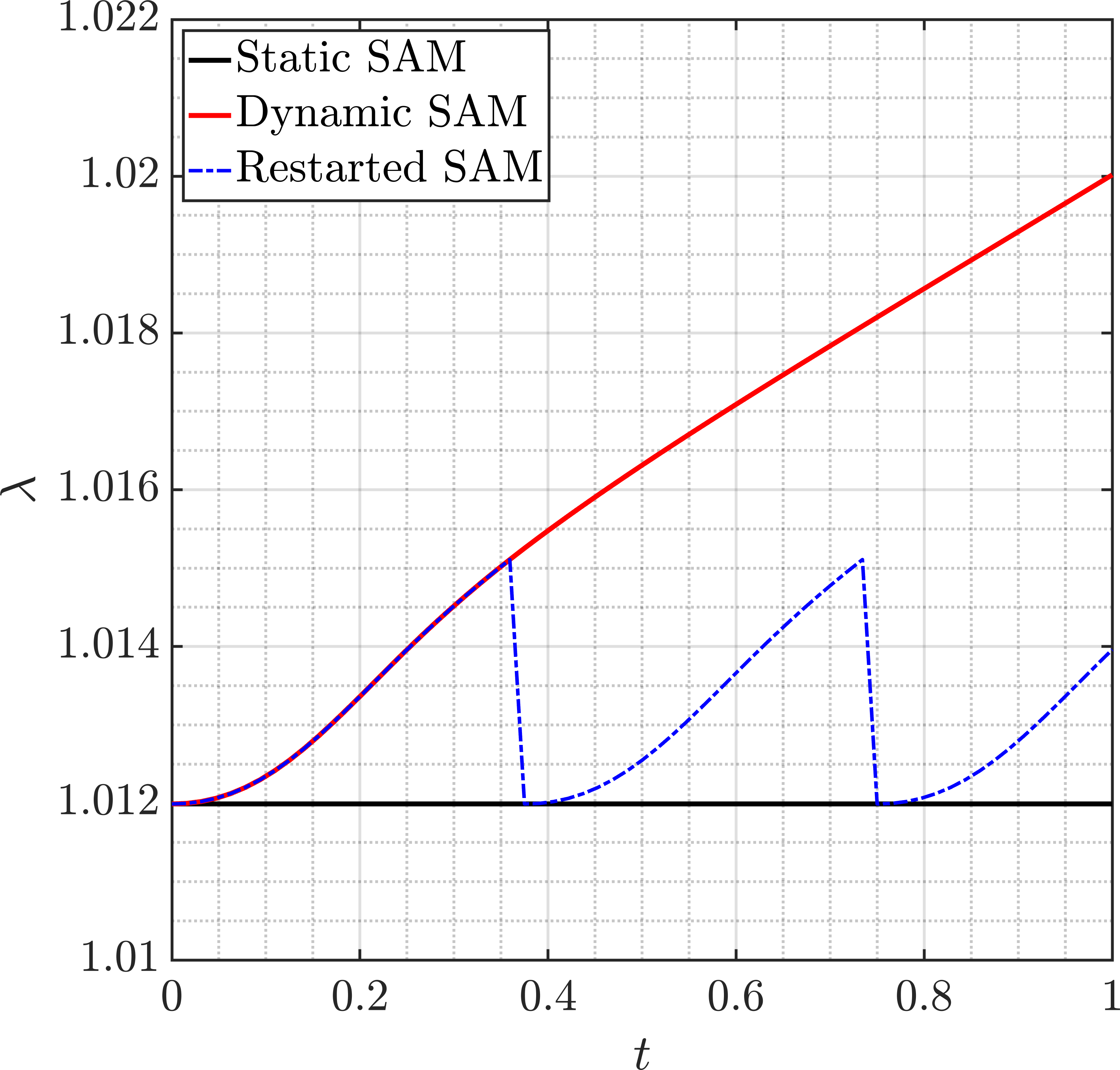}}
\caption{Test problem \ref{subsec:numerics-3D-helical}: 3$D$ swirling flow with the helical target function \eqref{3D-helical}. 
Shown are (a) $L^2$ Jacobian error $\mathcal{E}_2$ and (b) $L^1$ grid 
distortion history of the grids produced 
using static, dynamic, and restarted SAM at $N=128^3$. Restarted SAM controls the grid distortion associated with 
Lagrangian-type schemes.}
\label{fig:helix-t-compare}
\end{figure}

Plots of the time-history of the Jacobian errors $\mathcal{E}_2$ and $L^1$ distortion at $N=128^3$ are shown in  \Cref{fig:helix-t-compare}.  
For $0 \leq t \leq 0.5$, dynamic SAM produces the smallest errors, due to the greater accuracy with which the Poisson and 
transport problems are solved.  As expected, dynamic SAM meshes exhibit increasing grid distortion, which causes growth of the Jacobian 
error. The restart criterion in restarted SAM forces the mesh to reset two times during the simulation, which controls the growth of both 
the mesh distortion as 
well as the Jacobian error; for this example, $\Lambda = 1.003$.

\subsubsection{Resolution study}

\begin{figure}[ht]
\centering
\subfigure[$\log\log$ plot $\mathcal{E}_2$ vs $\Delta x$]{\label{fig:helix-compare1}\includegraphics[width=52.7mm]{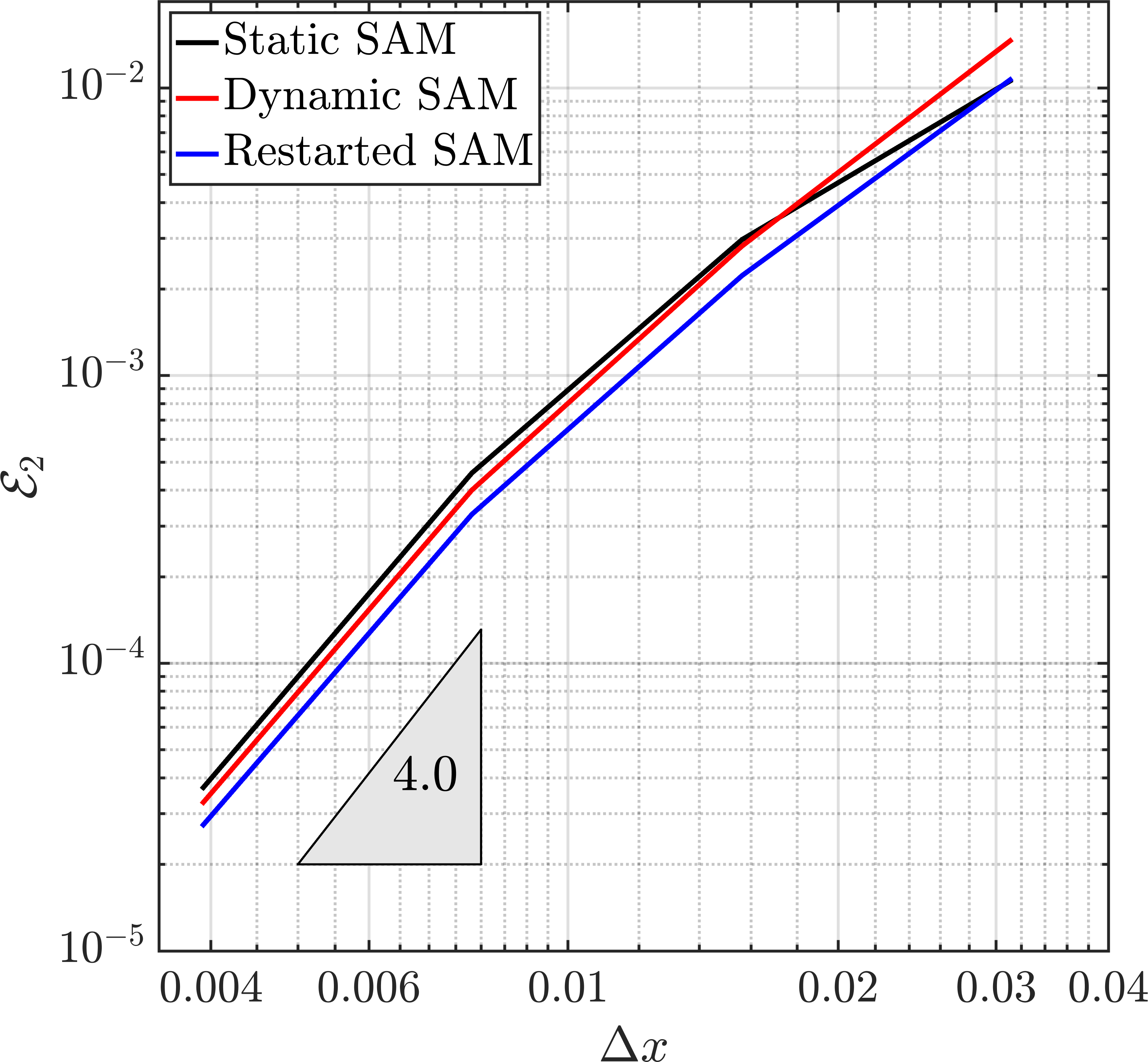}} 
\hspace{0em}
\subfigure[$L^1$ grid distortion vs $\Delta x$]{\label{fig:helix-compare2}\includegraphics[width=51.7mm]{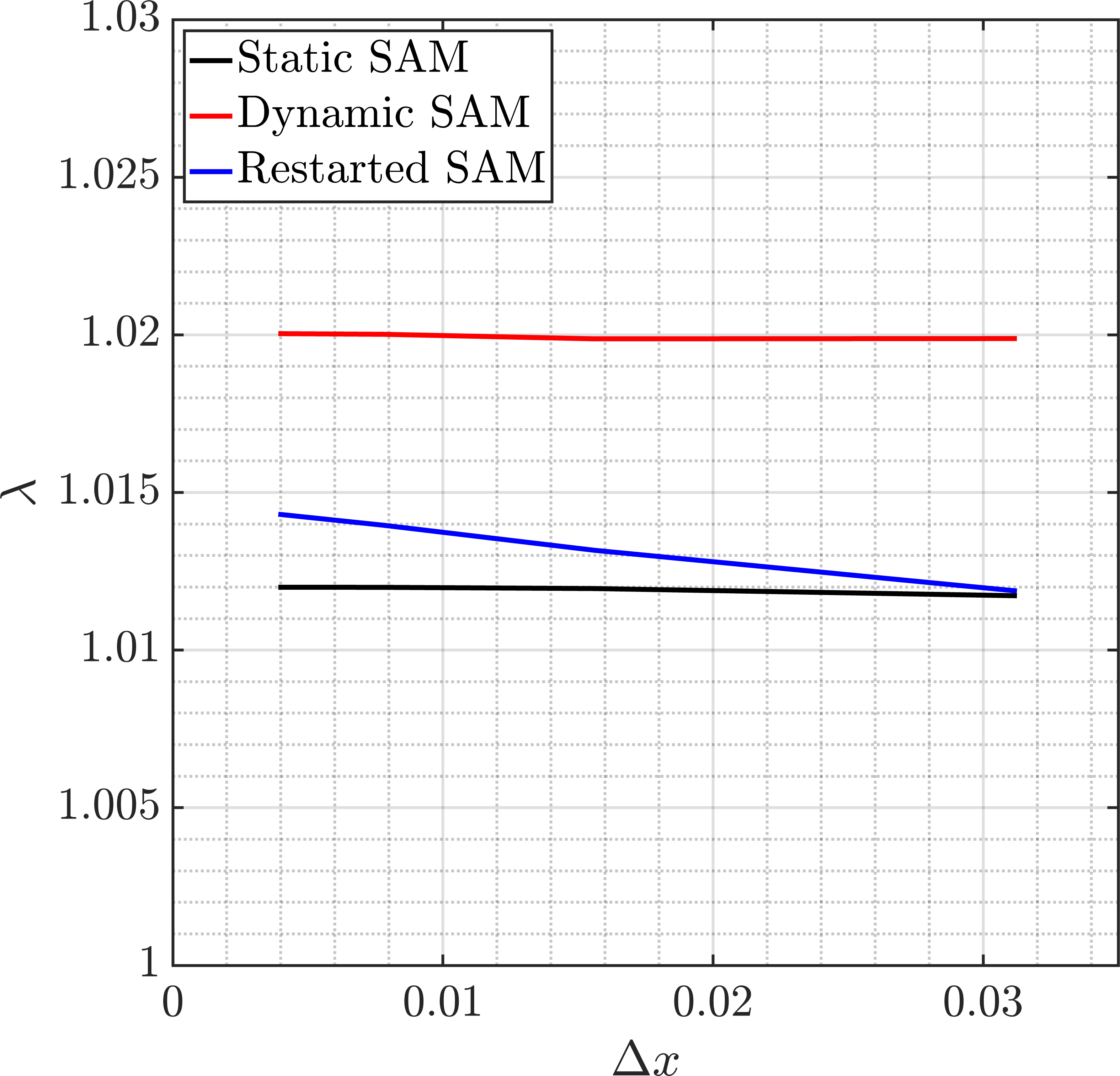}}
\hspace{0em}
\subfigure[$\log\log$ plot $T_{\mathrm{CPU}}$ vs $N$]{\label{fig:helix-compare3}\includegraphics[width=50mm]{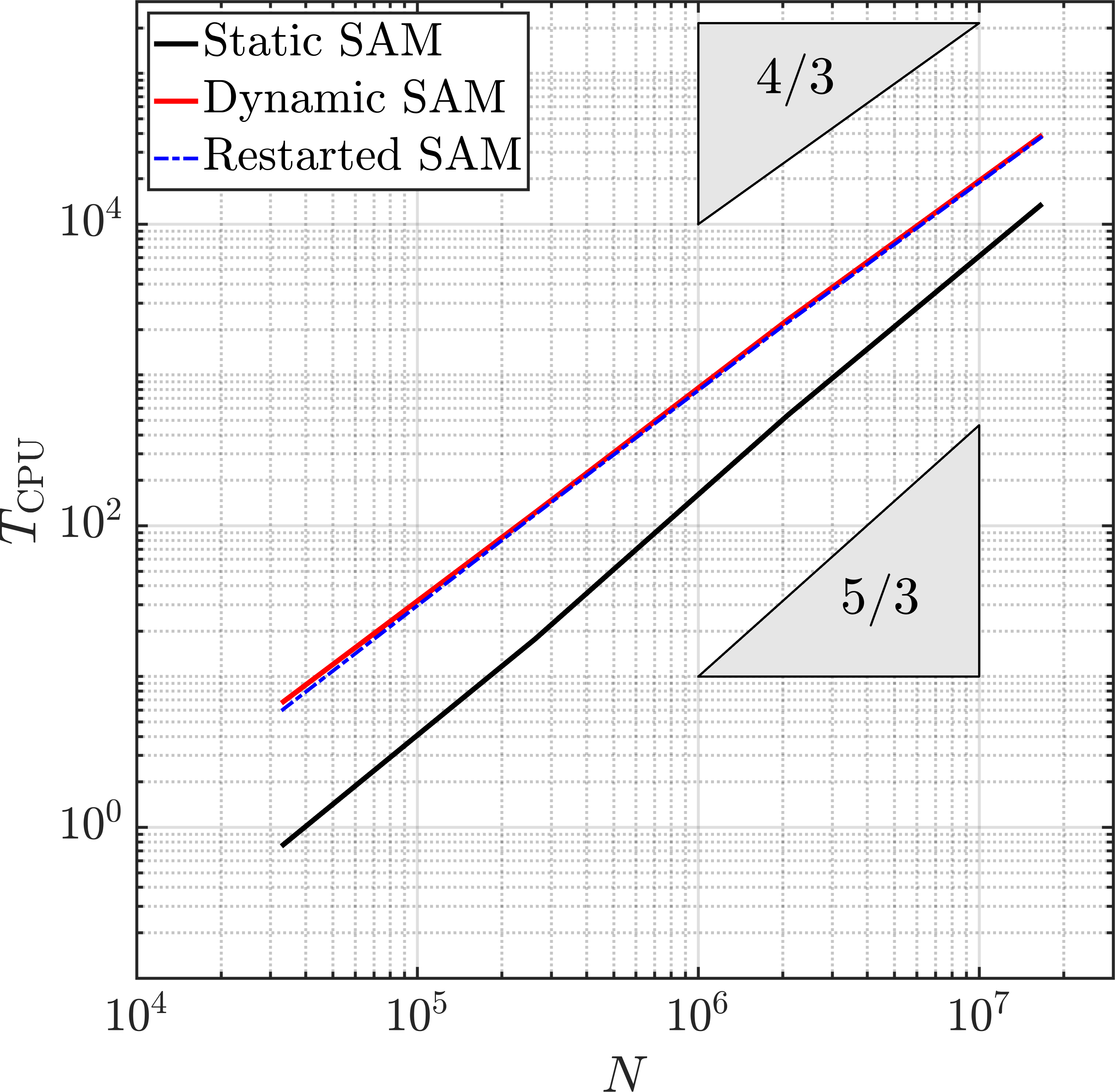}}
\caption{Test problem \ref{subsec:numerics-3D-helical}: 3$D$ swirling flow with the helical target function \eqref{3D-helical}. 
Shown are (a) $\log \log$ plots of $L^2$ Jacobian error $\mathcal{E}_2$ vs $\Delta x$,  (b) $L^1$ grid 
distortion vs $\Delta x$, and (c) $\log \log$ plots of total runtime $T_{\mathrm{CPU}}$ vs $N$ of the grids produced 
using static, dynamic, and restarted SAM for $N=32^3,\ldots,256^3$.}
\label{fig:helix-compare}
\end{figure}

Next, we provide in \Cref{fig:helix-compare} plots (as a function of the resolution $N$) 
of the Jacobian error, $L^1$ distortion, and CPU runtime at $t=1$.  
\Cref{fig:helix-compare1} shows that restarted SAM produces grids with the smallest Jacobian errors, but the errors for the 
various schemes are comparable for all the resolutions considered; as expected, we observe
4\textsuperscript{th} order convergence for all the schemes.  \Cref{fig:helix-compare2} shows that the $L^1$ distortion for 
both static and dynamic SAM is consistent across 
resolutions, with the grid distortion for restarted SAM bounded between the two.  Finally, \Cref{fig:helix-compare3} shows that, while 
static SAM is of complexity $\mathcal{O}(N \cdot N^{1/3} / \Delta t) = \mathcal{O}(N^{5/3})$, both dynamic and restarted SAM are of 
optimal complexity $\mathcal{O}(N/\Delta t) = \mathcal{O}(N^{4/3})$.   Based on this, we can estimate that, for this test, restarted 
SAM becomes more efficient than static SAM for $N > 735^3$.

\section{SAM-ALE scheme for gas dynamics} \label{sec:ALE}

We next couple our SAM scheme to a very simple FD WENO-based ALE scheme. The purpose of this section is to demonstrate the 
ability of SAM-ALE to reproduce high-resolution uniform runs using fewer cells and less total CPU time.  The numerical method for the 
ALE system of equations we use is highly simplified and not meant to be representative of the full class of ALE solvers. 
Nonetheless, even for the two very difficult test problems presented in 
\Cref{sec:ALE-sims}, the highly simplified scheme performs remarkably well.  
 
For the notation used in this section, we refer the reader to \Cref{sec:preliminaries}.  
  
\subsection{The 2$D$ ALE-Euler system}\label{subsec:Euler2D}

\subsubsection{Equations in Eulerian coordinates}

The 2$D$ compressible Euler system in Eulerian coordinates $y = (y^1,y^2) \in \Omega$ can be written in the following compact 
conservation-law form
\begin{subequations}
\label{Euler-2d}
\begin{alignat}{2}
\partial_t {{\mathsf{Q}}}+ D_i {{\mathsf{F}}}^i({{\mathsf{Q}}}) = 0,& && \ \ \  {(y,t)} \in \Omega \times (0,T),  \label{Euler-2d-motion} \\
{{{\mathsf{Q}}}}(y,0)  = { {{\mathsf{Q}}}}_0({y}),& && \ \ \ {(y,t)} \in \Omega \times \{0\}.  
\end{alignat}
\end{subequations}
Here, $\mathsf{Q}$ is the vector of conserved variables,  and ${\mathsf{F}}^1({\mathsf{Q}})$ and 
${\mathsf{F}}^2({\mathsf{Q}})$ are the \emph{flux functions}, defined as
\begin{equation}
{{{\mathsf{Q}}}} = \left ( \begin{array}{c} \uprho \\[0.15em] \uprho \ue^1 \\[0.15em] \uprho \ue^2 \\[0.15em] \Ee \end{array} \right ) \quad \text{ and } \quad 
{{{\mathsf{F}}}}^i({\mathsf Q}) = \begin{pmatrix}
		\uprho \ue^i \\[0.15em]
		\uprho \ue^1 \ue^i + \delta^i_1 \pe \\[0.15em]
		\uprho \ue^2 \ue^i + \delta^i_2 \pe \\[0.15em]
		\ue^i(\Ee+\pe)
		\end{pmatrix} \,. 
\end{equation}
The velocity vector is  $\ue = (\ue^1\,,\ue^2)$ with horizontal component $\ue^1$ and vertical
component $\ue^2$, $\uprho>0$ is the fluid density (assumed strictly positive),  $\Ee$ denotes the energy, and $\pe$ is the pressure 
defined by the ideal gas law, 
\begin{equation}\label{pres-eqn}
\pe = (\gamma-1)\left(\Ee - \frac{1}{2}\uprho | {\ue} |^2 \right)\,,
\end{equation}
where $\gamma$ is the adiabatic constant, which we will assume takes the value $\gamma=1.4$, unless
otherwise stated. 

\subsubsection{Equations in ALE coordinates}
Let $\Omegaref$ be the fixed reference domain with coordinates $(x^1,x^2)$, and assume that we have, for each $t \geq 0$,  
a smooth ALE map $\psi(\cdot,t) : \Omegaref \to \Omega$. Denote by the regular font $f$ the ALE counterpart 
to the Eulerian variable written with upright font $\mathsf{f}$ i.e. $f(x,t) = \fe \circ \psi (x,t)$.  
The 2$D$ ALE-Euler system can then be written in conservation law form as 
\begin{subequations}
\label{ALE-Euler-2d}
\begin{alignat}{2}
\partial_t {{{Q}}}+ \partial_j {{{F}}}^j({{{Q}}}) = 0,& && \ \ \  {(x,t)} \in \Omegaref \times (0,T),  \label{ALE-Euler-2d-motion} \\
{{{{Q}}}}(x,0)  = { {{{Q}}}}_0({x}),& && \ \ \ {(x,t)} \in \Omegaref \times \{0\},
\end{alignat}
\end{subequations}
where the conserved ALE variables ${Q}$ and {flux functions} ${{F}}^j({{Q}})$ are given as
\begin{equation}
{{{{Q}}}} = \left ( 
\begin{array}{c} 
\mathcal{J} \rho \\[0.15em] 
\mathcal{J} \rho u^1 \\[0.15em]
 \mathcal{J} \rho u^2 \\[0.15em]
  \mathcal{J}E \end{array} 
  \right ) \quad \text{ and } \quad 
{{{{F}}}}^j({Q}) = \begin{pmatrix}
		\rho a^j_i (u^i - \psi_t^i)  \\[0.15em]
		\rho u^1 a^j_i (u^i - \psi_t^i) + a^j_1 p \\[0.15em]
		\rho u^2 a^j_i (u^i - \psi_t^i) + a^j_2 p \\[0.15em]
		E a^j_i (u^i - \psi_t^i) + p a^j_i u^i
		\end{pmatrix}\,. 
\end{equation}
Here, $a^j_i$ denotes the components of the cofactor matrix defined by \eqref{deformation}, and $\psi_t^i$ is the $i$\textsuperscript{th} 
component of the mesh velocity. It is also convenient to introduce the \emph{ALE transport velocity} 
$v(x,t)$ with $j$\textsuperscript{th} component $v^j \coloneqq \frac{1}{\mathcal{J}} a^j_i (u^i - \psi_t^i)$. 
The 2$D$ ALE-Euler system \eqref{ALE-Euler-2d} is hyperbolic in the sense that each of $\nabla_Q F^j(Q)$ is diagonalizable with 
real eigenvalues (or wave speeds), which are given explicitly by
\begin{equation}
\lambda^{j,\pm} = \frac{1}{\mathcal{J}} ( v^j \pm c)  \quad \text{and} \quad \lambda^{j,0} = \frac{1}{\mathcal{J}} v^j \text{  (repeated)} \,,
\end{equation}
with $c = \sqrt{\gamma p /\rho}$ the sound speed. 

\subsubsection{Geometric conservation law and free-stream preservation}
An explicit computation shows that the Jacobian determinant $\mathcal{J}(x,t)$ satisfies the \emph{geometric conservation law} 
(GCL) \cite{ThLo1979}
\begin{equation}\label{GCL}
\partial_t \mathcal{J} - \partial_j (a^j_i \psi_t^i) = 0 \,.
\end{equation}
For \eqref{ALE-Euler-2d}, an equivalent property to the GCL is the \emph{free-stream preservation property}, which states that an initially 
uniform flow (i.e. $\mathsf{Q}_0 \equiv$ constant) is preserved under evolution by \eqref{ALE-Euler-2d-motion} i.e. $\mathsf{Q} \equiv$ constant 
for every $t > 0$.    
Numerical schemes that fail to preserve the free-stream produce unacceptably large 
errors that corrupt small-scale vortical structures \cite{Hindman1982,Visbal2002,CaLa2008,Nonomura2010,Jiang2014}.  

Finite difference schemes on static uniform meshes preserve the free-stream.  On dynamic adaptive meshes, however, this is no 
longer a given, and indeed many standard schemes (including WENO \cite{JiangShu1996}) fail to preserve the free-stream. 
As such, we design our numerical scheme to ensure free-stream preservation by explicitly incorporating \eqref{GCL} into the 
system of conservation laws to be solved \cite{Hindman1982,YaHuQi2012}.  
Specifically, we append to \eqref{ALE-Euler-2d} the equation \eqref{GCL} and 
consider the modified system 
\begin{subequations}
\label{ALE-Euler-2d-full}
\begin{alignat}{2}
\partial_t {{\tilde{Q}}}+ \partial_j {{\tilde{F}}}^j({{\tilde{Q}}}) = 0,& && \ \ \  {(x,t)} \in \Omegaref \times (0,T),  \label{ALE-Euler-2d-full-motion} \\
{{{\tilde{Q}}}}(x,0)  = { {{\tilde{Q}}}}_0({x}),& && \ \ \ {(x,t)} \in \Omegaref \times \{0\},
\end{alignat}
\end{subequations}
with 
\begin{equation}
{{{\tilde{Q}}}} = \left ( 
\begin{array}{c} 
\mathcal{J} \rho \\[0.15em] 
\mathcal{J} \rho u^1 \\[0.15em]
 \mathcal{J} \rho u^2 \\[0.15em]
  \mathcal{J}E \\[0.15em]
  \mathcal{J} \end{array} 
  \right )  \quad \text{ and } \quad 
{{{\tilde{F}}}}^j(\tilde{Q}) = \begin{pmatrix}
		\mathcal{J} \rho v^j  \\[0.15em]
		\mathcal{J} \rho u^1 v^j + a^j_1 p \\[0.15em]
		\mathcal{J} \rho u^2 v^j  + a^j_2 p \\[0.15em]
		\mathcal{J} E  v^j + p a^j_i u^i \\[0.15em]
		- a^j_i \psi_t^i
		\end{pmatrix}\,.  
\end{equation}
We emphasize that, while the cofactor matrix $a^j_i$ is computed directly from the map $\psi$ according to \eqref{deformation}, the Jacobian 
determinant $\mathcal{J}$ is computed (using the same numerical method used for the other equations in \eqref{ALE-Euler-2d-full}) via 
\eqref{GCL} and \emph{not} by the usual determinant formula 
$\mathcal{J} = \partial_1 \psi^1 \, \partial_2 \psi^2 - \partial_1 \psi^2 \, \partial_2 \psi^1$ \emph{except at the initial time} $t=0$.

\subsection{The $C$-method for 2$D$ ALE-Euler}

Next, we describe some aspects of our numerical framework for solving \eqref{ALE-Euler-2d-full}.  Specifically, we adapt the 
$C$-method, introduced in the Eulerian setting in \cite{RaReSh2019a,RaReSh2019b}, to the ALE setting. 
One of the key features of the $C$-method is space-time smooth tracking of shock/contact fronts and their geometries 
via so-called $C$-functions.  The $C$-functions are space-time smoothed versions of 
localized solution gradients, and are found as the solutions to auxiliary scalar reaction-diffusion equations. 
These $C$-functions in turn allow us to implement both directionally 
isotropic (for shock stabilization) and anisotropic (for contact stabilization) artificial 
viscosity schemes. In particular, the $C$-method is a PDE-level modification of \eqref{ALE-Euler-2d-full}. Consequently, the methods 
developed in  \cite{RaReSh2019a,RaReSh2019b} can be implemented in the ALE context in a straightforward manner. 
For the purposes of brevity, we omit some of the details here and refer the reader to \cite{RaReSh2019b} and 
Appendix \ref{appendix:C-method}.

\subsubsection{WENO-type reconstruction and computation of $a^j_i$}
We discretize the uniform mesh and index the nodes by $x_{r,s} = (x^1_r,x^2_s)$. 
At each $x_{r,s}$ we construct numerical flux functions $\hat{F}^1_{r+\frac{1}{2},s}$ and $\hat{F}^2_{r,s+\frac{1}{2}}$  that will be used to 
approximate the derivatives $\partial_1 \tilde{F}^1(\tilde{Q}) |_{x_{r,s}}$ and $\partial_2 \tilde{F}^2(\tilde{Q}) |_{x_{r,s}}$, respectively. 
We describe the procedure for $\hat{F}^1_{r+\frac{1}{2},s}$.   For ease of notation, we drop the superscript $1$ and let 
$\tilde{F}^1 \equiv \tilde{F}$.   
Decompose $\tilde{F} = \tilde{F}^{{v}} + \tilde{F}^p + \tilde{F}^E + \tilde{F}^{\mathcal{J}}$ with 
\begin{equation}
{{{\tilde{F}}}}^v= \begin{pmatrix}
		\mathcal{J} \rho v^j  \\[0.15em]
		\mathcal{J} \rho u^1 v^j  \\[0.15em]
		\mathcal{J} \rho u^2 v^j  \\[0.15em]
		\mathcal{J} E  v^j \\[0.15em]
		0
		\end{pmatrix}  \quad \text{ , } \quad 
{{{\tilde{F}}}}^p = \begin{pmatrix}
		0 \\[0.15em]
		 a^j_1 p \\[0.15em]
		a^j_2 p \\[0.15em]
		0 \\[0.15em]
		0
		\end{pmatrix} \quad \text{ , } \quad 
{{{\tilde{F}}}}^E = \begin{pmatrix}
		0 \\[0.15em]
		 0 \\[0.15em]
		0 \\[0.15em]
		p a^j_i u^i \\[0.15em]
		0
		\end{pmatrix} \quad \text{ , } \quad 
{{{\tilde{F}}}}^{\mathcal{J}} = \begin{pmatrix}
		0 \\[0.15em]
		 0 \\[0.15em]
		0 \\[0.15em]
		0 \\[0.15em]
		- a^j_i \psi_t^i
		\end{pmatrix}\,.  
\end{equation}
Each component of the advection term $\tilde{F}^v$ is approximated at the half-point $x_{r+\frac{1}{2},s}$ as
\begin{equation}\label{WENO-simple-reconstruct}
\hat{F}^v_{r+\frac{1}{2},s} = \mathrm{WENO}\left(q,{\mathcal{J}} v^j \right) 
\coloneqq q_{r+\frac{1}{2},s} (\mathcal{J} v^j )_{r+\frac{1}{2},s} \,,
\end{equation}
where $q$ denotes one of the variables $q \in \left\{ \rho,\rho u^1,\rho u^2,E \right\}$ and $q_{r+\frac{1}{2},s}$ is computed using a 
standard 5\textsuperscript{th} order WENO reconstruction \cite{Shu1998} of $q$ with upwinding based on the sign of 
$(\mathcal{J} v^j )_{r+\frac{1}{2},s}$.   The velocity $(\mathcal{J} v^j )_{r+\frac{1}{2},s}$ is computed according to the 4\textsuperscript{th} order 
average
\begin{equation}\label{vel-avg}
(w)_{r+\frac{1}{2},s} \coloneqq \frac{ -w_{r-1,s} + 7w_{r,s} + 7 w_{r+1,s} - w_{r+2,s} }{12} \,.
\end{equation}
The additional advection terms $\hat{F}^E_{r+\frac{1}{2},s} = \mathrm{WENO}(p,a^j_i u^i)$ and 
$\hat{F}^{\mathcal{J}}_{r+\frac{1}{2},s} = \mathrm{WENO}(1,-a^j_i \psi_t^i)$ can be approximated in a 
similar fashion to \eqref{WENO-simple-reconstruct}.  The pressure term $\hat{F}^p_{r+\frac{1}{2},s}$ is approximated by the 
4\textsuperscript{th} order average \eqref{vel-avg}.  Finally, the total flux is given by the sum 
$\hat{F}_{r+\frac{1}{2},s} = \hat{F}^v_{r+\frac{1}{2},s}  + \hat{F}^p_{r+\frac{1}{2},s}  + \hat{F}^E_{r+\frac{1}{2},s} 
 + \hat{F}^{\mathcal{J}}_{r+\frac{1}{2},s} $.   The semi-discrete scheme for \eqref{ALE-Euler-2d-full} then reads
 \begin{equation}\label{ALE-Euler-semi-discrete}
 \partial_t \tilde{Q}_{r,s} + \frac{\hat{F}^1_{r+\frac{1}{2},s} - \hat{F}^1_{r-\frac{1}{2},s}}{\Delta x^1} + 
 \frac{\hat{F}^2_{r,s+\frac{1}{2}} - \hat{F}^2_{r,s-\frac{1}{2}}}{\Delta x^2} = 0 \,.
 \end{equation}

For free-stream flows, we have that $q_{r,s} \equiv$ constant and the scheme becomes linear,  due to the linear averaging \eqref{vel-avg}. 
In particular, it is easy to verify that the free-stream is preserved, provided the components of the cofactor matrix $a^j_i$ are computed by 4\textsuperscript{th} order central differencing of the map $\psi$ as
\begin{equation}\label{gradpsi-CD4}
\left[ \partial_1 \psi^j \right]_{r,s} = \frac{ \psi^j_{r-2,s} - 8 \psi^j_{r-1,s} + 8 \psi^j_{r+1,s} - \psi^j_{r+2,s} }{12 \Delta x^1} \,,
\end{equation}   
and similarly for $\left[ \partial_2 \psi^j \right]_{r,s}$.  

To confirm this, we perform a free-stream test on the $50 \times 50$ time-dependent moving-mesh defined by 
 \begin{subequations}\label{free-stream-mesh}
\begin{align}[left = \empheqlbrace\,]
\psi^1(x^1,x^2,t) &= x^1 + 0.4 \sin \left( \frac{3 \pi t}{T} \right) \sin \left( \frac{3 \pi}{8} (x^2 + 8)  \right) \\
\psi^2(x^1,x^2,t) &= x^2 + 0.8 \sin \left( \frac{3 \pi t}{T} \right) \sin \left( \frac{3 \pi}{8} (x^1 + 8)  \right)  
\end{align}
\end{subequations} 
for $(x^1,x^2) \in [-8\,,+8]^2$ and $ 0 \leq t \leq T = 80$. The initial data is uniform $\mathsf{U}_0 \equiv 1$ and we employ periodic boundary 
conditions. The magnitude of the  density error at the final time $t=T$ is $||\uprho(\cdot,T)-1 ||_{L^\infty} = 9.10  \times 10^{-14}$ i.e. 
the scheme maintains free stream flows to machine precision. 

For non-smooth problems with shocks or contacts, it is necessary to add an artificial viscosity term to the right-hand side of 
\eqref{ALE-Euler-2d-full-motion}, and the semi-discrete scheme \eqref{ALE-Euler-semi-discrete} must be modified appropriately. 
The details of the particular form of artificial viscosity we use are provided in Appendix \ref{appendix:C-method}.

\begin{remark}
The simplified WENO-type reconstruction procedure outlined above is similar in some respects to the WENO schemes based 
on the so-called \emph{alternative flux formulation}, first introduced in \cite{ShuOsher1988} and explored extensively in several recent papers 
\cite{JiShZh2013,Jiang2014,NoTeAbFu2015,ChFeJiTa2018,LiJiZhLi2022}.  
In particular, both schemes define the flux $\hat{F}_{r+\frac{1}{2},s}$ by first reconstructing 
the variables $q_{r+\frac{1}{2},s}$.  On the other hand, the alternative flux formulation WENO schemes utilize characteristic 
decompositions and (exact or approximate) Riemann solvers. The resulting algorithms are more expensive but also more robust.  
Nonetheless, for simple problems, both the simplified WENO and alternative flux WENO schemes produce similar results 
 \cite{RaReSh2019a,RaReSh2019b}.  For more challenging problems, the simplified WENO scheme produces 
 oscillatory solutions; these oscillations can be suppressed with $C$-method artificial viscosity. 
\end{remark}

\subsubsection{Explicit interface tracking}

The $C$-method utilizes a simple method for 
 tracking of contact discontinuities which we first describe in the Eulerian setting i.e. for the system \eqref{Euler-2d}. 
Let $\ze : \mathcal{I} \times [0,T] \to \Omega$ be a parametrization 
of the material interface with parameter $\alpha \in \mathcal{I} \subset \mathbb{R}$, and 
with components $\ze = (\ze^1 \,, \ze^2)$.  
In many simulations, the contact discontinuity is a closed or periodic curve, and in this case 
we take $\mathcal{I} = [-\pi \,, \pi ]$.   Given an initial parametrization $\ze_0$ of the contact 
discontinuity, the interface $\ze(\alpha,t)$ is found as the solution to 
\begin{subequations}\label{interface-e}
\begin{alignat}{2}[left = \empheqlbrace\,]
\partial_t \ze (\alpha,t)  &=  \bar{\ue} \circ \ze (\alpha,t) \,, \quad && \alpha \in \mathcal{I} \text{ and } 0 < t \leq T    \label{interface-e-eqn1}  \\
\ze(\alpha,0) &= \ze_0(\alpha) \,, && \alpha \in \mathcal{I} \text{ and } t = 0    \label{interface-e-eqn2}
\end{alignat}
\end{subequations} 
Here, the velocity $\bar{\ue}$ is defined as the average 
$\bar{\ue} = \frac{1}{2} ( \ue^+ + \ue^-)$, with $\ue^\pm$ denoting the fluid velocity on 
either side of the interface. In a numerical implementation, the average $\bar{\ue}$ is 
approximated by bilinear interpolation of $\ue$ onto $\ze$.

The ALE analog of the (Lagrangian) interface tracking algorithm 
described above can be derived by defining the ALE interface parametrization 
$z : \mathcal{I} \times [0,T]$ as the image 
of $\ze$ under the action of the inverse ALE map 
$\psi^{-1} : \Omega \times [0,T] \to \Omegaref$ i.e. 
$$
z(\alpha,t) = \psi^{-1} \circ \ze (\alpha,t) \,.
$$
If the map $\psi$ resolves  mesh points around $\ze$, then the ALE interface $z$  represents a ``zoomed-in'' 
version of $\ze$ that magnifies small scale structures c.f. \Cref{fig:RT-adaptive-z-ALE}. 

A chain rule computation shows that $z$ is the solution to
\begin{subequations}\label{interface}
\begin{alignat}{2}[left = \empheqlbrace\,]
\partial_t z (\alpha,t)  &=  \bar{v} \circ z (\alpha,t) \,, \quad && \alpha \in \mathcal{I} \text{ and } 0 < t \leq T    \label{interface-eqn1}  \\
z(\alpha,0) &= z_0(\alpha) \,, && \alpha \in \mathcal{I} \text{ and } t = 0    \label{interface-eqn2}
\end{alignat}
\end{subequations} 
where $\bar{v} = \frac{1}{2} \left( v^+ + v^- \right)$. The initial interface $z_0$ is defined by 
\begin{equation}\label{initial-condition}
z_0(\alpha) = \psi^{-1} \circ \ze (\alpha,0) \,.
\end{equation}
In a numerical implementation, the initial ALE interface $z$ can be computed as the roots of $\psi_0(z_0) = \ze_0$ using e.g. 
Newton's method.

\subsection{Coupled SAM-ALE algorithm}

Our SAM algorithm is coupled to the ALE $C$-method by defining an appropriate target Jacobian function $\Ge_k$. In this work, for 
simplicity, we shall assume that $\Ge_k$ is explicitly defined, either by some particular formula (as in the Noh test), or via the 
interface $\ze_k$ (for the RT test).  Future work will investigate coupling of SAM-ALE by means of balanced 
monitoring of solution gradients \cite{DaZe2010}.  
In the case of RT instability, it is important to use the interface $\ze$ to control adaptation since it 
allows high mesh concentration in KH roll up zones, in contrast to the balanced monitoring approach in which
the magnitudes of solution gradients decrease in KH zones due to mixing \cite{TaTa2003}. 

The complete SAM-ALE algorithm is provided in \Cref{alg:SAM-ALE}.

\begin{algorithm}[ht]
\caption{\textbf{\textsc{: coupled SAM-ALE}}}\label{alg:SAM-ALE}
\begin{enumerate}[itemsep=0.0em,leftmargin=2.0cm,label=\textbf{Step \arabic* :}]
\setcounter{enumi}{-1}

\item \textbf{Initialization $t=0$.}
\begin{enumerate}[itemsep=0.0em,leftmargin=0.0cm]
\item Define the initial Eulerian data $\Qe_0$ on the uniform mesh $\mathcal{U} \subset \Omega$ and the initial interface parametrization 
$\ze_0(\alpha)$. 
\item Define the initial target Jacobian function $\Ge_0$ on $\mathcal{U}$. Compute the initial ALE map $\psi_0 : \Omegaref \to \Omega$ and 
adaptive mesh $\mathcal{T}_0 = \psi_0(\Tref) \subset \Omega$ using static SAM \Cref{static-alg1}. 
\item Define the initial ALE data $Q_0$. Compute the initial ALE interface $z_0$ using Newton's method.
\end{enumerate}

\item \textbf{Time-stepping $t = t_k \geq 0$.} Assume that we are given all quantities at $t=t_k$. 
\begin{enumerate}[itemsep=0.0em,leftmargin=0.0cm]
\item Define the target Jacobian function $\Ge_{k+1}$ and compute the map $\psi_{k+1}$ and 
adaptive mesh $\mathcal{T}_{k+1}$ according 
to restarted dynamic SAM \Cref{dynamic-alg2}. 
\item Compute the cofactor matrix $a^j_i$ using \eqref{gradpsi-CD4}. 
Define the mesh velocity $\partial_t \psi_{k+1} = \frac{\psi_{k+1}-\psi_k}{\Delta t}$. 
\item Compute the ALE variables $\tilde{Q}_{k+1}$ and $z_{k+1}$ using the $C$-method and RK4 time-stepping. The mesh, cofactor matrix, and 
mesh velocity are kept fixed over the time step. 
\item Compute the interface $\ze_{k+1} = \psi_{k+1} \circ z_{k+1}$. 
\item If $t_{k+1} = T$, then stop; else, set $t=t_{k+1}$ and return to \textbf{Step 1}(a). 
\end{enumerate}

\end{enumerate}
\end{algorithm}

\section{SAM-ALE simulations of gas dynamics}\label{sec:ALE-sims}

\subsection{Noh implosion}
The first test is the 2$D$ Noh implosion: an initially cold gas is directed towards the origin with speed 1 and instantaneously 
implodes  at the origin, resulting in a radially symmetric infinite strength shock propagating outwards with speed $1/3$. 
This is an extremely difficult test problem and almost all codes report errors in the form of wall heating, lack of symmetry, incorrect 
shock speeds, or even failure to run \cite{LiWe2003}. This is the case for both Lagrangian-type codes with artificial viscosity 
\cite{LipnikovShashkov2010,Breil2016,Cook2013}, as well as AMR codes such as RAGE \cite{RAGE2008}. 
Extensive numerical testing in \cite{Timmes2005} showed that catastrophic 
anomalies occur in AMR solutions, with the anomalies persisting, or even worsening as the grid is refined. These anomalies occur 
due to spurious wave reflections on discontinuous grids \cite{VichnevetskyTurner1991,FLASH2000}. 

\subsubsection{Problem description} 
The domain as $\Omega = [0,1]^2$,  the adiabatic constant is $\gamma = 5/3$, and the initial data is
\begin{equation}\label{noh_initialdata1}
\begin{bmatrix}
\uprho_0 \\ (\uprho \ue^1)_0 \\ (\uprho \ue^2)_0 \\ \Ee_0 
\end{bmatrix}
=
\begin{bmatrix}
1 \\ -\cos(\theta) \\ -\sin(\theta) \\ 0.5+10^{-6}/(\gamma-1)
\end{bmatrix} \chi_{r > 0} +  
\begin{bmatrix}
1 \\ 0 \\ 0 \\ 0.5+10^{-6}/(\gamma-1)
\end{bmatrix} \chi_{r = 0}  \,,
\end{equation}
where $r =|y|$ is the radial coordinate,  $\theta \in [0,\frac{\pi}{2})$ is the polar angle, and $\chi_A$ is the indicator function on the set $A$. 
We employ reflecting boundary conditions on the 
left and bottom boundaries and use the exact solution to impose the boundary conditions at the top and right boundaries. 
The problem is run until the final time $T=2$. 

\subsubsection{Uniform mesh simulations}
\begin{figure}[ht]
\centering
\subfigure[$N = 50 \times 50$]{\label{fig:Noh-50x50}\includegraphics[width=46mm]{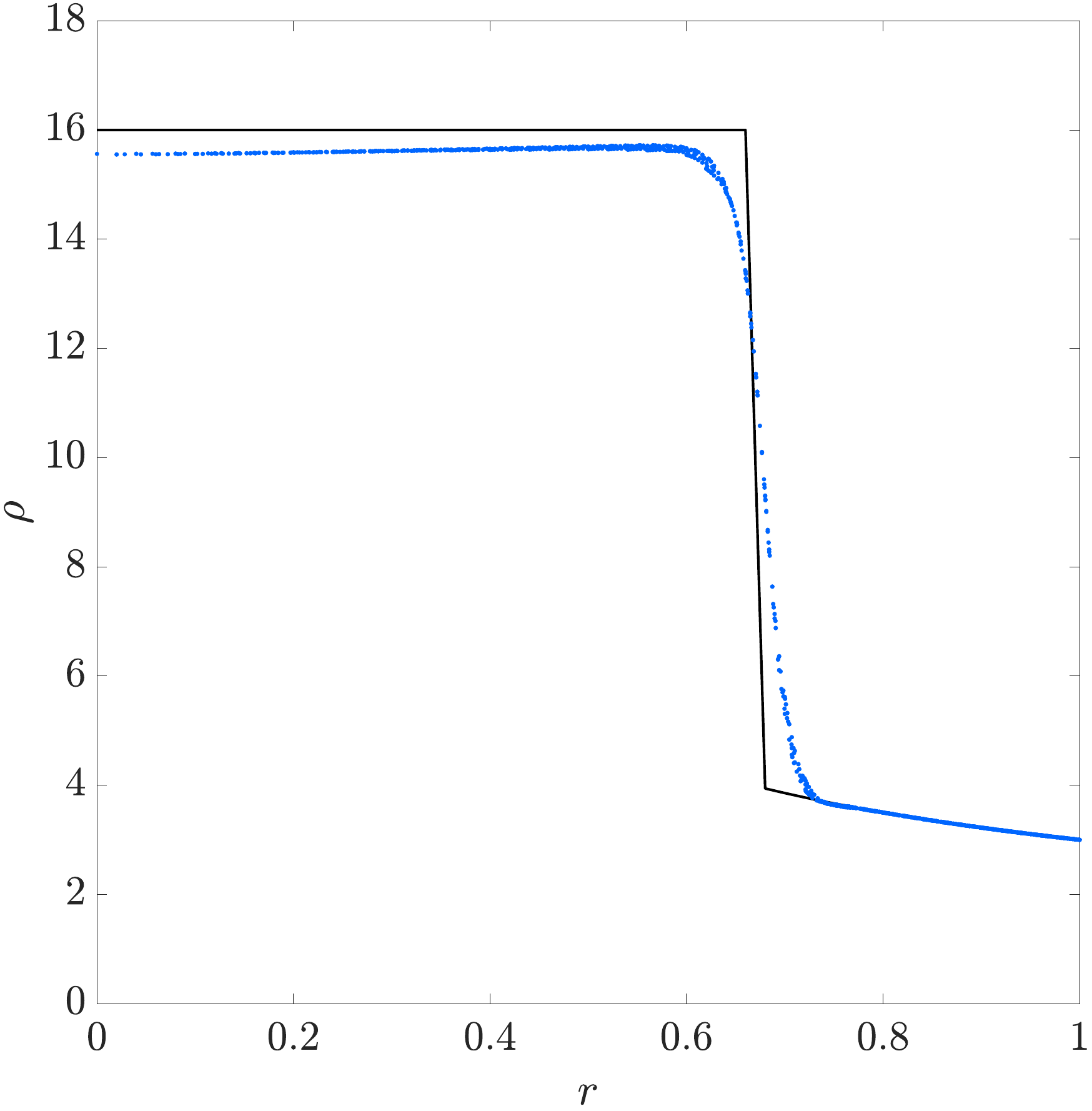}} 
\hspace{2em}
\subfigure[$N = 100 \times 100$]{\label{fig:Noh-100x100}\includegraphics[width=42.5mm]{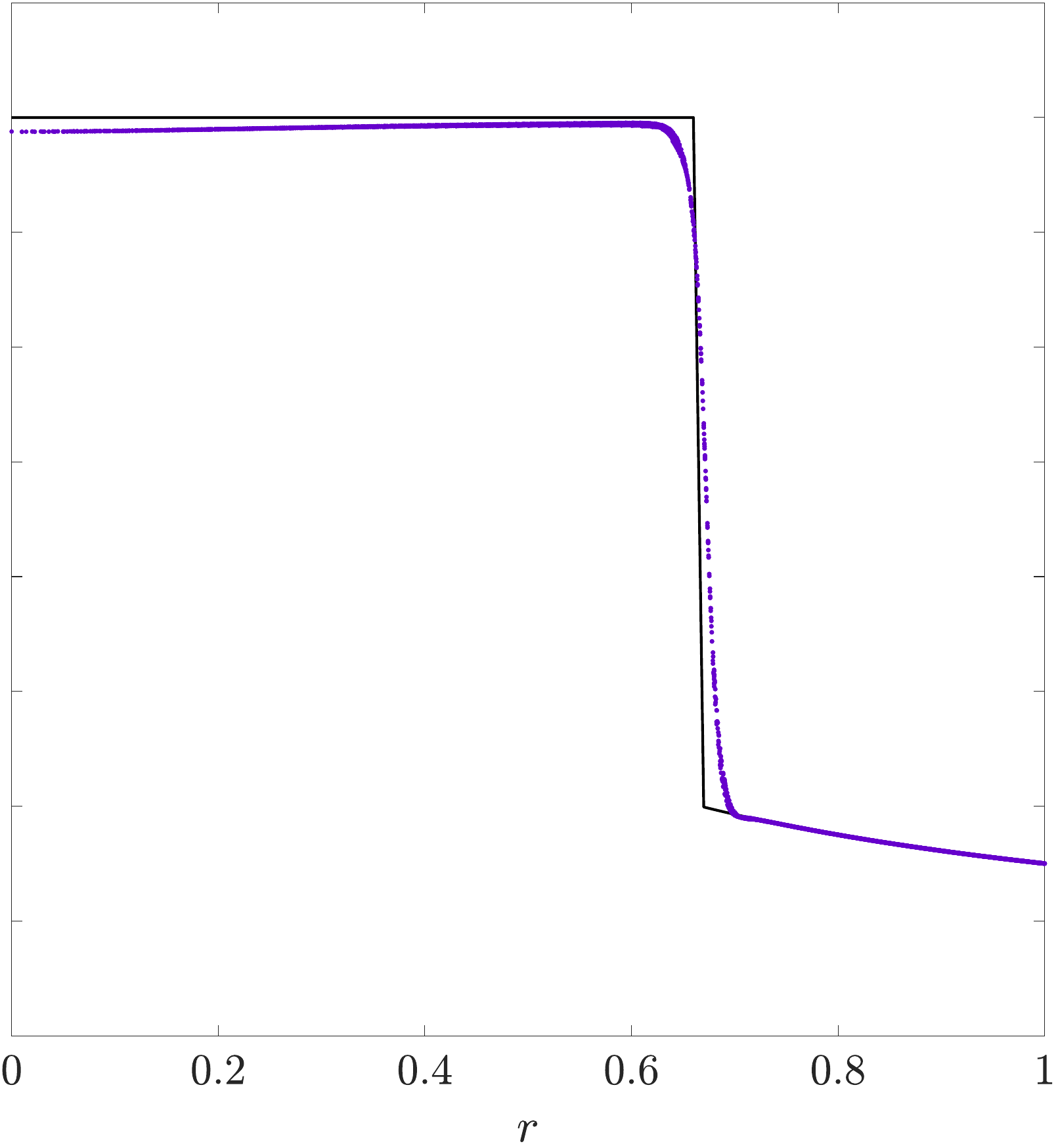}} 
\hspace{2em}
\subfigure[$N = 200 \times 200$]{\label{fig:Noh-200x200}\includegraphics[width=42.5mm]{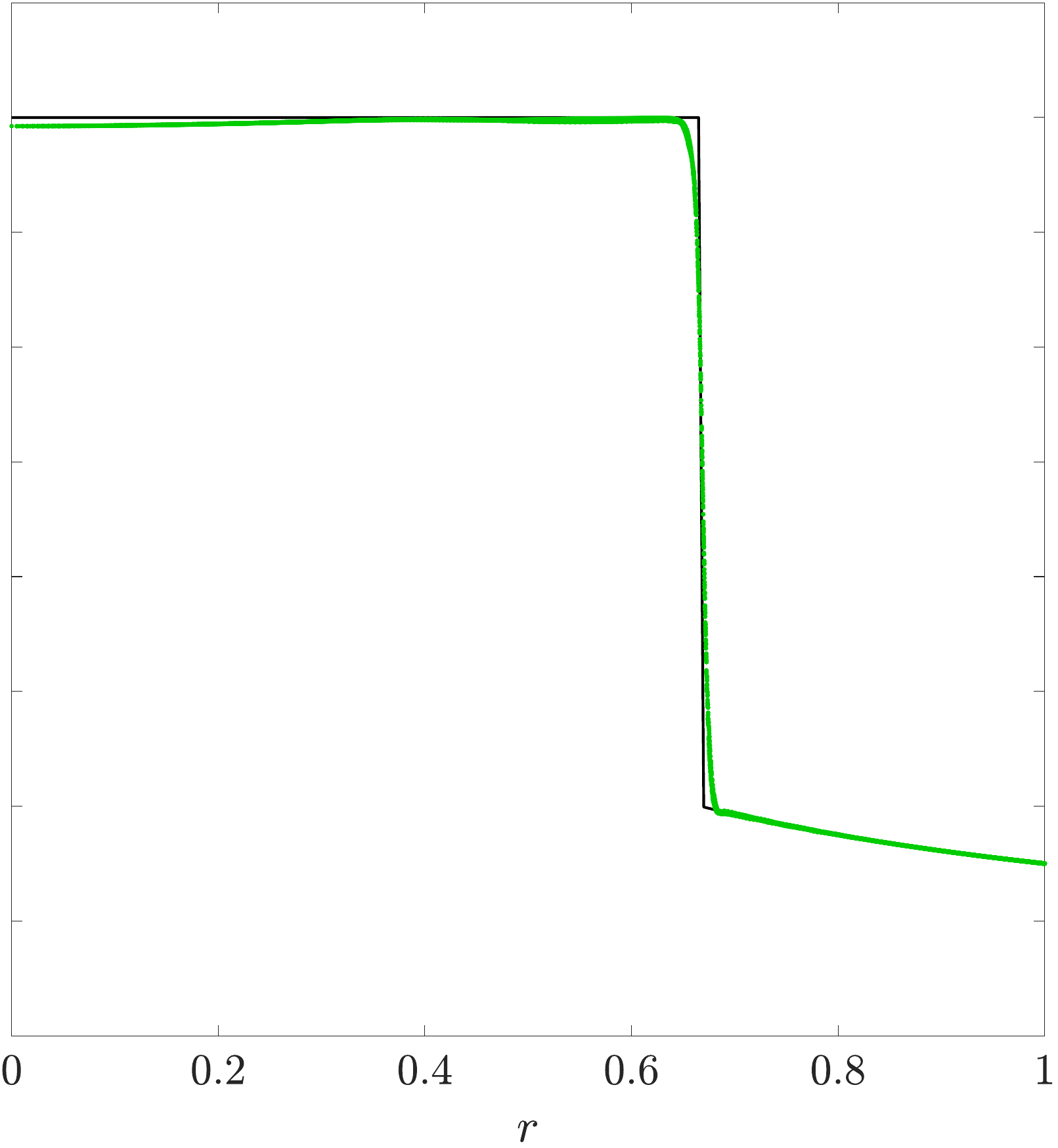}} 
\caption{Uniform mesh runs for the 2$D$ Noh problem. Shown are the density scatter plots vs radial coordinate $r$. The black curve in 
each subfigure is the exact solution. The shock fronts are sharp and the solutions free of the spurious 
asymmetry, wall-heating, oscillation, and shock-racing errors associated 
with the majority of numerical methods for this test. }
\label{fig:Noh_uniform}
\end{figure}

We apply the $C$-method as described in \cite{RaReSh2019b} on $50 \times 50$, $100 \times 100$, and $200 \times 200$ meshes with 
time step $\Delta t$ set so that $\mathrm{CFL} \approx 0.2$.  The $C$-method artificial viscosity coefficients in \eqref{euler2d-ALE2}
 are fixed as $\beta_u = 0.35$,  $\beta_E = 2.5$, and $\mu=0$. 
The scatter plots of density vs $r$ in \Cref{fig:Noh_uniform} show that the $C$-method produces stable non-oscillatory solutions that 
maintain radial symmetry.  Moreover, the smooth artificial viscosity almost entirely removes the 
wall-heating error  in the higher resolution runs. 

\subsubsection{SAM-ALE simulations}
Next, we apply SAM-ALE on a $50 \times 50$ dynamic adaptive mesh. For simplicity, we choose a 
specially designed forcing function $\Ge$ for the mesh generation, defined as 
\begin{align}
C_\psi(y^1,y^2,t) &= \exp \left[  -400 \left(  r^2 - t^2/9 \right)  \right] \,, \nonumber \\
\bar{\Ge}(y,t) &= \frac{1}{1 + \frac{\kappa}{1-\kappa}\frac{C_\psi(y,t)}{\int_{\Omega} C_\psi(y,t) \, \mathrm{d}y}} \,. \label{G-for-Noh}
\end{align}
This forcing function is  designed, using the known analytical solution, to track the moving shock.  In the future, a shock-tracking 
scheme analogous to the $\ze$-type advection \eqref{interface-e} for contract tracking will be employed to define $\bar{\Ge}$. 
The $\ze$-type advection  can track the shock with high accuracy, and the resulting $\bar{\Ge}$ is 
almost exactly the same as \eqref{G-for-Noh}. As such, for simplicity we
 use the specially designed function \eqref{G-for-Noh} in this work, with the understanding that  
 similar results can be obtained when  $\ze$-type shock tracking is used instead.  
The particular normalization used to define $\bar{\Ge}$ is motivated by the 
balanced monitoring  method \cite{DaZe2010}. We set $\kappa=0.3$ and the time-step as $\Delta t = 5 \times 10^{-4}$, which yields 
$\mathrm{CFL} \approx 0.2$, and choose artificial viscosity 
parameters $\beta_u = 0.1$,  $\beta_E = 0.7$, and $\mu = 0$. 

\begin{figure}[ht]
\centering
\subfigure[$N = 50 \times 50$ SAM-ALE mesh]{\label{fig:Noh-mesh}\includegraphics[width=50.5mm]{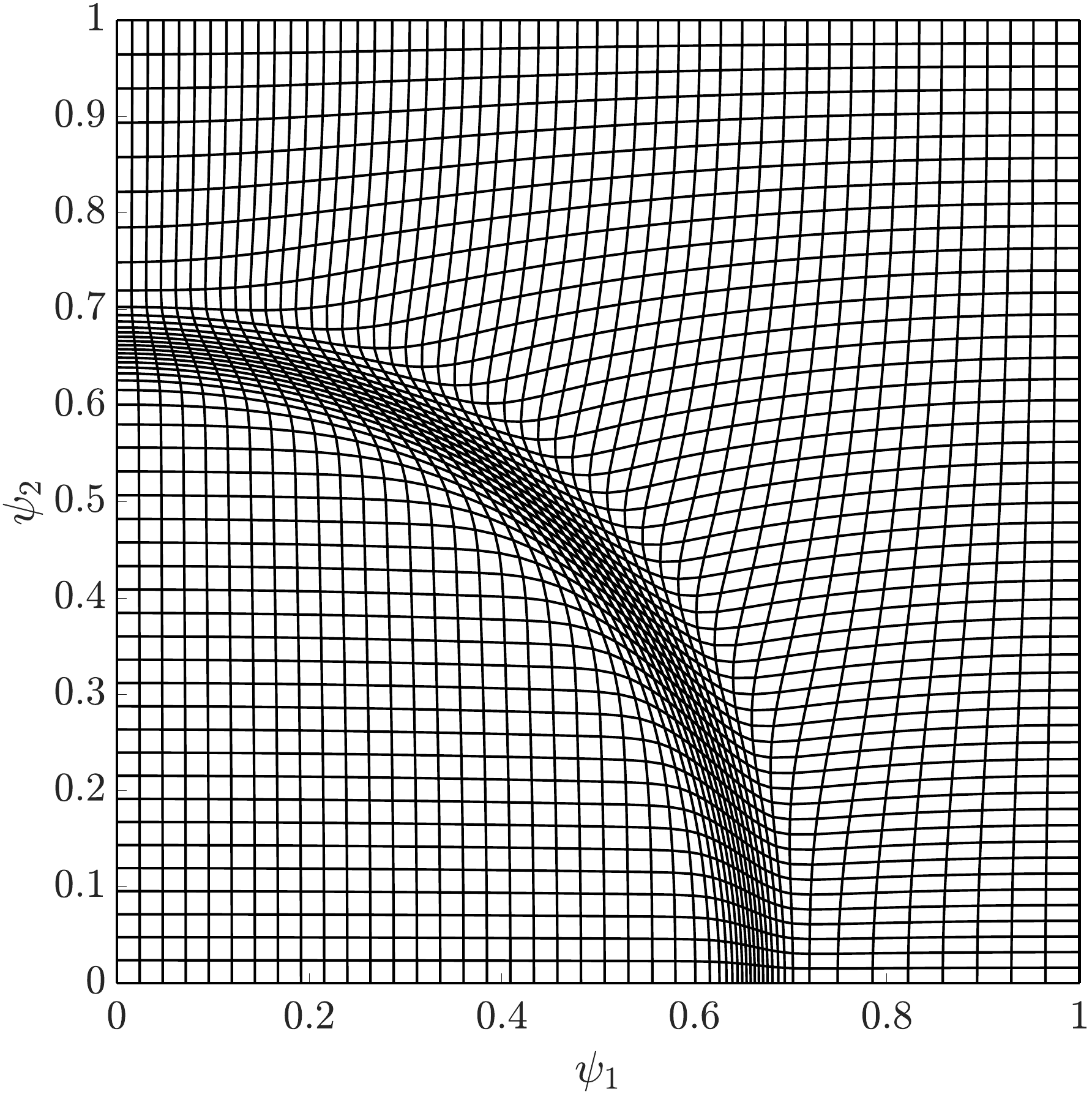}} 
\hspace{1em}
\subfigure[Density scatter plot]{\label{fig:Noh-rho}\includegraphics[width=50mm]{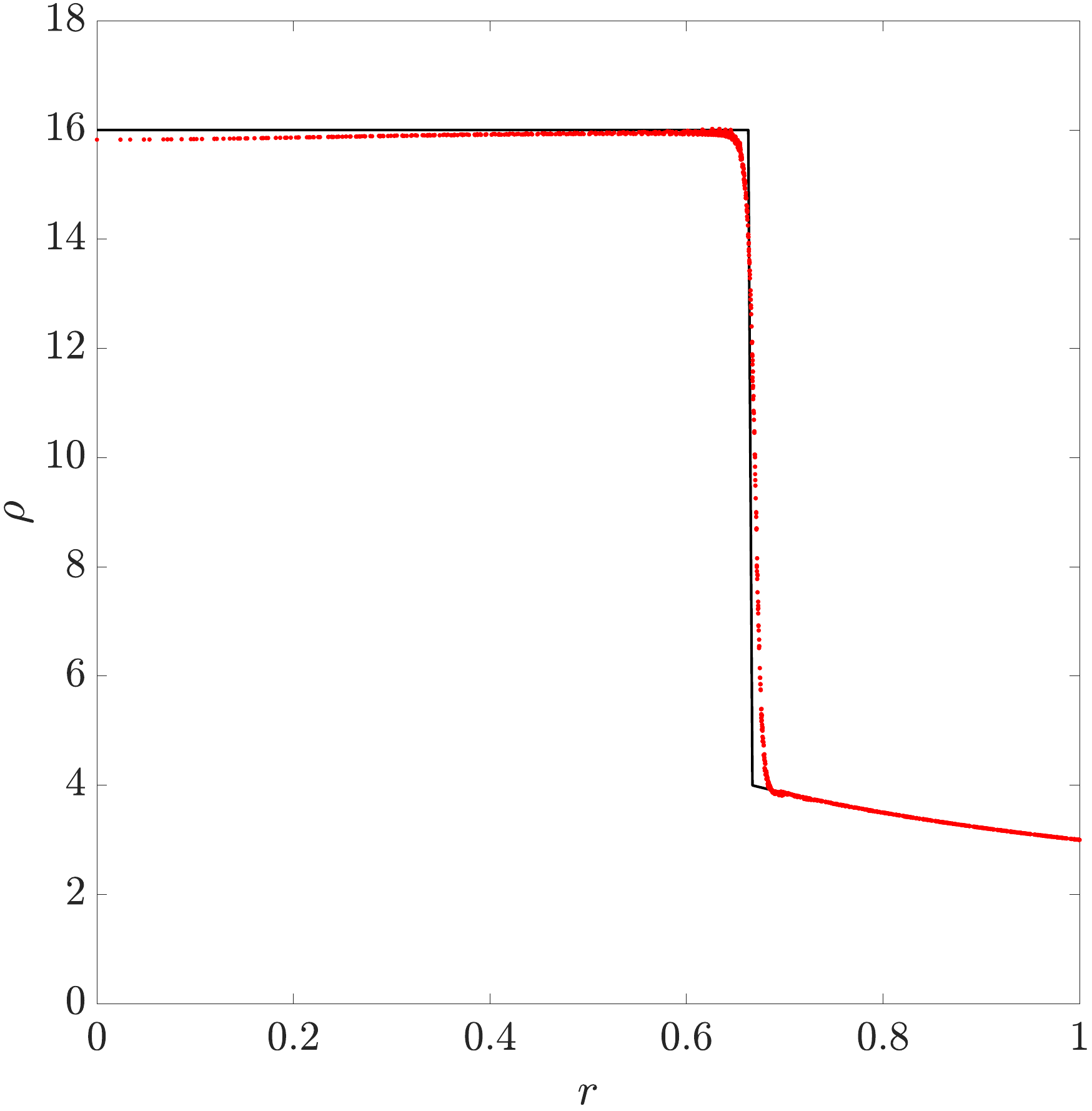}} 
\hspace{1em}
\subfigure[Density comparison]{\label{fig:Noh-comparison}\includegraphics[width=51.5mm]{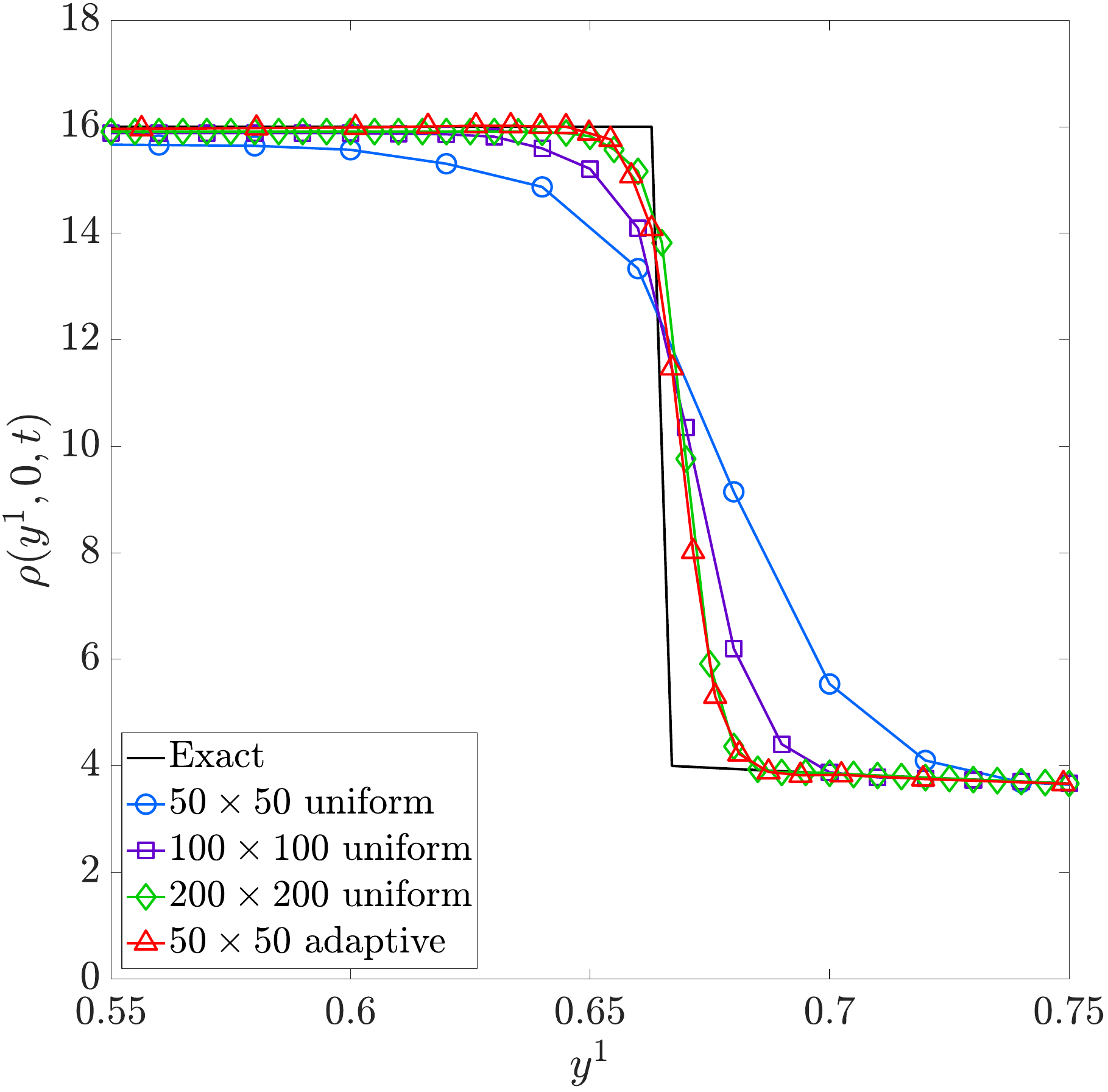}} 
\caption{SAM-ALE simulations of the Noh implosion. Shown are (a) adaptive mesh $\mathcal{T}$, (b) density scatter plot, and 
(c) comparison of uniform vs SAM-ALE density zoom-in at the shock. The smooth concentration and alignment of the mesh 
in the vicinity of the shock front allows for a sharp shock representation in the SAM-ALE solution, comparable 
to the high-resolution $200 \times 200$ uniform mesh solution.}
\label{fig:Noh_adaptive}
\end{figure} 

The results are shown in \Cref{fig:Noh_adaptive}. The shock front is sharp, the wall-heating error is very small, 
and solution symmetry is well preserved. The latter is a consequence of both $C$-method artificial viscosity as well as grid alignment with the 
shock front.   The density cross sections 
$\uprho(y^1,0,t)$ along the $y^1$-axis for the various simulations are shown in 
\Cref{fig:Noh-comparison}, which clearly shows that the $50 \times 50$ adaptive simulation outperforms the low-res and mid-res uniform
simulations, and is comparable to the high-res uniform simulation. The wall heating error is smallest for the adaptive simulation, and the sharpness 
of the shock fronts for the $200 \times 200$ uniform and $50 \times 50$ adaptive simulations are comparable. 
As shown in \Cref{table:adaptive-vs-uniform}, the 
adaptive mesh simulation produces the solution with the smallest $L^2$ error in the density. Moreover, the adaptive simulation 
is approximately 6 times faster than the high-res uniform simulation, and requires roughly the same amount of memory as the lowest-resolution uniform run.

\begin{table}[H]
\centering
\renewcommand{\arraystretch}{1.0}
\scalebox{0.8}{
\begin{tabular}{|c|c|c|c|}
\toprule
\midrule
\multirow{2}{*}{\textbf{Simulation}}  & \multicolumn{3}{c|}{\textbf{Simulation statistic}}\\

{}  & $L^2$ density error  &  CPU time (secs)  & Memory usage (MBs)   \\
\midrule
$50 \times 50$ uniform  &  $1.019 \times 10^{0}$  & $4.3$  & $7.5$ \\
\midrule
$100 \times 100$ uniform  &  $6.917 \times 10^{-1}$  & $35.3$  & $14.6$ \\
\midrule
$200 \times 200$ uniform  &  $5.406 \times 10^{-1}$  & $289$  & $43.7$ \\
\midrule
$50 \times 50$ adaptive  &  $4.897 \times 10^{-1}$  & $45.6$  & $7.8$ \\
\midrule
\bottomrule
\end{tabular}}
\caption{Comparison of simulation statistics for the uniform and adaptive mesh $C$-method simulations for the Noh problem. 
The low-res SAM-ALE simulation is more accurate than the high-res uniform simulation, while running 6 times faster and requiring 
only 18\% as much memory.}
\label{table:adaptive-vs-uniform}
\end{table}

\subsection{Rayleigh-Taylor instability}\label{subsec:RT}

Our second test problem is the classical RT instability. This test poses a huge challenge for Lagrangian and ALE methods due to the complex 
geometry of the evolving unstable interface. As such, limited RT ALE simulations are available in the literature (but see 
 \cite{ZeBoTa2005,Loubere2010,DoKoTzRi2012,GuPoSa2020} for some examples).  
 In fact, the RT problem is  so challenging for ALE codes that very often the goal is simply to perform a simulation that runs until the final 
 time without excessive mesh tangling, at which point the simulation breaks down \cite{Loubere2010,BaMaRiRiSh2016}. 

\subsubsection{Problem description}
We add the source term 
$\tilde{S}(x,t) = (0,0,-\mathcal{J}\rho g,-\mathcal{J}\rho g u^2,0,0)^\mathcal{T}$ to the right-hand side of 
\eqref{ALE-Euler-2d-full-motion}.  
 The domain is $\Omega =  [-0.25,0.25] \times [0,1]$ and we apply periodic and free-flow conditions in the $y^1$ 
 and $y^2$ directions \cite{RaSh2020a}. 
The initial data is $\ue_0 = 0$, and
\begin{subequations}
\label{CASTRO-initial}
\begin{alignat}{1}
 \pe_0 &= \begin{cases}
 	   5 - \uprho^{-}gy^2 \, &, \text{ if } y^2 < 0.5 \\
 	   5 - 0.5\uprho^{-}g - \uprho^{+}g(y^2-0.5) &, \text{ if } y^2 \geq 0.5 \\	
 	   \end{cases} \,,  \\
\uprho_0(y^1,y^2) &= \uprho^{-} + \frac{\uprho^{+}-\uprho^{-}}{2} \left[  1+ \tanh \left(\frac{y^2 - \eta_0(y^1)}{h} \right) \right]\,,
\end{alignat}
\end{subequations}
where $\uprho^{+}=2$ and $\uprho^{-}=1$,  $\eta_0(y^1) = 0.5-0.01 \cos (4 \pi y^1)$,   $h=0.005$, and $g=1$.  The problem is run 
until the final time $T = 2.5$. 

\subsubsection{Uniform mesh simulations} 
\begin{figure}[ht]
\centering
\subfigure[$N=64 \times 128$]{\label{fig:RT-uniform-64x128}\includegraphics[width=42.13mm]{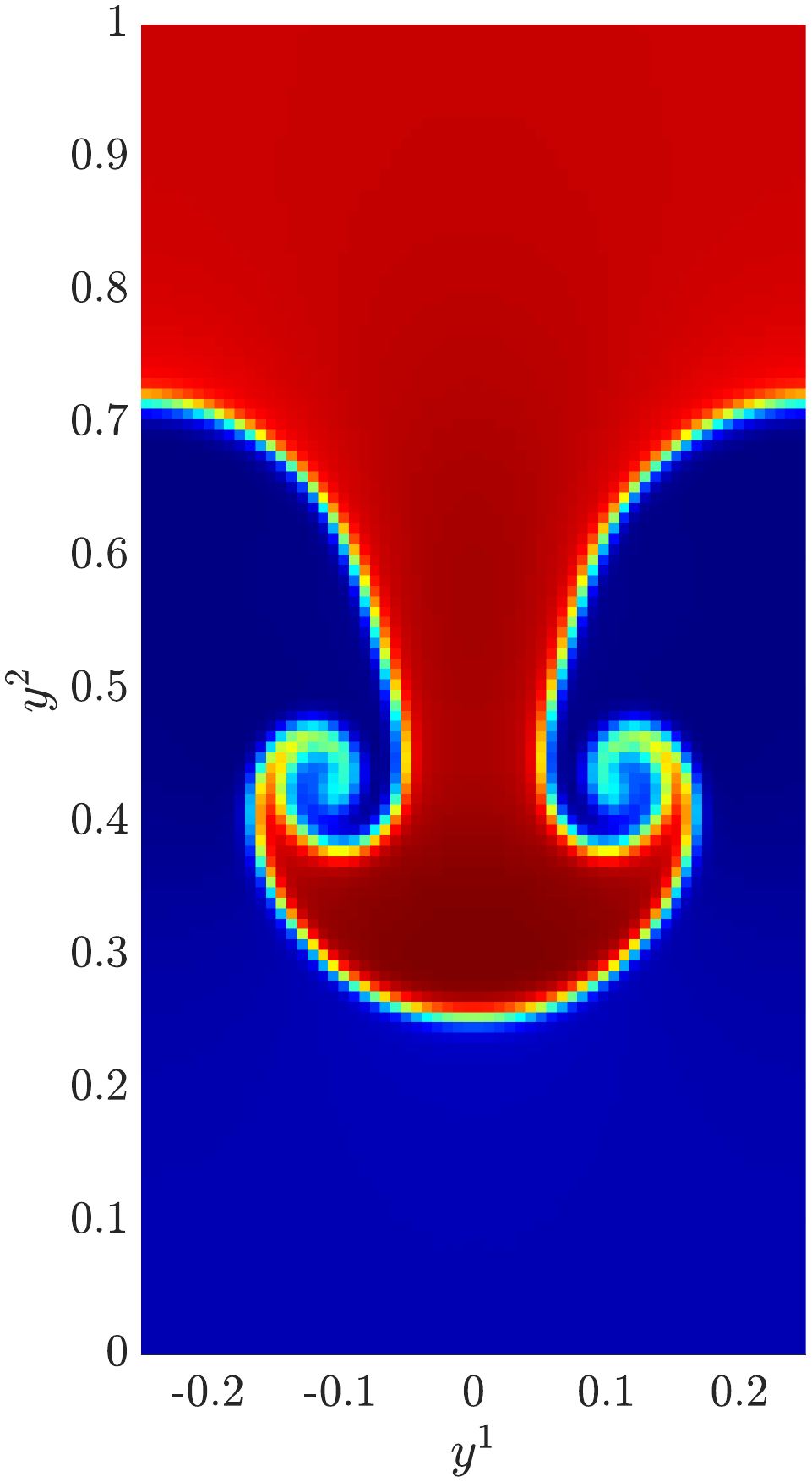}} 
\hspace{0em}
\subfigure[$N=128 \times 256$]{\label{fig:RT-uniform-128x256}\includegraphics[width=35mm]{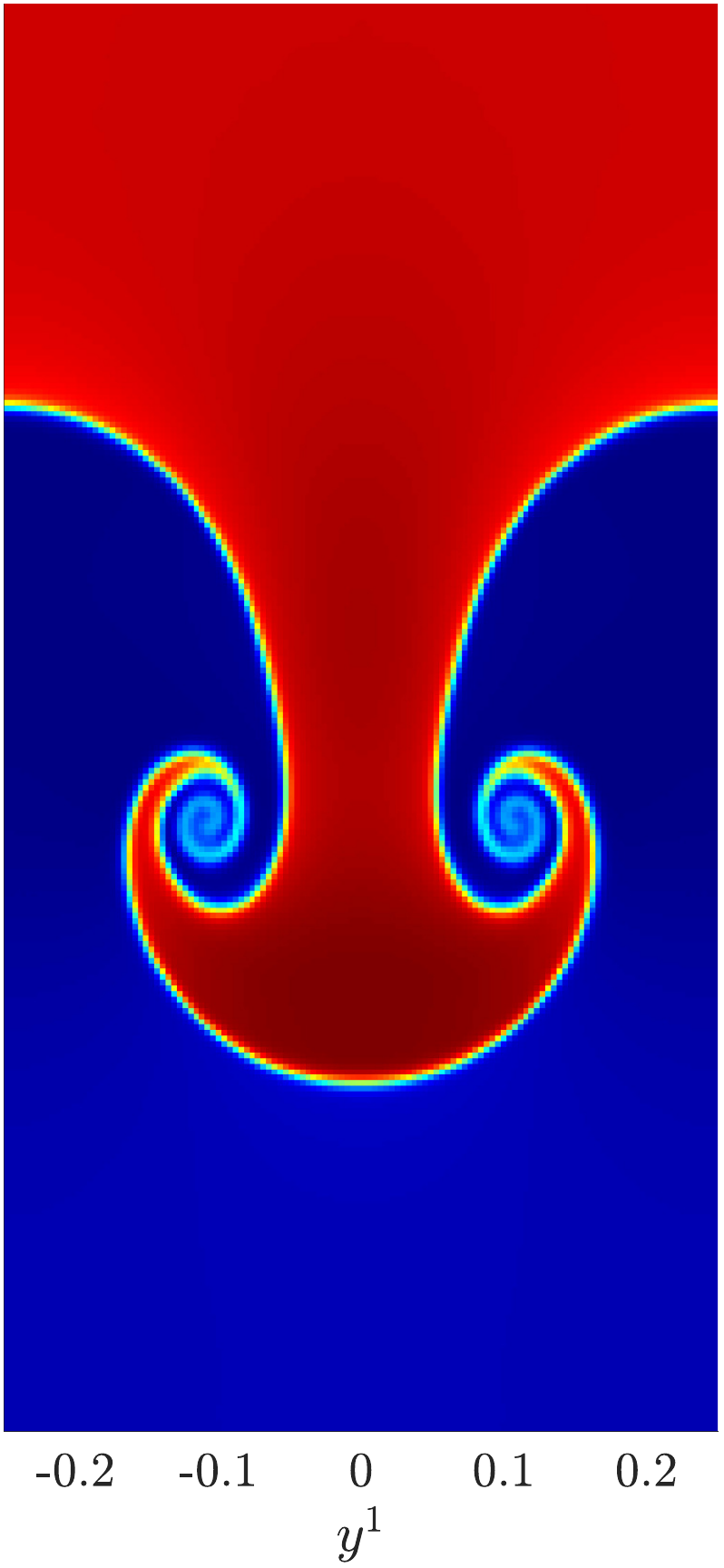}} 
\hspace{0em}
\subfigure[$N=256 \times 512$]{\label{fig:RT-uniform-256x512}\includegraphics[width=35mm]{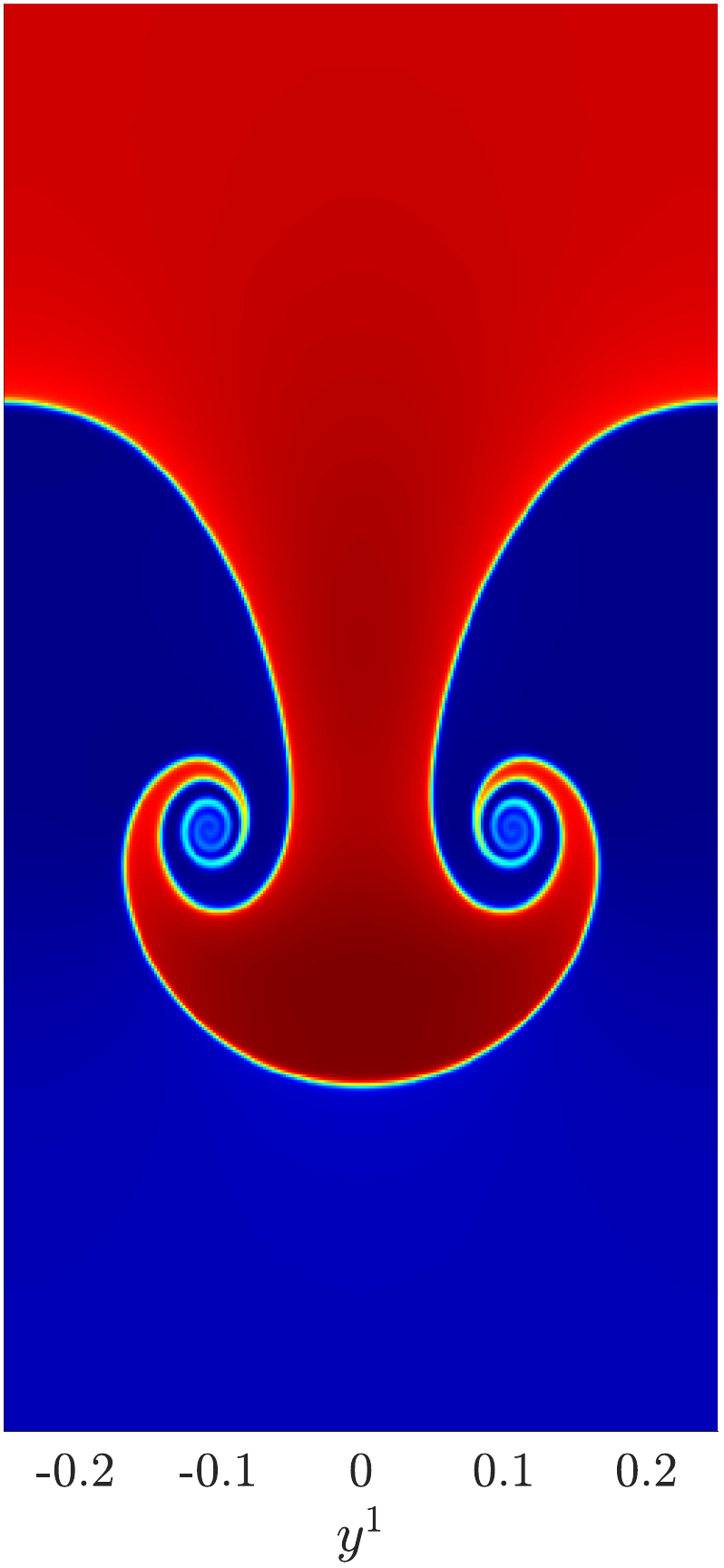}} 
\hspace{0em}
\subfigure[$N=512 \times 1024$]{\label{fig:RT-uniform-512x1024}\includegraphics[width=35mm]{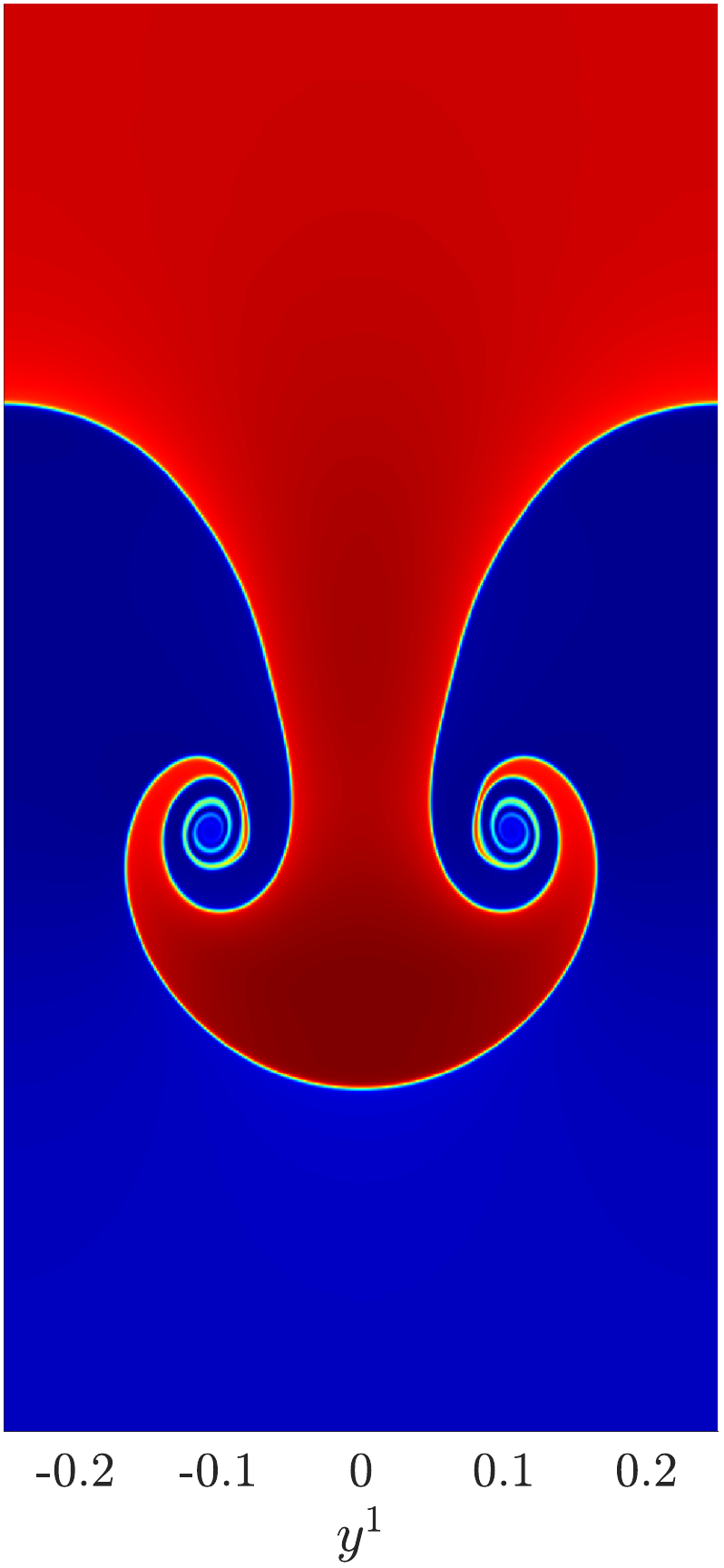}} 
\caption{Uniform mesh simulations of RT instability with sharper fronts and more small scale structure in the KH zone as the 
resolution increases.}
\label{fig:RT-uniform-comparison}
\end{figure}

We compute a sequence of uniform mesh simulations for resolutions $N = 64 \times 128$ through $N=512 \times 1024$ with 
$\mathrm{CFL} \approx0.45$. The artificial viscosity parameters are set as $\mu = 7.5 \times 10^{-4}$ and 
$\beta_u = \beta_E = 0$, and we
show heatmap plots of the density in \Cref{fig:RT-uniform-comparison}.  As the resolution is increased, more small-scale structure can be 
seen in the main KH roll up region. The artificial viscosity term suppresses further secondary instabilities that usually occur with other 
dimensionally split numerical methods \cite{LiWe2003,AlmgrenEtAl2010}. 

\subsubsection{Mesh generation with large zoom-in factor}
\begin{figure}[ht]
\centering
\subfigure[$\delta = 0.85$]{\label{fig:RT-non-convex}\includegraphics[width=42.5mm]{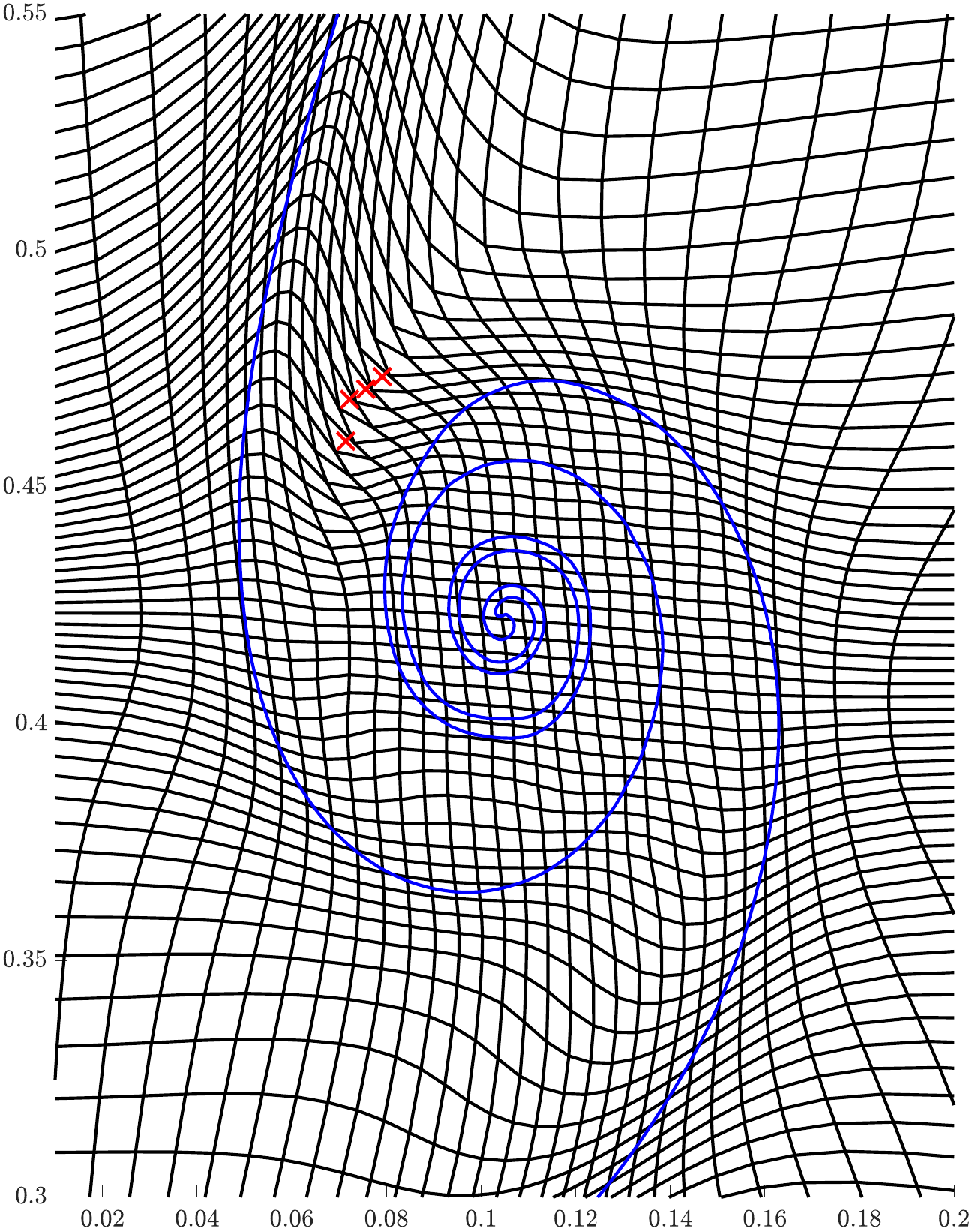}} 
\hspace{2em}
\subfigure[$\delta = 0.92$]{\label{fig:RT-non-convex-error}\includegraphics[width=42.5mm]{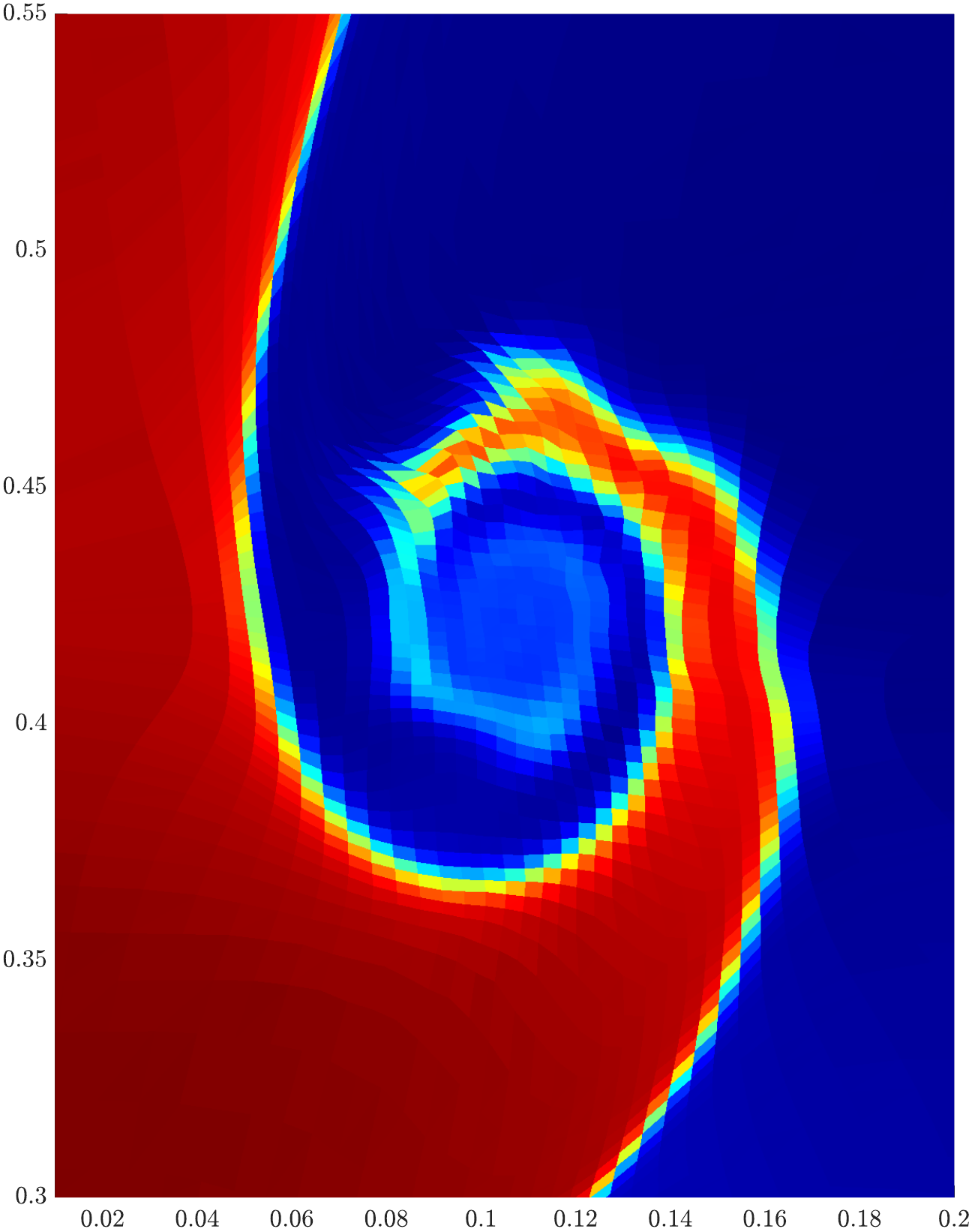}} 
\hspace{2em}
\subfigure[$\delta = 0.97$]{\label{fig:RT-mesh-large-zoom-in}\includegraphics[width=42.5mm]{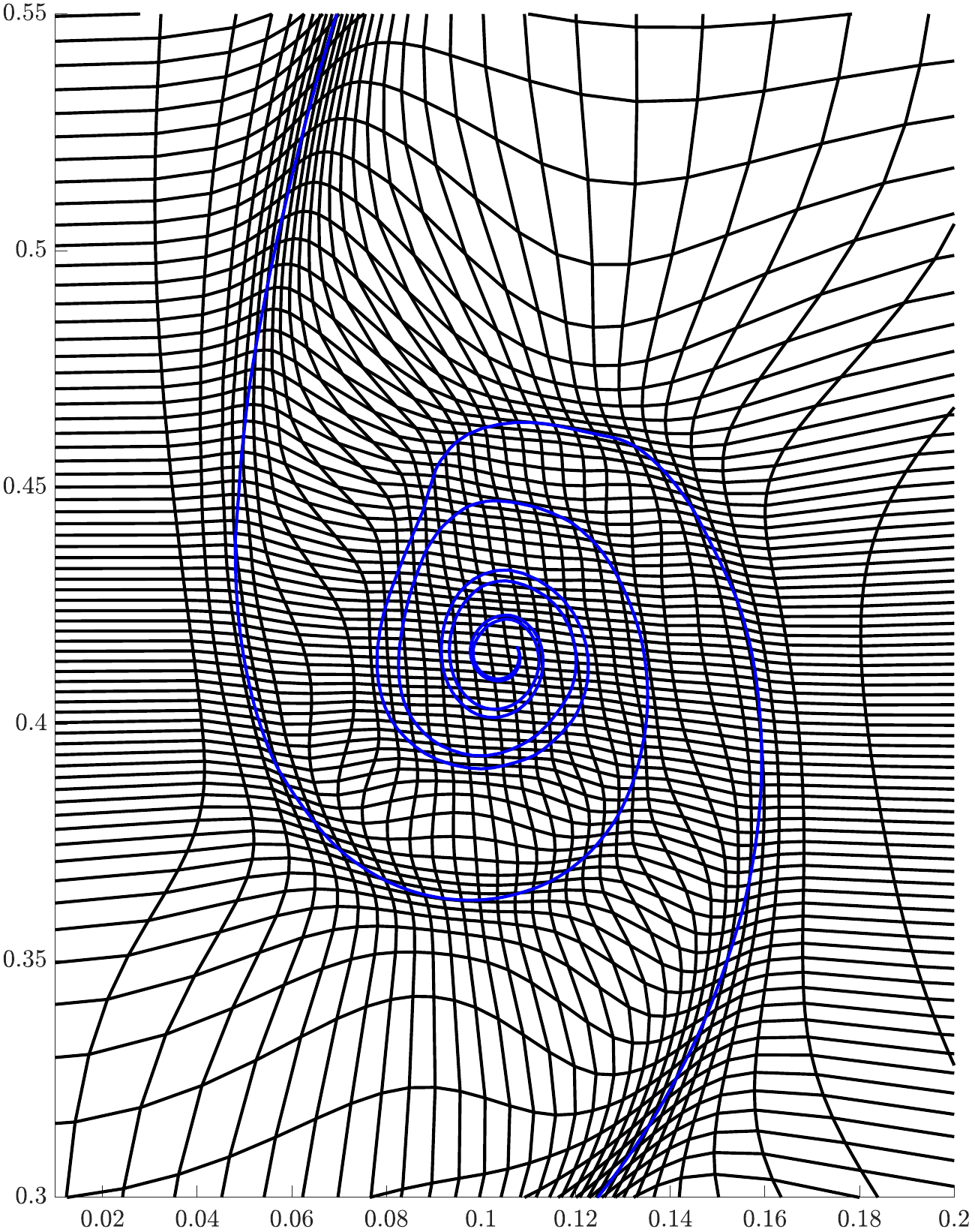}} 
\caption{$64 \times 128$ adaptive mesh simulations of RT with 
large zoom-in factor. Figure (a) is a zoom-in of the mesh computed with restarted 
SAM and $\delta=0.85$. The interface $\ze$ is shown as the blue curve, and the non-convex elements are indicated by red crosses. Figure 
(b) is a zoom-in of the density with $\delta = 0.92$. The non-convex elements cause spurious instabilities along the interface. 
Figure (c) shows the mesh computed with the large zoom-in algorithm; all the elements are convex and the mesh is smooth.}
\label{fig:RT-mesh-non-convex}
\end{figure}

Next, we aim to produce a $64 \times 128$ adaptive mesh with large zoom-in factor that resolves 
around the material interface $\ze$ and define 
a target Jacobian function as 
\begin{equation}\label{target-jacobian-z}
\Ge_{\delta}(y,t) = 1 - \delta \exp \left( - \left| \sigma  \min_{\alpha} |y - \ze(\alpha,t)| \right|^2  \right) \,,
\end{equation}
with $\sigma = 25$. 
For this resolution, the meshes produced with dynamic SAM contain non-convex elements for $\delta$ larger than approximately 0.85, as 
shown in \Cref{fig:RT-non-convex}. These non-convex elements arise due to a strong cusp-type flow in the region between the 
``stem'' of the mushroom and the roll up region.  
The choice $\delta=0.85$ produces a mesh with smallest cell size only approximately 3.8 times smaller 
than a uniform mesh cell.  Increasing the value of $\delta$ further produces a mesh with more non-convex elements, which in turn
 causes spurious errors in the computed numerical solution as shown in \Cref{fig:RT-non-convex-error}.  
 
A simple technique to resolve this issue is to use the large zoom-in algorithm described in \Cref{subsec:large-zoom-in}.  Specifically, we use 
the large zoom-in algorithm (with 25 sub time steps) in combination with restarted dynamic SAM.  The $64 \times 128$ adaptive mesh 
with $\delta = 0.97$ is shown in \Cref{fig:RT-mesh-large-zoom-in}, from which it can be seen that the mesh is smooth and all elements are 
convex. The smallest cell size in the mesh is approximately 13 times smaller than a uniform cell.  
The large zoom-in algorithm is applied only when the mesh resets, and the increase in CPU runtime is therefore 
negligible. 

\subsubsection{Comparison of adaptive and uniform simulations}

\begin{figure}[ht]
\centering
\subfigure[Adaptive mesh $\mathcal{T}$]{\label{fig:RT-adaptive-mesh}\includegraphics[width=42.2mm]{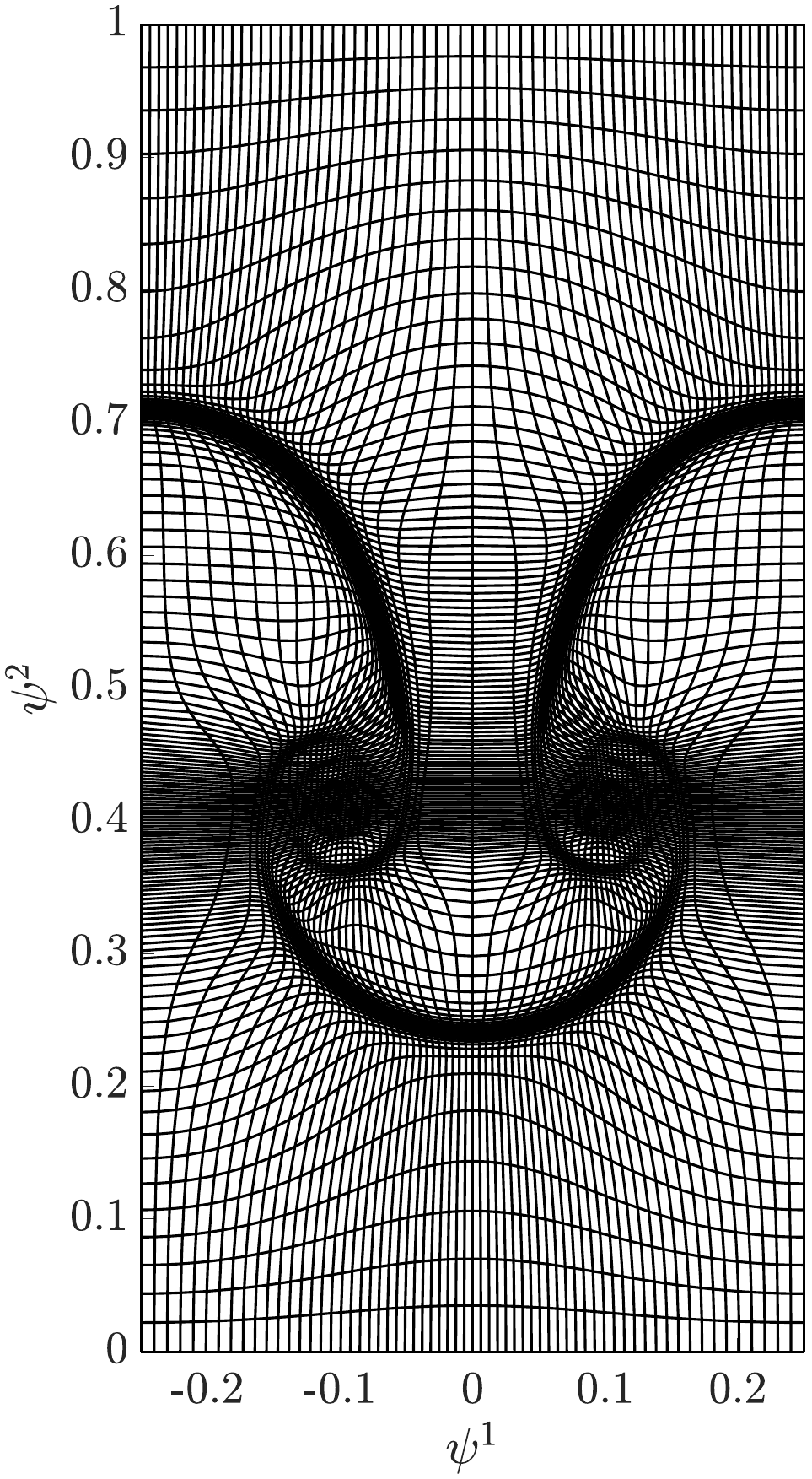}} 
\hspace{0em}
\subfigure[Density $\rho$ on $\mathcal{T}$]{\label{fig:RT-adaptive-rho}\includegraphics[width=35mm]{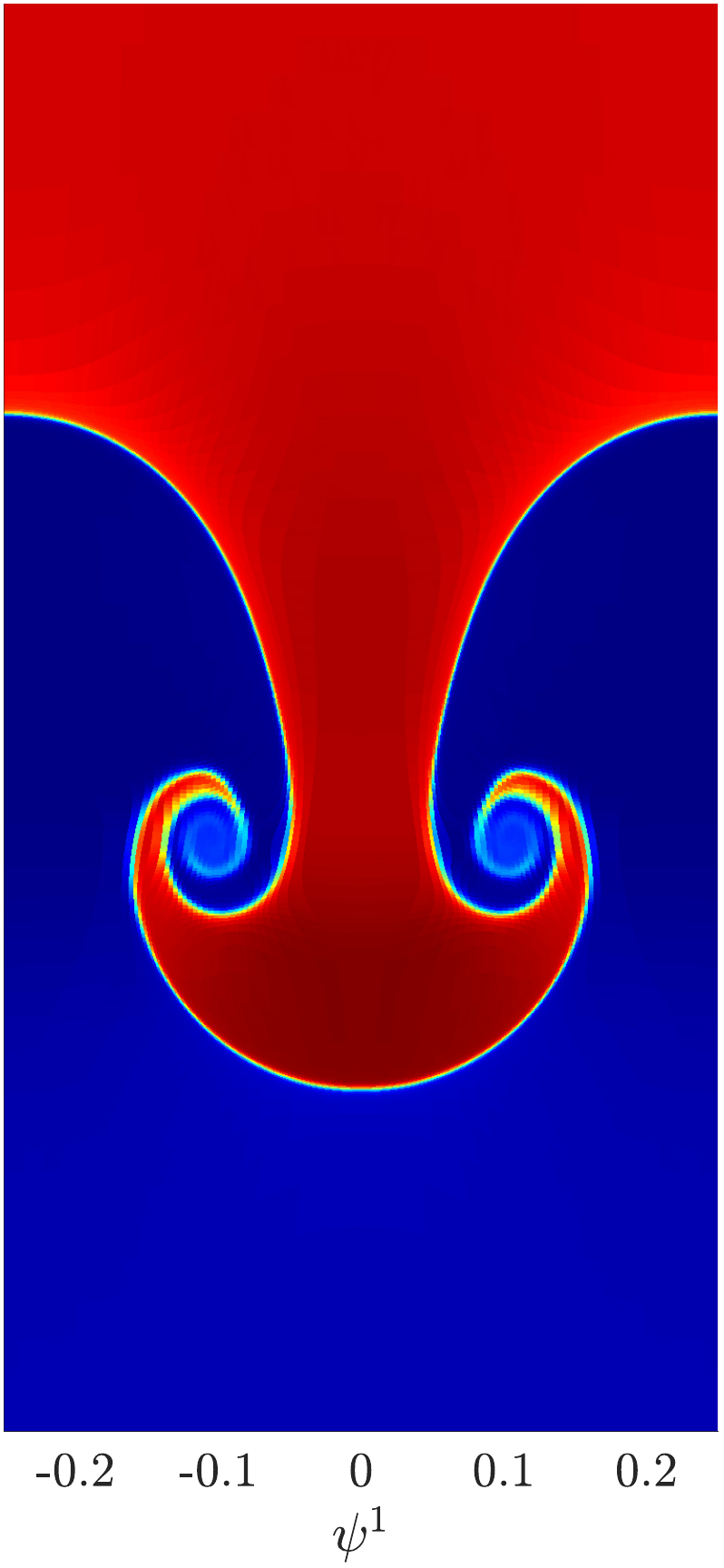}} 
\hspace{0em}
\subfigure[Interface $\ze$]{\label{fig:RT-adaptive-z}\includegraphics[width=35mm]{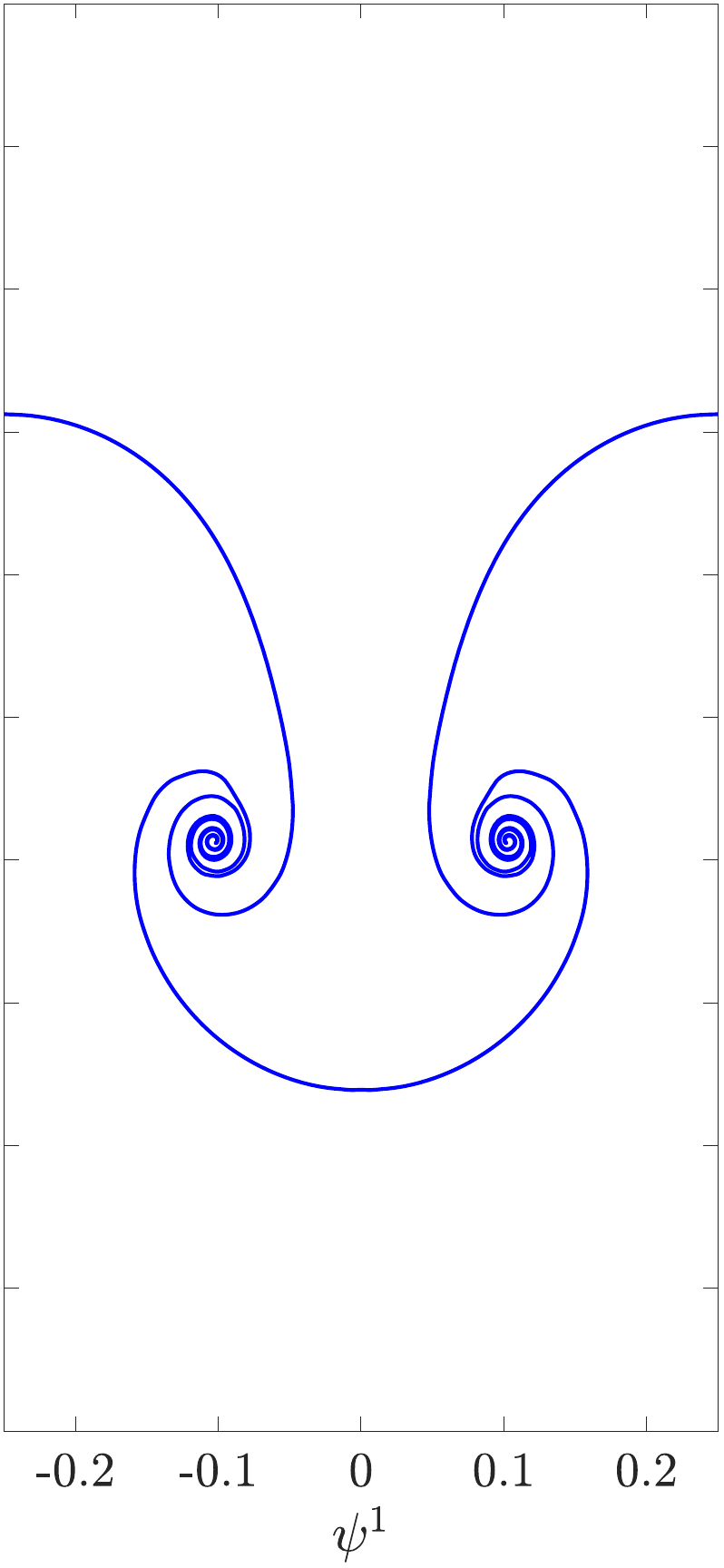}} 
\hspace{0em}
\subfigure[ALE interface $z$]{\label{fig:RT-adaptive-z-ALE}\raisebox{0.45mm}{\includegraphics[width=35mm]{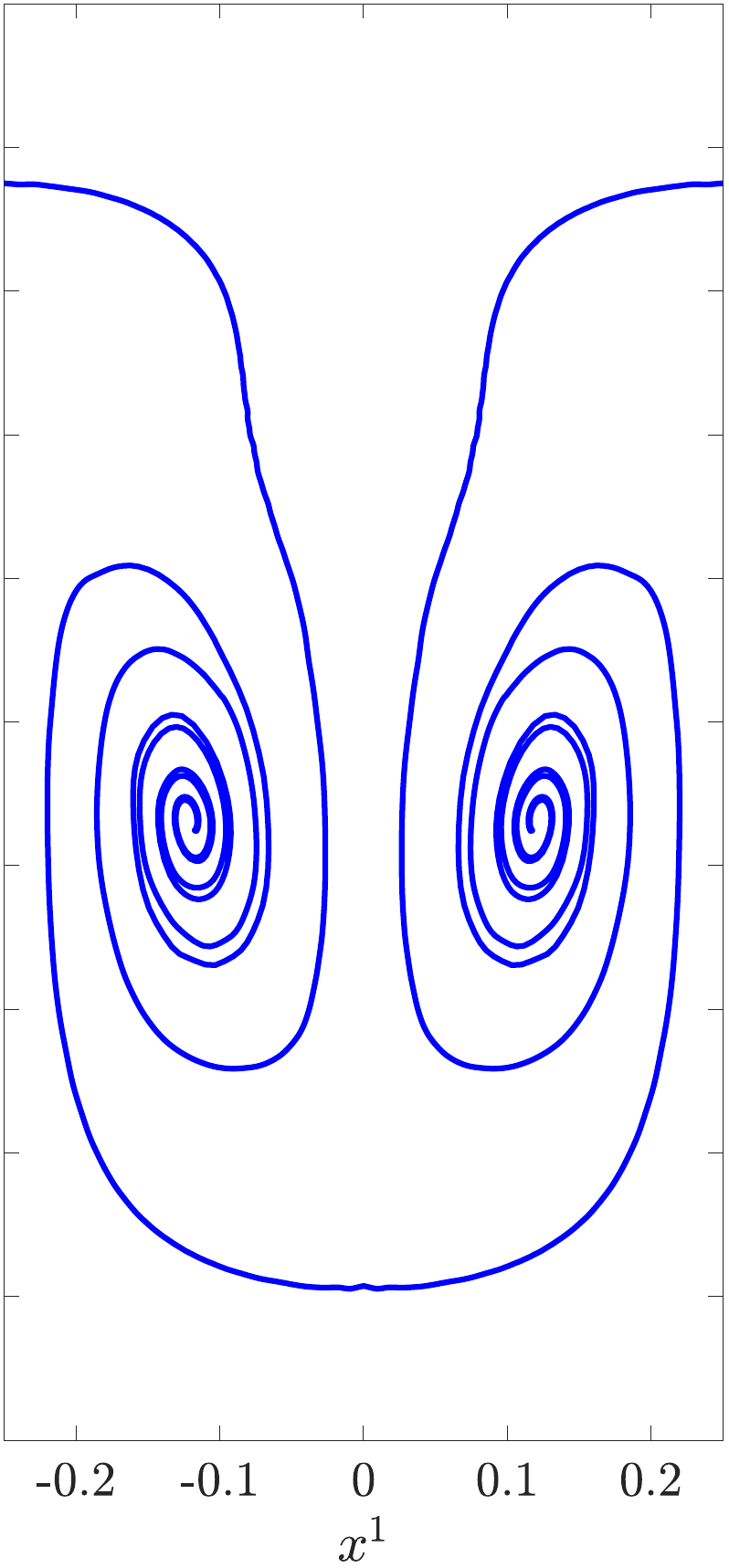}}} 
\caption{$64 \times 128$ SAM-ALE simulation of RT instability with $\delta = 0.97$.}
\label{fig:RT-adaptive}
\end{figure}

We perform a $64 \times 128$ cell SAM-ALE simulation with zoom-in parameter $\delta = 0.97$ and  $\Delta t = 1.5625 \times 10^{-4}$. 
Plots of the adaptive mesh and density heatmap  are provided in  \Cref{fig:RT-adaptive-mesh} and  \Cref{fig:RT-adaptive-rho}, and we refer to 
\Cref{fig:RT-mesh-large-zoom-in} for the mesh zoom-in. A comparison with the uniform mesh simulations in 
\Cref{fig:RT-uniform-comparison} shows that the $64 \times 128$ SAM-ALE simulation has a much sharper interface and exhibits more 
small-scale roll-up than the $64 \times 128$ uniform simulation, and is roughly comparable to the $N = 256 \times 512$ simulation. 
However, some of the small-scale structure is not observed in the SAM-ALE density.  
Interestingly, this roll up is captured by the interface $\ze$, shown 
in \Cref{fig:RT-adaptive-z}.  This suggests that a more robust ALE solver (e.g. WENO with alternative flux formulation) may produce 
improved results\footnote{See also \cite{DaZe2010} for a comparison of Lax-Friedrichs vs low dissipation 
HLLC flux reconstruction in the FV framework.}.  
The ALE interface $z$ is shown in \Cref{fig:RT-adaptive-z-ALE} and is clearly a zoomed-in version of $\ze$, with 
the small scale KH zones magnified and represented over a much larger region.  

\begin{table}[ht]
\centering
\renewcommand{\arraystretch}{1.0}
\scalebox{0.8}{
\begin{tabular}{|r|ccccc|}
\toprule
\midrule
\multirow{2}{*}{\textbf{Runtime (sec)}}  & \multicolumn{5}{c|}{\textbf{Cells}}\\

 & $64 \times 128$   & $128 \times 256$    & $256 \times 512$   & $512 \times 1024 $ & $64 \times 128$ SAM-ALE \\
\midrule
 $T_{\mathrm{CPU}}$ & 
$2.21 \times 10^1$  & $1.67 \times 10^2$  & $1.37 \times 10^3$ & $1.21 \times 10^4$  & $1.38 \times 10^2$ \\
\midrule
\bottomrule
\end{tabular}}
\caption{Total CPU runtime for uniform and adaptive simulations of RT instability.}
\label{table:RT-runtime}
\end{table}

The CPU runtimes of the various simulations 
are provided in \Cref{table:RT-runtime}, from which we see that the SAM-ALE simulation is approximately 10 times and 88 times faster than 
the $256 \times 512$ and $512 \times 1024$ uniform runs, respectively. For this problem, the CPU time spent on mesh generation is 
roughly the same as the time spent on ALE calculations.  Since SAM is roughly 100-200 times faster than MK mesh generation, it is clear that 
an MK-ALE scheme cannot provide a speed-up over uniform mesh simulations.  On the other hand, the use of a more robust ALE 
solver can only improve the relative efficiency of SAM-ALE, since the main computational expense will be the ALE calculations rather than 
mesh generation.  

\begin{figure}[ht]
\centering
\subfigure[$ ||\omega(\cdot,t)||_{L^2}$ vs $t$]{\label{fig:RT-curlu_L2}\includegraphics[width=65mm]{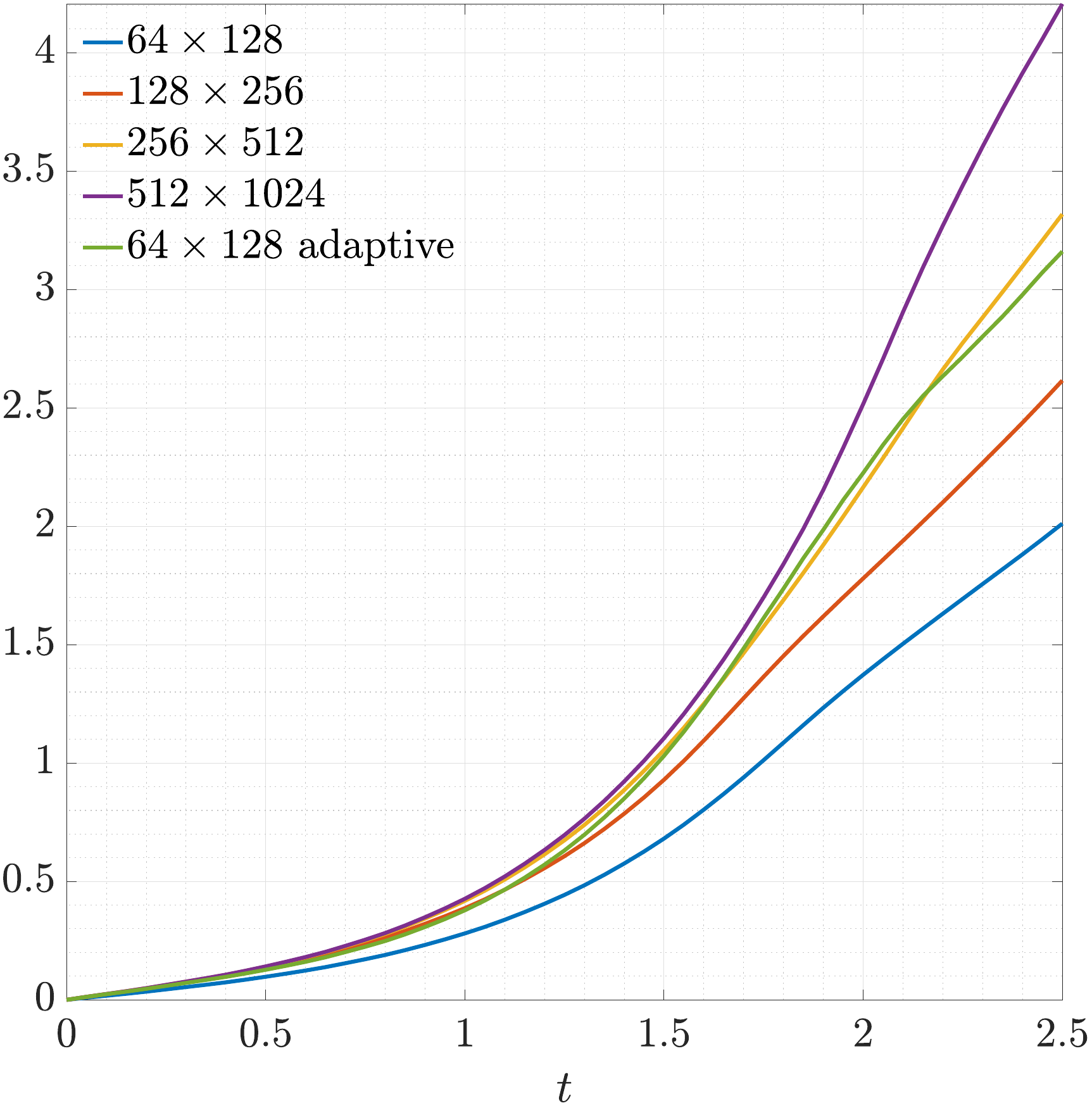}} 
\hspace{4em}
\subfigure[$|| \omega(\cdot,t) ||_{L^\infty}$ vs $t$]{\label{fig:RT-curlu_Linf}\includegraphics[width=65mm]{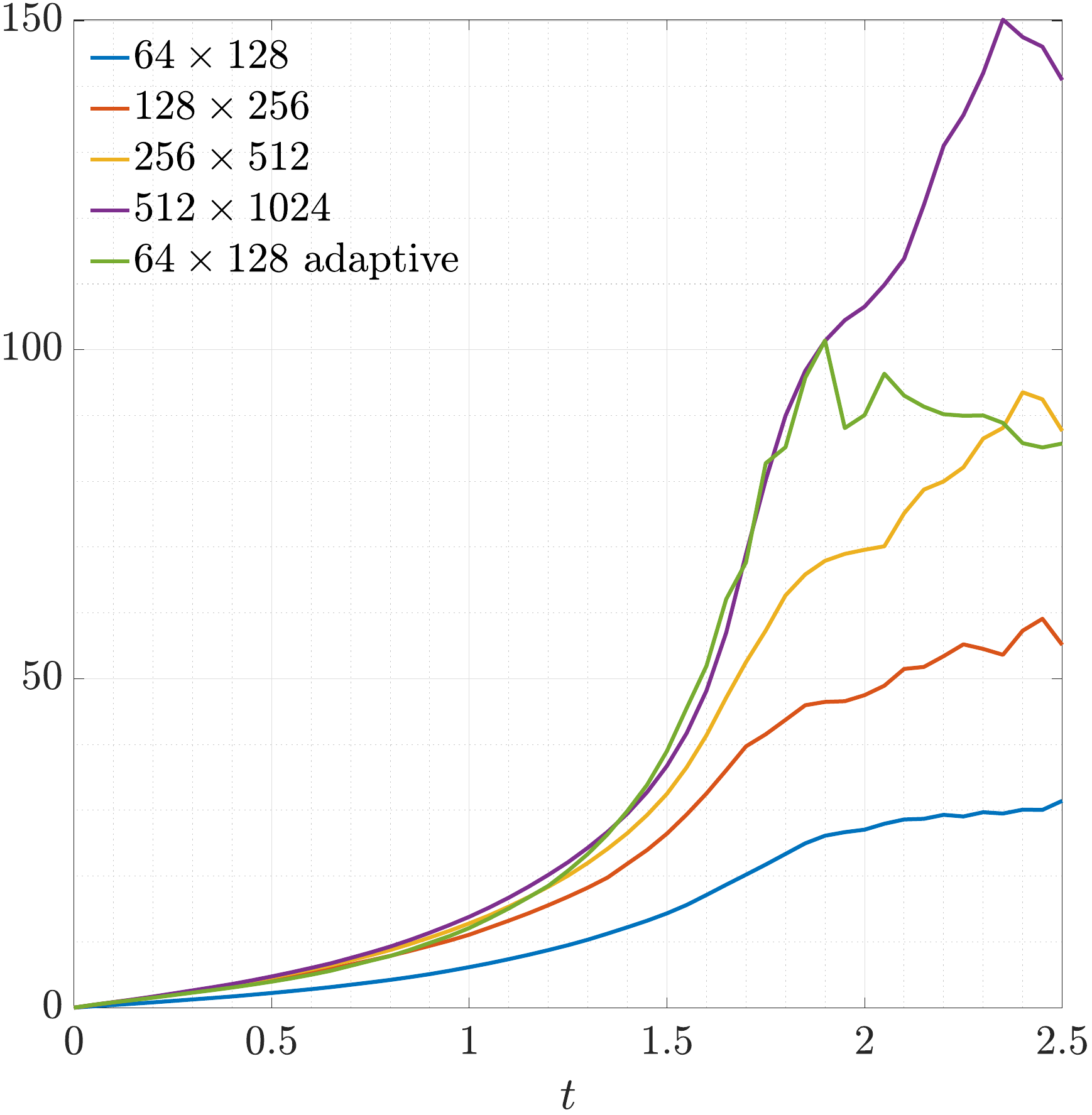}} 
\caption{Time history of the $L^2$ and $L^\infty$ norms of the vorticity for uniform and adaptive mesh simulations of RT instability.}
\label{fig:RT-curlu}
\end{figure}

The time histories of the $L^2$ and $L^\infty$ norms of the vorticity $\omega$ for the uniform and adaptive mesh simulations 
are shown in \Cref{fig:RT-curlu}. These figures confirm that the $64 \times 128$ SAM-ALE run is comparable 
to the $256 \times 512$ uniform run.  In fact, for $t \leq 1.75$, when the mesh zoom-in factor is 
approximately 20 times, the $64 \times 128$ SAM-ALE run closely 
approximates the $512 \times 1024$  uniform run.  For $t > 1.75$, the mesh zoom-in factor decreases due to the stretching of the 
interface and the adaptive mesh is no longer able to capture the smallest scales that are present in the $512 \times 1024$ run. 
The decrease in the mesh zoom-in factor is a consequence of the fact that the number of cells in the mesh are fixed. So-called 
$h$-$r$ adaptive mesh methods \cite{DoKnKoMiKeTo2021} are a way to overcome this issue; the simplicity of our 
algorithmic framework suggests that a dynamic 
$h$-$r$ method based on SAM can be readily formulated and implemented, and this will be investigated in future work.  

\section{Concluding remarks}\label{sec:conclusion}
In this work, we developed a new Smooth Adaptive Meshing (SAM) algorithm based on a new perturbation formulation and 
implementation of the deformation method.  The resulting numerical algorithm is simple, stable, automated, high-order accurate, and 
able to generate smooth and untangled meshes resolving around complex multi-$D$ flows.  We coupled SAM to a simple ALE 
scheme for gas dynamics and presented adaptive-simulation speed-up results for the challenging Noh and 
Rayleigh-Taylor problems. 

\begin{figure}[ht]
\centering
\includegraphics[width=100mm]{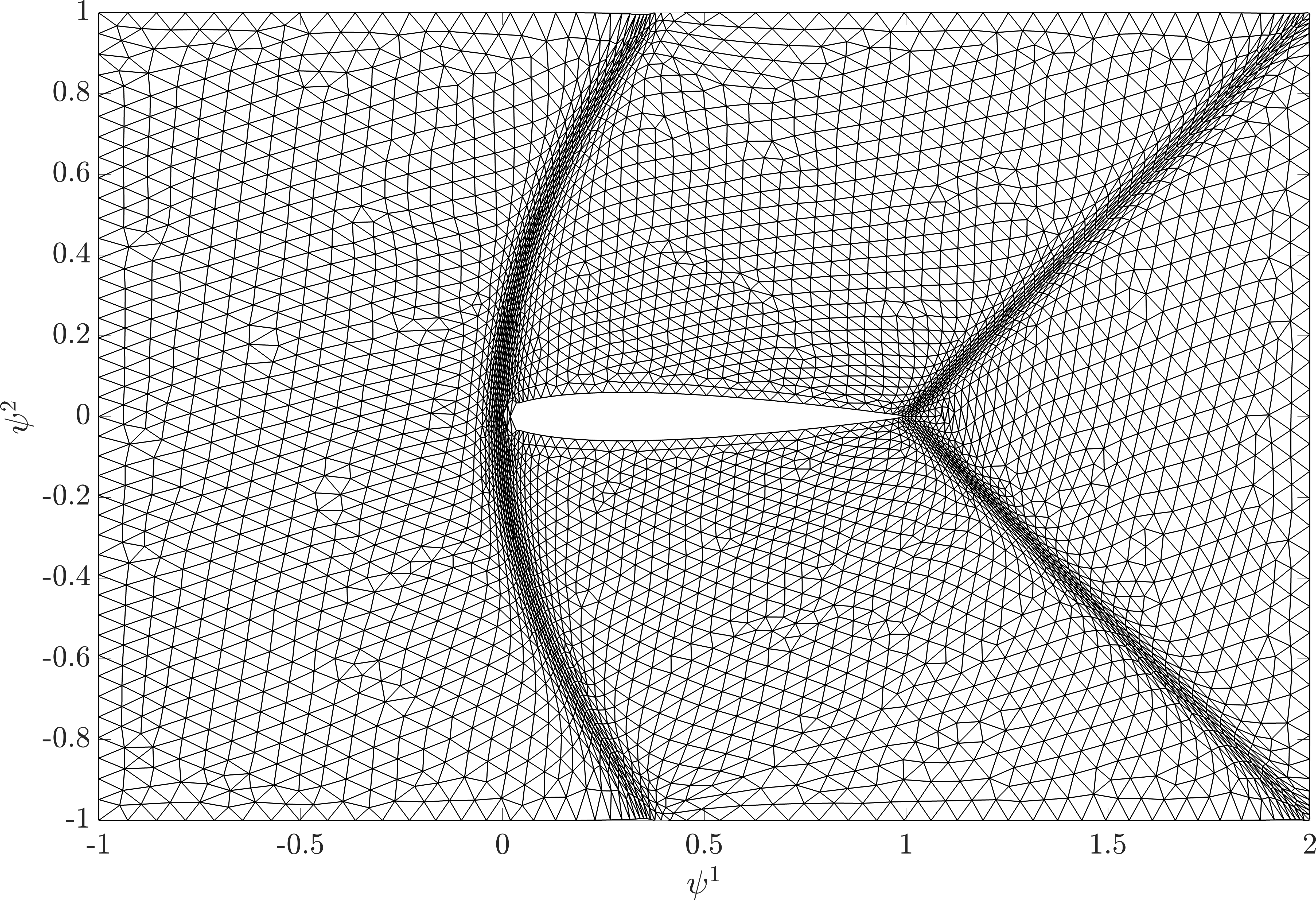}
\caption{Unstructured mesh modeling compressible flow past an airfoil. The mesh is constructed using (a preliminary version of) 
unstructured SAM in the Firedrake framework \cite{Firedrake}.}
\label{fig:SAM-unstructured}
\end{figure}

Several aspects of our SAM formulation and algorithm require further investigation and improvement.  As discussed in 
\Cref{subsec:RT}, we are interested in developing an $h$-$r$-refinement scheme based on SAM and, more generally, a dynamic 
SAM algorithm on general unstructured meshes.  The numerical implementation of unstructured SAM is obviously more delicate than 
the simple uniform-mesh scheme presented in the current paper, and will be thoroughly 
investigated in future work.  Nonetheless, we provide in \Cref{fig:SAM-unstructured}
a preliminary result showing an unstructured SAM mesh that models compressible flow past an airfoil. 
This mesh was produced\footnote{We express our gratitude to Dr. Mariana Clare for her assistance with writing the code and generating 
the result shown in \Cref{fig:SAM-unstructured}.} within the finite-element based Firedrake code \mbox{\cite{Firedrake}}.  
In the future, we will investigate the theoretical properties of SAM solutions on general domains, and their connections to 
the regularity of the discrete mesh $\mathcal{T}$. 

%%% Acknowledgements
\subsubsection*{Acknowledgements} \addcontentsline{toc}{section}{Acknowledgements}
Research reported in this publication was supported by NSF grant DMS-2007606 and   DTRA grant HDTRA11810022.  This research was also
supported by  Defense Nuclear Nonproliferation, NA-22 and NA-24;
we note that the views of the authors do not necessarily reflect the views of the USG. This work was also supported by the Laboratory Directed Research and Development Program of the Los Alamos National Laboratory, which
is under the auspices of the National Nuclear Security Administration of the U.S. Department of Energy under
DOE Contracts W-7405-ENG-36 and LA-UR-10-04291.   

We would like to thank the UNM Center for Advanced Research Computing, supported in part by the National Science Foundation, for 
providing high performance computing resources used in this work.

We would like to express our gratitude to the anonymous referees for their numerous suggestions that have greatly improved the manuscript.

%%% Appendices

\begin{appendices}\label{sec-appendices}

\section{The $C$-method for 2$D$ ALE-Euler}\label{appendix:C-method}

We provide a brief review of the $C$-method for adding space-time smooth 
artificial viscosity to shocks and contacts \cite{RaReSh2019b}. The most important 
feature of the $C$-method is smooth tracking of shock/contact fronts and their geometries 
via so-called $C$-functions.  The $C$-functions are space-time smoothed versions of 
localized solution gradients, and are found as the solutions to auxiliary scalar reaction-diffusion equations.  
Specifically, we use ${C}$ to denote a smoothed 
shock tracking function, and ${\tau}$ to denote the vector-valued function 
${\tau} = (\tau^1 \,, \tau^2 )$. The 
function $\vec{\tau}$ is a smoothed version of the tangent vector to an evolving 
contact discontinuity.  These $C$-functions allow us to implement both directionally 
isotropic (for shock stabilization) and anisotropic (for contact stabilization) artificial 
viscosity schemes. 

To summarize the method, it is convenient to introduce 
advection, artificial viscosity, and $C$-equation operators as follows. 

\subsubsection{ALE advection operator}
For a scalar function $Q : \Omegaref \to \mathbb{R}$, and a vector-valued 
function $v : \Omegaref \to \mathbb{R}^2$, define 
\begin{equation}\label{opALE-adv}
\mathscr{A} \left[ Q \,; v \right] \coloneqq \partial_k \left( Q a_l^k v^l \right) \,.
\end{equation}

\subsubsection{ALE isotropic artificial viscosity operator}
For a scalar function $Q : \Omegaref \to \mathbb{R}$, define
\begin{equation}\label{opALE-diff-isotropic}
	\mathscr{D} \left[ Q \,; \beta  \right] \coloneqq  \partial_k \left( \tilde{\beta} \rho {C} \, a_i^k  a_i^l  \, \partial_l Q  \right)  \,,
\end{equation}
with
$$
\tilde \beta = \frac{  |\Delta x|^2 }{\max {C}} \beta \,.
$$ 
The constant $\beta$ is an isotropic artificial viscosity parameter for shock 
stabilization.   

\subsubsection{ALE anisotropic artificial viscosity operator}
For a scalar function $Q: \Omegaref \to \mathbb{R}$, we define
	\begin{equation}\label{opALE-diff}
	\mathscr{D}^{\tau} \left[ Q \,; \mu  \right] \coloneqq  \partial_k \left[ \tilde{\mu} \rho \, {\tau^i} {\tau^j} a^k_i   a^l_j \,  \partial_l Q  \right]  \,,
	\end{equation}
	with 
	\begin{equation}\label{beta-ALE}
	\tilde{\mu} = \frac{ |\Delta x|^2  }{\alpha^2} \mu  \,.
	\end{equation}	
	Here, $\mu$ is the anisotropic artificial viscosity parameter for 
	contact discontinuity stabilization and  $\alpha = \max_x \{ |{\tau^1}| \,, |{\tau^2}| \}$.

\subsubsection{ALE $C$-equation operator}
For a scalar function $H : \Omegaref \to \mathbb{R}$ and 
scalar forcing function $Q : \Omegaref \to \mathbb{R}$, let
	\begin{equation}\label{opALE-Ceqn}
	\mathscr{L} \left[H   \,; Q  \right] \coloneqq  \frac{\mathcal{S}}{  \varepsilon | \Delta x| } \left(  Q - H  \right) + \kappa \mathcal{S}  |\Delta x| \Delta H \,.
	\end{equation}
  	
\subsubsection{The complete ALE-Euler-$C$ system}

Now, we can write the full ALE-Euler-$C$ system as 
\begin{subequations}
\label{euler2d-ALE2}
\begin{alignat}{3}
\partial_t (\mathcal{J} \rho ) + \mathscr{A}  \left[ \rho \,; u - \psi_t   \right] &= 0 \,, \label{euler2d-ALE2-eqn1} \\
\partial_t (\mathcal{J} \rho u^r) + \mathscr{A}   \left[ \rho u^r \,; u - \psi_t \right] &=  \mathscr{D}^{\tau} [  u^r \,; \mu ] +  \mathscr{D} [  u^r \, ; \beta_u ]   - \partial_{j} (a^j_r p)  \,,  \qquad && \text{for } r=1,2,  \label{euler2d-ALE2-eqn2}\\ 
\partial_t (\mathcal{J} E) +  \mathscr{A}   \left[ E \,; u - \psi_t \right] +  \mathscr{A}   \left[ p \,; u \right] &= \mathscr{D} [  E/\rho \, ; \beta_E ]  \,, \label{euler2d-ALE2-eqn3} \\
\partial_t \mathcal{J}  - \mathscr{A}  \left[ 1 \,; \psi_t   \right] &= 0 \,, \label{euler2d-ALE2-eqn4} \\
\partial_t C - \mathscr{L} \left[ C \,; F \right] &= 0  \,, \label{euler2d-ALE2-eqn6} \\ 
\partial_t {\tau^r} - \mathscr{L} \left[ {\tau^r} \,; F^r \right] &= 0 \,, && \text{for } r=1,2.  \label{euler2d-ALE2-eqn7}
\end{alignat}
\end{subequations}
The forcing functions for  \eqref{euler2d-ALE2-eqn6} and 
\eqref{euler2d-ALE2-eqn7} are defined as follows. The shock ${C}$ forcing function is 
given by
\begin{equation}\label{Cshock-forcing-ALE}
\hat{{F}} =  \frac{|\frac{1}{\mathcal{J}} a^j_i \partial_j \rho |}{\max |\frac{1}{\mathcal{J}} a^j_i \partial_j \rho|} \,,
\end{equation}
while the components of the forcing to the contact tangent vector ${\tau}$ equations are
defined by 
\begin{equation}\label{Ccont-forcing-ALE2}
{{F}}^{1} = - \frac{1}{\mathcal{J}} a^j_2 \partial_{j} \rho \qquad \text{and} \qquad  {{F}}^{2} = \frac{1}{\mathcal{J}} a^j_1 \partial_{j} \rho   \,.
\end{equation}
The initial conditions for $C$ and $\tau$ are defined by solving the time-independent versions of \eqref{euler2d-ALE2-eqn6} and 
\eqref{euler2d-ALE2-eqn7}.

\section{Boundary smoothing for non-Neumann functions}\label{sec-appendices-a}
Herein, we describe a simple boundary smoothing technique for non-Neumann functions. 
Let $x^r_{\mathrm{mid}} = \frac{1}{2} \left( x^r_{\mathrm{min}} + x^r_{\mathrm{max}} \right)$, for $r=1,2$. Define 
smooth cutoff functions
\begin{align*}
\phi^1(\xi) &= \frac{1}{2} \left[  \tanh \left( \frac{\xi - (x^1_{\mathrm{min}} + d_1)}{\varepsilon} \right) - \tanh \left( \frac{\xi - (x^1_{\mathrm{max}} - d_1)}{\varepsilon} \right)  \right] \,, \\
\phi^2(\eta) &= \frac{1}{2} \left[  \tanh \left( \frac{\eta - (x^2_{\mathrm{min}} + d_2)}{\varepsilon} \right) - \tanh \left( \frac{\eta - (x^2_{\mathrm{max}} - d_2)}{\varepsilon} \right)  \right] \,,
\end{align*}
where $\varepsilon$ is a smoothing parameter, which we choose as $\varepsilon = 0.02$.  The function $\phi^1$ is equal to 1 in the interior 
of the domain, then smoothly decreases to 0 at a distance $d_1$ near the left and right boundaries. The function $\phi^2$ behaves similarly. 
We set $d_r = 0.05(x^r_{\mathrm{max}} - x^r_{\mathrm{min}})$. 

Given a non-Neumann function $\Ge$, we first compute the derivatives $D_1 \Ge$, $D_2 \Ge$, and $D_{12} \Ge$. We then 
compute 
\begin{align*}
\mathcal{I}^{(1)}(y^1) &= \int_{x^1_{\mathrm{mid}}}^{y^1} \phi^1(\xi) D_1 \Ge(\xi,x^2_{\mathrm{mid}}) \,\mathrm{d}\xi \,, \\
\mathcal{I}^{(2)}(y^2) &= \int_{x^2_{\mathrm{mid}}}^{y^2} \phi^2(\eta) D_2 \Ge(x^1_{\mathrm{mid}},\eta) \,\mathrm{d}\eta \,, \\
\mathcal{I}^{(3)}(y^1,y^2) &= \int_{x^2_{\mathrm{mid}}}^{y^2} \int_{x^1_{\mathrm{mid}}}^{y^1} \phi^1(\xi) D_{12} \Ge(\xi,\eta) \,\mathrm{d}\xi \mathrm{d}\eta \,,
\end{align*}
and define 
$$
\Ge^*(y^1,y^2) \coloneqq \Ge(x^1_{\mathrm{mid}},x^2_{\mathrm{mid}}) + \mathcal{I}^{(1)}(y^1) + \mathcal{I}^{(2)}(y^2) + \mathcal{I}^{(3)}(y^1,y^2) \,.
$$
The function $\Ge^*$ then satisfies $D \Ge^* \cdot \nu = 0$ on $\partial \Omega$. 

\section{The MK scheme}\label{sec:MK}

The MK scheme solves for the 
unique \cite{Brenier1991,Caffarelli1990} diffeomorphism $\psi$ satisfying 
\eqref{MA-eqn} that minimizes the $L^2$ displacement  $|| \psi(x) - x ||_{L^2}$. The MK formulation is developed by writing 
$\psi = x + \nabla \Psi$, where $\Psi$ is a scalar potential. 
The equation governing $\Psi$ is found by minimizing a 
functional consisting of the $L^2$ displacement and a local Lagrange multiplier, where the latter is 
used to enforce the Jacobian constraint \eqref{MA-eqn}. 
The resulting equation for $\Psi$ is fully nonlinear, and the MK scheme uses an iterative 
Newton-Krylov solver with multigrid preconditioning to 
find an approximation to the solution $\Psi$, within some error tolerance $\epsilon$. 

\subsection{Machine comparison}

To reliably compare the runtimes of our static SAM \Cref{static-alg1} with the MK scheme as listed in \cite{Delzanno2008}, we need 
to account for the different machines on which these codes were run.  
Therefore, we perform the following machine comparison experiment. In \cite{Delzanno2008}, the authors also report the CPU runtimes for a 
deformation method of \citet{LiaoAnderson1992}, whose description is provided in the Appendix of \cite{Delzanno2008}. We coded a 
numerical implementation of this method, which we refer to as LA, and ran the numerical experiments from \cite{Delzanno2008} on 
our machine.  The runtimes for LA on our machine, along with the LA runtimes from Table 3 of \cite{Delzanno2008}, are shown in 
\Cref{table:machine-comparison}. 
These data show that our machine runs approximately 2.2 times faster than the machine on which 
the MK simulations in \cite{Delzanno2008} were performed.

\begin{table}[ht]
\centering
\renewcommand{\arraystretch}{1.0}
\scalebox{0.8}{
\begin{tabular}{|lr|ccccc|}
\toprule
\midrule
\multirow{2}{*}{\textbf{Scheme}} &  & \multicolumn{5}{c|}{\textbf{Cells}}\\

{}  & & $16 \times 16$   & $32 \times 32$    & $64 \times 64$   & $128 \times 128 $ & $256 \times 256$ \\
\midrule
{LA on \cite{Delzanno2008} machine} & $T_{\mathrm{CPU}}$ & 
$0.2$  & $0.9$  & $3.4$ & $13.6$  & $55.0$ \\
\midrule
\multirow{2}{*}{LA on our machine} &  $T_{\mathrm{CPU}}$ & 
$0.12$  & $0.41$  & $1.53$ & $6.22$  & $24.16$ \\
				    & speed-up factor & 1.7 & 2.2   & 2.2  & 2.2 & 2.3 \\
\midrule
\bottomrule
\end{tabular}}
\caption{CPU runtimes for the LA scheme on the machine from \cite{Delzanno2008} and the LA scheme on our machine. 
The data for the LA scheme in the top row is taken from Table 3 of \cite{Delzanno2008}.}
\label{table:machine-comparison}
\end{table}

\end{appendices}

\bibliography{references}

\end{document}